\documentclass[11pt]{article} 
\usepackage[utf8]{inputenc}
\usepackage{multirow}
\usepackage{amsfonts}
\usepackage{amsmath}
\usepackage{amssymb}
\usepackage{amsthm}
\usepackage{caption}
\usepackage[dvipsnames]{xcolor}
    \definecolor{darkgreen}{rgb}{0,0.5,0}
    \definecolor{darkblue}{rgb}{0,0,0.6}
    \definecolor{purple}{rgb}{0.4,.2,0.7}
\usepackage[margin = 2.5cm]{geometry}
\usepackage{graphicx}
\usepackage[hyperfootnotes = false, colorlinks = true, linkcolor = darkblue, citecolor = purple]{hyperref}
\usepackage{subcaption}
\usepackage{ytableau}

\usepackage{csquotes}

\usepackage[numbers,sort&compress]{natbib}
\usepackage{tikz}
\usetikzlibrary{decorations.pathmorphing}
\usetikzlibrary{decorations.markings}
\definecolor{mathred}{RGB}{180,44,37}
\definecolor{mathblue}{RGB}{39,94,190}
\tikzset{>=latex} 
\tikzset{ photon/.style={decorate, decoration={snake}, draw=black}}

\usepackage{comment}

\usepackage{physics}

\newcommand{\be}{\begin{equation}}
\newcommand{\ee}{\end{equation}}
\newcommand{\bea}{\begin{eqnarray}}
\newcommand{\eea}{\end{eqnarray}}

\def\tr{\mathrm{tr}}

\usepackage{tikz}
\newcommand{\fourpt}{%
\begin{tikzpicture}[baseline=-0.6ex,scale=0.5,every node/.style={font=\small},
  leg/.style={line width=0.9pt},
  rib/.style={line width=0.6pt}]
  \def\dy{0.09}
  \draw[rib] (0,\dy) -- (2.6,\dy); \draw[rib] (0,-\dy) -- (2.6,-\dy);
  \node[above=5pt] at (1.3,\dy) {$(N_1,s_1)$};
  \draw[leg] (0,0) -- (-1.0,1.0)  node[above left=-3pt] {$p_2$};
  \draw[leg] (0,0) -- (-1.0,-1.0) node[below left=-3pt] {$p_1$};
  \draw[leg] (2.6,0) -- (3.6,1.0)  node[above right=-3pt] {$p_3$};
  \draw[leg] (2.6,0) -- (3.6,-1.0) node[below right=-3pt] {$p_4$};
  \fill (0,0) circle (2.6pt); \fill (2.6,0) circle (2.6pt);
\end{tikzpicture}}

\newcommand{\sixpt}{%
\begin{tikzpicture}[baseline=-0.6ex,scale=0.5,every node/.style={font=\small},
  leg/.style={line width=0.9pt},
  rib/.style={line width=0.6pt}]
  \def\dy{0.09}
  \foreach \xa/\xb/\lab in {0/3.4/{$(N_1,s_1)$},3.4/6.8/{$(N_2,s_2)$},6.8/10.2/{$(N_1,s_1)$}}{
    \draw[rib] (\xa,\dy) -- (\xb,\dy); \draw[rib] (\xa,-\dy) -- (\xb,-\dy);
    \node[above=5pt] at ({(\xa+\xb)/2},\dy) {\lab};
  }
  \draw[leg] (0,0) -- (-1.0,1.0)  node[above left=-3pt] {$p_2$};
  \draw[leg] (0,0) -- (-1.0,-1.0) node[below left=-3pt] {$p_1$};
  \draw[leg] (3.4,0) -- (3.4,1.8)  node[above] {$p_3$};
  \draw[leg] (6.8,0) -- (6.8,-1.8) node[below] {$p_4$};
  \draw[leg] (10.2,0) -- (11.2,1.0)  node[above right=-3pt] {$p_5$};
  \draw[leg] (10.2,0) -- (11.2,-1.0) node[below right=-3pt] {$p_6$};
  \foreach \x in {0,3.4,6.8,10.2}{ \fill (\x,0) circle (2.6pt); }
\end{tikzpicture}}

\newcommand{\vertexA}{%
\begin{tikzpicture}[baseline=-0.6ex,scale=0.5,every node/.style={font=\small},
  leg/.style={line width=0.9pt},
  rib/.style={line width=0.6pt}]
  \def\dy{0.09}
  \draw[rib] (0,\dy) -- (4.0,\dy); \draw[rib] (0,-\dy) -- (4.0,-\dy);
  \draw[rib] (4.0,\dy) -- (8.0,\dy); \draw[rib] (4.0,-\dy) -- (8.0,-\dy);
  \node[above=5pt] at (1.9,\dy) {$(N_1,s_1)$};
  \node[above=5pt] at (6.1,\dy) {$(N_2,s_2)$};
  \node[left=2pt] at (0,0) {$i$};
  \node[right=2pt] at (8.0,0) {$j$};
  \draw[leg] (4.0,0) -- (4.0,1.8);
  \fill (4.0,0) circle (2.6pt);
\end{tikzpicture}}

\newcommand{\vertexB}{%
\begin{tikzpicture}[baseline=-0.6ex,scale=0.5,every node/.style={font=\small},
  leg/.style={line width=0.9pt},
  rib/.style={line width=0.6pt}]
  \def\dy{0.09}
  \draw[rib] (0,\dy) -- (4.0,\dy); \draw[rib] (0,-\dy) -- (4.0,-\dy);
  \draw[rib] (4.0,\dy) -- (8.0,\dy); \draw[rib] (4.0,-\dy) -- (8.0,-\dy);
  \node[above=5pt] at (1.9,\dy) {$(N_2,s_2)$};
  \node[above=5pt] at (6.1,\dy) {$(N_1,s_1)$};
  \node[left=2pt] at (0,0) {$i$};
  \node[right=2pt] at (8.0,0) {$j$};
  \draw[leg] (4.0,0) -- (4.0,-1.8);
  \fill (4.0,0) circle (2.6pt);
\end{tikzpicture}}

\newcommand{\eightpt}{%
\begin{tikzpicture}[baseline=-0.6ex,scale=0.5,every node/.style={font=\small},
  leg/.style={line width=0.9pt},
  rib/.style={line width=0.6pt}]
  \def\dy{0.09}
  \foreach \xa/\xb/\lab in {0/3.2/{$(N_1,s_1)$},3.2/6.4/{$(N_2,s_2)$},6.4/9.6/{$(N_1,s_1)$},9.6/12.8/{$(N_2,s_2)$},12.8/16.0/{$(N_1,s_1)$}}{
    \draw[rib] (\xa,\dy) -- (\xb,\dy); \draw[rib] (\xa,-\dy) -- (\xb,-\dy);
    \node[above=5pt] at ({(\xa+\xb)/2},\dy) {\lab};
  }
  \draw[leg] (0,0) -- (-1.0,1.0)  node[above left=-3pt] {$p_2$};
  \draw[leg] (0,0) -- (-1.0,-1.0) node[below left=-3pt] {$p_1$};
  \draw[leg] (3.2,0)  -- (3.2,1.8)   node[above] {$p_3$};
  \draw[leg] (6.4,0)  -- (6.4,-1.8)  node[below] {$p_4$};
  \draw[leg] (9.6,0)  -- (9.6,1.8)   node[above] {$p_5$};
  \draw[leg] (12.8,0) -- (12.8,-1.8) node[below] {$p_6$};
  \draw[leg] (16.0,0) -- (17.0,1.0)  node[above right=-3pt] {$p_7$};
  \draw[leg] (16.0,0) -- (17.0,-1.0) node[below right=-3pt] {$p_8$};
  \foreach \x in {0,3.2,6.4,9.6,12.8,16.0}{ \fill (\x,0) circle (2.6pt); }
\end{tikzpicture}}

\usepackage{tikz}
\usetikzlibrary{arrows.meta,decorations.markings}
\tikzset{
  hl/.style={baseline={([yshift=-0.6ex]current bounding box.center)}, line width=0.6pt, font=\scriptsize},
  leg/.style={postaction={decorate}, decoration={markings, mark=at position 0.6 with {\arrow{Stealth[length=4pt]}}}},
  boson/.style={double, double distance=1.6pt},
  vtx/.style={circle, fill, inner sep=1.1pt},
}
\newcommand{\halfladderfive}{%
\begin{tikzpicture}[hl]
  \coordinate (L) at (0,0); \coordinate (M) at (1.3,0); \coordinate (R) at (2.6,0);
  \draw[boson] (L) -- node[above] {$s_1$} (M);
  \draw[boson] (M) -- node[above] {$s_2$} (R);
  \draw[leg] (L) -- (-0.8,-0.55) node[below left, inner sep=1pt] {$p_1$};
  \draw[leg] (L) -- (-0.8, 0.55) node[above left, inner sep=1pt] {$p_2$};
  \draw[leg] (M) -- (1.3,0.8)    node[above, inner sep=1.5pt]   {$p_3$};
  \draw[leg] (R) -- (3.4, 0.55)  node[above right, inner sep=1pt] {$p_4$};
  \draw[leg] (R) -- (3.4,-0.55)  node[below right, inner sep=1pt] {$p_5$};
  \node[vtx] at (L) {}; \node[vtx] at (M) {}; \node[vtx] at (R) {};
  \node[below=2pt] at (M) {$p$};
\end{tikzpicture}}

\newcommand{\halfladderfour}{%
\begin{tikzpicture}[hl]
  \coordinate (L) at (0,0); \coordinate (R) at (1.4,0);
  \draw[boson] (L) -- node[above] {$s$} (R);
  \draw[leg] (L) -- (-0.8,-0.55) node[below left, inner sep=1pt] {$p_1$};
  \draw[leg] (L) -- (-0.8, 0.55) node[above left, inner sep=1pt] {$p_2$};
  \draw[leg] (R) -- (2.2, 0.55)  node[above right, inner sep=1pt] {$p_3$};
  \draw[leg] (R) -- (2.2,-0.55)  node[below right, inner sep=1pt] {$p_4$};
  \node[vtx] at (L) {}; \node[vtx] at (R) {};
\end{tikzpicture}}

\begin{document}

\thispagestyle{empty}
\begin{center}
    ~\vspace{5mm}

  \vskip 2cm 
  
   {\LARGE \bf 
    Half-ladder partial waves for all $n,d$ with applications to string theory
   }

   \vspace{0.5in}
     
   {\bf Jeffrey V. Backus\footnote{\tt jvabackus@princeton.edu}
   }

    \vspace{0.5in}

    Joseph Henry Laboratories, Princeton University, Princeton, NJ 08544, USA
                
    \vspace{0.5in}

    \vspace{0.5in}

\end{center}

\vspace{0.5in}

\begin{abstract} 
Given any four-point scattering amplitude, the partial wave expansion resolves its residues into contributions from individual spin-$s$ internal particles, where each coefficient in the expansion must be non-negative due to unitarity. In this paper, we introduce a partial wave expansion tailored to ``half-ladder'' Feynman diagrams at generic multiplicity $n$ and spacetime dimension $d$, with scalars on the external lines and massive bosons on the internal lines. Derived by enforcing that the half-ladder residue is an eigenvector of each internal particle's angular momentum, each basis element is written as a sum over products of special functions related to the Gegenbauer polynomials, and we give a set of simple graphical rules for explicitly building any of them at generic $n$ and $d$. To put the formalism into action, we apply our technology to deformed ``string theory'' amplitudes, defined \textit{a priori} in any dimension and with arbitrary Regge intercept $\alpha_0$. By constructing multi-particle positivity tests for the half-ladder and ``twisted'' half-ladder diagrams, we show that the only amplitudes consistent with unitarity are those belonging to open bosonic string theory ($\alpha_0 = -1$) below the critical dimension $d \leq 26$. In particular, we confirm the expectation that $Z$-theory ($\alpha_0 = 0$) fails to define a unitary string theory at finite $\alpha'$. Finally, for the open bosonic string, we show how knowledge of the partial wave expansion allows us to resolve the contributions to the half-ladder diagram from individual degenerate states at a given level and spin in a purely on-shell analysis. As an ancillary file, we include a Python script \texttt{combwaves.py} which implements the partial wave expansion given a generic (twisted) half-ladder residue.

\end{abstract}

\vspace{1in}

\pagebreak

\setcounter{tocdepth}{3}
{\hypersetup{linkcolor=black}\tableofcontents}

\section{Introduction}\label{sec:intro}

Imagine a generic $2 \to 2$ planar scattering amplitude $\mathcal{A}_4(s,t)$ with external scalars of mass $m$ in generic dimension $d$. Now, consider setting the internal particle of the $s$-channel diagram on-shell so that $s = -M^2$, and then taking the residue of $\mathcal{A}_4(s,t)$. The \textit{partial wave expansion} is an orthogonal basis choice of what comes out of this process, where each basis element corresponds to an intermediate particle with spin $J$~\cite{Jacob:1959at}:
\begin{equation}
    \Res_{s = -M^2} \mathcal{A}_4(s,t) = \sum_{J = 0}^\infty y_J C_J^{\left( \frac{d-3}{2}\right)}\left( 1 - \frac{2t}{M^2 - 4m^2 }\right)
\end{equation}
Above, the $y_J$ are the partial wave coefficients, and $C_s^{\left( \frac{d-3}{2}\right)}$ is the \textit{Gegenbauer} polynomial.

Throughout the history of quantum mechanics, this particular expansion has proved to be a crucial tool in gaining a deeper understanding of the physics of scattering. It was first used in a non-relativistic setting in the theory of phase shifts, the effective-range expansion, and the optical theorem~\cite{Faxen:1927,Bethe:1949yr}. As a practical engine, it was (and is) critical to collider experiments--- allowing us to extract resonances, spins, and parities from production and scattering data~\cite{Peters:2004qw,Stoks:1993tb}. It was paramount to the development of Regge theory~\cite{Regge:1959mz,Gribov:1961fr,Collins:1977jy} and to the Froissart-Martin bound~\cite{Froissart:1961ux,Martin:1962rt}, and, by expressing the amplitude in this basis, one can place perturbative unitarity bounds on $2 \to 2$ scattering, such as the Lee-Quigg-Thacker bound on the maximal mass of the Higgs boson~\cite{Lee:1977eg,Lee:1977yc}.

It is therefore perhaps no surprise that the partial wave expansion continues to find itself, now in the $21^{\mathrm{st}}$ century, to be of importance to the study of scattering amplitudes. In its current incarnation, it is a crucial part of the modern S-matrix bootstrap program~\cite{Paulos:2017fhb,Correia:2020xtr,Kruczenski:2022lot}, where it is used as an input for dispersive bounds on Effective Field Theories (EFTs)~\cite{Adams:2006sv,deRham:2017avq,Arkani-Hamed:2020blm,deRham:2022hpx}. This has led to many far-reaching results--- including bounds (sometimes two-sided) on certain Wilson coefficients~\cite{Caron-Huot:2020cmc,Tolley:2020gtv,Sinha:2020win,Chiang:2021ziz,Bellazzini:2020cot}, bounds on higher-curvature couplings in gravity~\cite{Cheung:2016wjt,Hamada:2018dde,Bern:2021ppb,Caron-Huot:2021rmr}, consistency regions in massive gravity theories~\cite{Cheung:2016yqr} and inflationary EFTs~\cite{Baumann:2015nta}, positivity constraints on SMEFT operators~\cite{Zhang:2018shp,Remmen:2019cyz}, and the string theory bootstrap~\cite{Huang:2020nqy,Berman:2023jys,Haring:2023zwu,Albert:2024yap,Berman:2024wyt,Cheung:2024uhn,Cheung:2025tbr,Elvang:2026pmc}.

One limiting factor of the analysis has been its restriction to $2 \to 2$ scattering. In recent years, there has been a push to include --- in the bootstrap and beyond --- constraints from higher-point amplitudes. Work has approached this question from a variety of directions, with authors exploring: the positivity of higher-point contact operators in the forward limit~\cite{Chandrasekaran:2018qmx}; sum rules from consistent factorization of higher-point residues~\cite{Arkani-Hamed:2023jwn,Basile:2026gnd,Elvang:2026pmc,Berman:2026ezk}; multi-positivity bounds in soft and complex-forward kinematics~\cite{Cheung:2025nhw,Cheung:2026lpv,Jeong:2026xzk}; ``splitting'' constraints at five-points~\cite{Berman:2025owb}; positivity of six-point Wilson coefficients from tree-level UV completions~\cite{Kremminger:2026edr}; and explicit five-point partial waves in arbitrary dimension $d$~\cite{Saha:2026ftv}. All of these results are also predated by much earlier expeditions into helicity expansions of multi-Regge theory~\cite{Toller:1969vt,Brower:1974yv}. However, one missing component in this story has been the analog of the partial wave expansion for $2 \to 2$ scattering extended to generic multiplicity $n$. This is the topic we pursue in the present manuscript.

Note that it is, of course, always possible to expand the amplitude (and, for that matter, \textit{any} function) in a random spanning basis --- even an orthogonal one --- consisting of some generic functions of the kinematic invariants. So, at first glance it may seem like there is not much to learn from a partial wave expansion at higher-points: you can just pick any basis you like! However, recall that the key to the partial wave basis at four-points is that each element corresponds precisely to an internal mode of definite spin. Thus, in extending the construction to all $n$, we are after a basis which does the same thing: an orthogonal set of functions for an $n$-point tree-level Feynman diagram with all internal lines on-shell, in arbitrary spacetime dimension, designed specifically to resolve the spin of every internal mode as well as the structure of every internal vertex. This means that, given any $n$-point Feynman diagram as a polynomial in dot products of external momenta, one can proceed via orthogonality and determine exactly how much each internal configuration contributes to the total, just as we do at four-points with the classical partial wave expansion.

A few things make the general-$n$ situation different from the standard $2 \to 2$ case. For $2 \to 2$ scattering, each basis element in the partial wave expansion is given by sewing together two three-point amplitudes, both of which are uniquely known from Lorentz invariance:
\begin{equation}
R^{(4)}_{s} = \halfladderfour
  = \sum_{I}\mathcal A(1,2,\epsilon_{I})\,\mathcal A(\epsilon_{I},3,4).
\end{equation}
Upon performing the sum over $I$, one derives that each basis element is a Gegenbauer polynomial labeled by the internal spin $s$. However, starting at five-points, there is a \textit{new} type of vertex, buried inside of the five-point Feynman diagram, connecting two particles of spin $s_1$ and $s_2$ to the external particle (which we take to be a scalar):
\begin{equation}
R^{(5)}_{s_1,s_2,b} = \halfladderfive
  = \sum_{I_1,I_2}\mathcal A(1,2,\epsilon_{I_1})\,\mathcal A(\epsilon_{I_1},3,\epsilon_{I_2})\,\mathcal A(\epsilon_{I_2},4,5),
\end{equation}
where the middle vertex ranges over $\min(s_1,s_2) + 1$ allowed Lorentz-invariant structures:
\begin{equation}\label{eq:int-3-particle}
\mathcal A_3(\epsilon_{I_1},3,\epsilon_{I_2})
 = i\sum_{b=0}^{\min(s_1,s_2)} g_b\,
   (\epsilon_{I_1}\!\cdot\epsilon_{I_2})^{b}\,
   (\epsilon_{I_1}\!\cdot p_3)^{s_1-b}\,
   (\epsilon_{I_2}\!\cdot p_3)^{s_2-b}.
\end{equation}
So the five-point partial wave basis element will be labeled by $3$ quantum numbers: the two spins $s_1$ and $s_2$, and a new ``vertex'' quantum number $b$ telling us which three-point structure is being used for the middle amplitude. Additionally, at four-points, the partial wave expansion depended only on one variable: the angle between incoming and outgoing lines in the center-of-mass frame of the internal particle. At five-points, the count jumps to $3$ variables, and at general $n$ it is $(n-3)(n-2)/2$. So one must decide what the most convenient kinematic parameterization is for constructing the partial wave basis. At five-points, we can solve the problem by judiciously selecting three angular variables --- two ``polar'' angles (corresponding to each internal line) and one overall ``azimuthal'' angle.

In this paper, we give a solution to these problems by proposing and proving a general partial-wave expansion for the ``half-ladder diagram'' (shown in Fig.~\ref{fig:n-pt-half-ladder}), valid at arbitrary $n,d$, with generically-massive scalar modes on the external lines. The basis elements $R_{\vec{s},\vec{\mu}}$ --- labeled by the spins $\vec{s} = (s_1, \ldots, s_{n-3})$ and ``vertex'' quantum numbers $\vec{\mu} = (\mu_1, \ldots, \mu_{n-4})$ of the internal modes --- are derived from requiring that, in the rest frame of each internal particle $I_i$, they are eigenfunctions of the angular momentum in general dimension $d$:
\begin{equation}
    \hat{L}^2_{I_i} R = s_i (s_i + d - 3) R.
\end{equation}
Each basis element resulting from these constraints takes the form of a particular sum over products of special functions, all of which are related directly to the Gegenbauer polynomials and are given explicitly in the main text. In turn, each special function depends on a single angular variable, which together parameterize the full orientation of the half-ladder diagram. For example, at six-points we find that the basis elements are
\begin{equation}\label{eq:6pt-intro-1}
R^{(6)}_{s_1 s_2 s_3;\,\mu_1\mu_2}
 = N^{(6)}\,A^{d}_{s_1,\mu_1}(\theta_{2,3})\;
   F^{d}_{s_2,\mu_1,\mu_2}(\theta_{3,4},\theta_{2,4},\theta_{3,5},\theta_{2,5})\;
   A^{d}_{s_3,\mu_2}(\theta_{4,5})
\end{equation}
where
\begin{equation}\label{eq:6pt-intro-2}
F^{d}_{s_2,\mu_1,\mu_2}
 = \sum_{l=0}^{\min(\mu_1,\mu_2)}(-1)^l\sqrt{b_{\mu_1,l}\,b_{\mu_2,l}}\;
   T^{d}_{s_2,\mu_1,\mu_2,l}(\theta_{3,4})\,
   A^{d-1}_{\mu_1,l}(\theta_{2,4})\,A^{d-1}_{\mu_2,l}(\theta_{3,5})\,
   C^{(\frac{d-5}{2})}_{l}(\cos\theta_{2,5}),
\end{equation}
with some overall normalization $N^{(6)}$. We define all of these functions, angular variables, and constants in great detail in Sec.~\ref{sec:six-pt}.

You'll notice that we are now labeling the internal vertices by quantum numbers $\mu_i$, which corresponds to an orthogonal choice of basis for the internal three-particle amplitudes~\eqref{eq:int-3-particle}, instead of $b_i$. As we discuss in the text, we can rotate back to the $b$-basis in Eq.~\eqref{eq:int-3-particle} straightforwardly after we compute them in the $\mu$-basis, but orthogonality is critical to the extraction of the partial wave coefficients.

Next, we apply this technology to the case of ``string'' amplitudes, given at all $n,d$ by the formula
\begin{equation}
\mathcal A_n = \int_0^\infty \prod_{j=3}^{n-1}\frac{dy_{1,j}}{y_{1,j}}
  \prod_{i < j} u_{i,j}^{\,\alpha'X_{i,j}+\alpha_0},
\end{equation}
with variable Regge intercept $\alpha_0$. The $u$-variables shown above each satisfy a ``$u$-equation''~\cite{Arkani-Hamed:2019mrd,Arkani-Hamed:2019plo,Arkani-Hamed:2024nzc}, which, when solved in terms of the $y$-variables, allows us to straightforwardly extract the half-ladder residue of this amplitude~\cite{Arkani-Hamed:2024nzc}. We then can use our general $n,d$ partial wave expansion to construct multi-particle positivity checks from unitarity on this object, as well as uncover how individual states at a given $(N,s)$ couple to each other and to external modes through on-shell three-point amplitudes.

The set-up of the paper is as follows. In Sec.~\ref{sec:review}, we review what is known about the partial wave expansion in general $d$, which amounts to the four- and five-point expansions. We re-derive these bases by enforcing that each element is an eigenvector of each internal particle's angular momentum, and in doing so correct an error in the literature about the general-$d$ extrapolation of the five-point expansion~\cite{Saha:2026ftv}. Next, in Sec.~\ref{sec:halfladder}, we first introduce in Sec.~\ref{sec:kinematics} the angular variables $\theta_{i,j}$ that parameterize the half-ladder diagrams, and we write them in a totally Lorentz-invariant manner in terms of dot products of external, on-shell momenta. Then, in Sec.~\ref{sec:six-pt}, we work through the proof of the expansion at $n = 6$ in arbitrary dimension $d \geq 5$ (shown in Eqs.~\eqref{eq:6pt-intro-1} and~\eqref{eq:6pt-intro-2}), where we introduce the full cast of special functions that appear at any $n,d$. We then move on to the general proof (styled as an induction argument, with $n = 6$ serving as the base) when $d \geq n - 1$ in Sec.~\ref{sec:proof}. While these sections are quite technical, we follow them by Sec.~\ref{sec:writing}, where we introduce a set of simple, graphical rules for writing down any basis element. Finally, in Sec.~\ref{sec:gram}, we show how to obtain the expansion when $d < n - 1$ (when the fact that the Gram determinant of momenta vanishes reduces the number of degrees of freedom) as a limit from the general expansion.

In Sec.~\ref{sec:string-theory}, we apply this formalism to the ``string'' amplitudes referenced above. Localizing on the half-ladder contribution, we derive the partial wave expansion for these objects at low levels $N$. We show in Sec.~\ref{sec:unitarity} that the requirement of multi-particle positivity when all internal lines have the same $(N,s)$ restricts the string intercept to be either that of $Z$-theory~\cite{Broedel:2013tta,Carrasco:2016ldy} at $\alpha_0 = 0, d \leq 10$ or that of open bosonic string theory~\cite{Green:1987sp} at $\alpha_0 = -1, d \leq 26$. In Sec.~\ref{sec:violation}, we further rule out $Z$-theory as a unitary string theory at finite $\alpha'$ by checking multi-particle positivity for the so-called ``twisted'' half-ladder diagram with alternating internal modes $(2,1)$ and $(1,0)$. In Sec.~\ref{sec:degen-1}, for the remaining unitary theory (the open bosonic string at $\alpha_0 = -1$), we initiate an effort to uncover, at each valid string intercept, the contributions of \textit{individual} states at a given $(N,s)$ to the half-ladder diagram. We conclude in Sec.~\ref{sec:discussion}.

With our manuscript, we include a Python script \texttt{combwaves.py} which, if fed a generic half-ladder diagram as a polynomial in external momenta, extracts the partial wave coefficients; an introduction to the code is included in App.~\ref{sec:code}.

\section{Review: Four- and Five-Point Partial Waves}\label{sec:review}

Let us begin by giving a review of what is known about the partial wave expansion, at four- and five-points in general dimension $d$~\cite{Jacob:1959at,Collins:1977jy,Paulos:2017fhb,Correia:2020xtr}. For simplicity, here we will consider scalar modes of generic masses $m_i$ on the external lines, and on the internal line(s) we will place massive bosons of spin $s_i$ and mass $M_i$. We work here in the mostly-plus metric convention and use the all-outgoing statement of momentum conservation:
\begin{equation}
    \sum_{i = 1}^{n} p_i^\mu = 0.
\end{equation}
Additionally we will parameterize kinematic space by the \textit{planar Mandelstam invariants} $X_{i,j}$ given by
\begin{equation}
    X_{i,j} = (p_i + \cdots + p_{j-1})^2,
\end{equation}
with $X_{i,i+1} = p_i^2 = -m_i^2$.

\begin{figure}[t]
    \centering
    \includegraphics[width=1.0\linewidth]{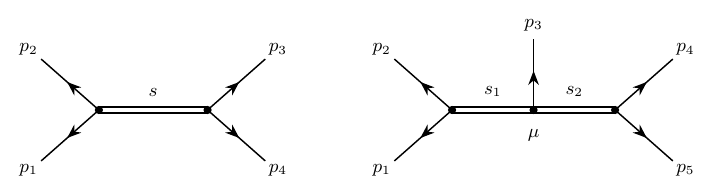}
    \caption{(Left) The four-point half-ladder diagram with internal boson labeled by spin $s$. (Right) The five-point half-ladder diagram with internal bosons of spin $s_1, s_2$. The middle three-point vertex is not uniquely determined by Lorentz invariance and instead has $\min(s_1, s_2) + 1$ possible structures, a freedom which we label by $\mu$. All diagrams are taken here to have external scalars.}
    \label{fig:4-5-half-ladder}
\end{figure}

At four points, the story is classic. Consider the $X_{1,3}$-channel diagram shown on the left-hand-side of Fig.~\ref{fig:4-5-half-ladder}. Localizing on the residue $X_{1,3} \to -M^2$ and fixing an internal spin $s$, the residue of the four-point amplitude may be computed by ``sewing'' together the two three-point amplitudes that make up the $X_{1,3}$-channel diagram:
\begin{equation}\label{eq:4-pt-res}
    R_s^{(4)} = \Res_{X_{1,3} = -M^2} \mathcal{A}_4 = \sum_I \mathcal{A}_3(1,2,\epsilon_I) \mathcal{A}_3(\epsilon_I, 3,4),
\end{equation}
where the sum is over all states $I$ of the spin-$s$ irrep of the $SO(d-1)$ massive little group. The three-point amplitudes $\mathcal{A}_3$ are, in turn, uniquely fixed by Lorentz invariance~\cite{Arkani-Hamed:2017jhn,Kravchuk:2016qvl} to be
\begin{equation}\label{eq:gluing}
    \mathcal{A}_3(1,2,\epsilon_I) = i c_s (p_1 - p_2)^{\mu_1} \cdots (p_1 - p_2)^{\mu_s} \epsilon_{I_{\mu_1, \mu_2, \ldots, \mu_s}},
\end{equation}
where $c_s$ is its coupling constant. One can then plug Eq.~\eqref{eq:gluing} into Eq.~\eqref{eq:4-pt-res}, which, upon performing the spin-sum over $I$, leads us to a particular polynomial in $(p_1 - p_2)^2$, $(p_3 - p_4)^2$, and $(p_1 - p_2) \cdot (p_3 - p_4)$. Our goal is to understand what this polynomial is.

To derive the result, one must first move into the rest frame of the internal state. In that frame, the single degree-of-freedom is the angle $\theta_{2,3}$ between the incoming particles $1$ and $2$ and the outgoing particles $3$ and $4$. Hence, we will take $R_s^{(4)} = R_s^{(4)}(\theta_{2,3})$. It is then clear that Eq.~\eqref{eq:4-pt-res} must be an eigenstate of the angular momentum operator, since the internal particle has definite spin $s$ in its center-of-mass frame. Hence, the residue must satisfy
\begin{equation}\label{eq:I1-ang-mom}
    \hat{L}_{I}^2 R_s^{(4)} = - \Delta_{S^{d-2}} R_s^{(4)} = s (s + d - 3) R_s^{(4)},
\end{equation}
which is the generalization of the familiar angular momentum eigenvalue equation from quantum mechanics (corresponding to $d = 4$) to arbitrary $d$, where the operator $\Delta_{S^{d-2}}$ may be written as
\begin{equation}
    \Delta_{S^{d-2}} = \partial_{\theta_{2,3}}^2 + (d - 3) \cot \theta_{2,3} \partial_{\theta_{2,3}}.
\end{equation}
Therefore, we find that $R_s^{(4)}$ satisfies
\begin{equation}
    -R'' - (d-3) \cot \theta_{2,3} R' = s ( s + d - 3) R.
\end{equation}
The solution to this equation is famous: the \textit{Gegenbauer polynomial} $C_s^{(\lambda)}(\cos \theta_{2,3})$, with $\lambda = (d - 3) /2$. The Gegenbauer polynomial is given by the generating function:
\begin{equation}
    \frac{1}{\left(1-2xt+t^{2}\right)^{\lambda}}
    \;=\;\sum_{s=0}^{\infty} C^{(\lambda)}_{s}(x)\,t^{s}.
\end{equation}
In $d = 3$ ($\lambda = 0$), it reduces (as a limit, with appropriate normalization) to the Chebyshev polynomials $\cos(s \theta)$, and in $d = 4$ ($\lambda = 1/2$) they are the Legendre polynomials $P_s(\cos \theta) $. Additionally, they obey an orthogonality condition given in Eq.~\eqref{eq:gegen-ortho} of App.~\ref{sec:ortho}.

Thus each partial wave basis element $R_s^{(4)}$ is given by a Gegenbauer polynomials~\cite{Paulos:2017fhb,Correia:2020xtr,Arkani-Hamed:2022gsa}, and so the \textit{entire} expansion is
\begin{equation}
    R^{(4)}(\cos\theta_{2,3}) = \sum_{s = 0}^\infty y_s C_s^{(\lambda)}(\cos \theta_{2,3}),
\end{equation}
where each coefficient $y_s$ can be derived from $R^{(4)}$ via orthogonality~\eqref{eq:gegen-ortho}. Finally, to write the expansion in a completely Lorentz-invariant manner, we must express the angular variable $\theta_{2,3}$ in terms of the momenta. This is a straightforward computation, where we find that
\begin{equation}\label{eq:costheta-form}
    \cos\theta_{2,3}= -
    \frac{2M^{2}\big(X_{2,4}+m_1^{2}+m_4^{2}\big)
    -\big(M^{2}+m_1^{2}-m_2^{2}\big)\big(M^{2}+m_4^{2}-m_3^{2}\big)}
    {\sqrt{\lambda(M^{2},m_1^{2},m_2^{2})\,\lambda(M^{2},m_3^{2},m_4^{2})}} ,
\end{equation}
where $\lambda(a,b,c)=a^2+b^2+c^2-2ab-2ac-2bc$ is the familiar Källén function. Therefore, using Eq.~\eqref{eq:costheta-form}, if we are given the four-point amplitude of \textit{any} theory (written, as usual, in terms of momenta), we have a systematic method to derive its partial wave expansion.

Before moving on to the case at five-points, let us make one comment about \textit{positivity} conditions, since it will be important later on in this paper. From the perspective of ``sewing'' three-particle amplitudes as in Eqs.~\eqref{eq:4-pt-res} and~\eqref{eq:gluing}, the partial wave coefficient $y_s$ can be understood as the \textit{sum} of all three-point couplings $c_s^2$ for all particles appearing in the $X_{1,3}$-channel:
\begin{equation}
    y_s = \sum_{i = 1}^{N_s} (c_s^{(i)})^2 \geq 0,
\end{equation}
with $N_s$ the number of species at spin-$s$ flowing through the internal line. This is a famous result at four-points which is a direct result of unitarity, that is, of the fact that the four-point diagram is given by sewing together all possible three-particle amplitudes. In the literature, it has been frequently used, often in bootstrap settings where one wishes to constrain the available space of theories from simply imposing basic axioms and conditions of quantum field theory~\cite{Adams:2006sv,deRham:2017avq,Caron-Huot:2020cmc,Tolley:2020gtv,Sinha:2020win,Bern:2021ppb,Arkani-Hamed:2020blm,Caron-Huot:2021rmr,Huang:2020nqy,deRham:2022hpx,Kruczenski:2022lot}. In later sections of this manuscript, we will likewise make use of this and higher-point generalizations of positivity in the context of simple string amplitudes and their deformations.

Having understood the story at four-points, let us turn to five-points, which has a contribution from the half-ladder diagram shown on the right-hand-side of Fig.~\ref{fig:4-5-half-ladder}. In this case, we wish to take the double residue $X_{1,3} \to -M_1^2$, $X_{1,4} \to -M_2^2$ in order to put both internal lines on-shell. This is equivalent to gluing together three three-point vertices, the middle one being new with respect to the four-point problem:
\begin{equation}
    R_{s_1,s_2}^{(5)} = \Res_{X_{1,3} = -M_1^2, X_{1,4} = -M_2^2} \mathcal{A}_5 = \sum_{I_1,I_2} \mathcal{A}(1,2,\epsilon_{I_1}) \mathcal{A}(\epsilon_{I_1}, 3, \epsilon_{I_2}) \mathcal{A}(\epsilon_{I_2}, 4, 5).
\end{equation}
Unlike the left and right vertices, the middle vertex is \textit{not} uniquely determined by Lorentz invariance. It is instead composed of all the ways one can fully contract $\epsilon_{I_1}$, $\epsilon_{I_2}$, and $p_3^\mu$. One can work out that there are exactly $\min(s_1,s_2) + 1$ ways of doing it; in general, it is given by the expansion
\begin{equation}\label{eq:3pt-expansion}
    \mathcal A_3(\epsilon_{I_1},3,\epsilon_{I_2})
    =i\sum_{b=0}^{\min(s_1,s_2)} g_b\,
    \big(\epsilon_{I_1}\!\cdot\epsilon_{I_2}\big)^{b}
    \big(\epsilon_{I_1}\!\cdot p_3\big)^{s_1-b}
    \big(\epsilon_{I_2}\!\cdot p_3\big)^{s_2-b}.
\end{equation}
So, as shown in Fig.~\ref{fig:4-5-half-ladder}, the partial wave expansion acquires an additional quantum number at $n = 5$, corresponding to the choice of internal three-point vertex, which we call $0 \leq \mu \leq \min(s_1,s_2)$. As we will show in the following discussion, the basis parameterizing the label $\mu$ is not the same as the basis parameterizing the label $b$ shown in Eq.~\eqref{eq:3pt-expansion}.

At five-points, there are three independent kinematic variables, and we will again use angles to represent them. The first two are \textit{polar} angles $\theta_{2,3}$ and $\theta_{3,4}$, defined by moving into the rest frame of $I_1$ ($I_2$) and measuring the angle between the incoming and outgoing modes. The invariant formulae relating these angles to the external momenta are
\begin{align}
\cos\theta_{2,3}&=-\,
\frac{2M_1^{2}\big[X_{2,4}+m_1^{2}+M_2^{2}\big]
-\big(M_1^{2}+m_1^{2}-m_2^{2}\big)\big(M_1^{2}+M_2^{2}-m_3^{2}\big)}
{\sqrt{\lambda(M_1^{2},m_1^{2},m_2^{2})\,\lambda(M_1^{2},M_2^{2},m_3^{2})}},
\\[6pt]
\cos\theta_{3,4}&=-
\frac{2M_2^{2}\big[X_{3,5}+M_1^{2}+m_5^{2}\big]
-\big(M_2^{2}+M_1^{2}-m_3^{2}\big)\big(M_2^{2}+m_5^{2}-m_4^{2}\big)}
{\sqrt{\lambda(M_2^{2},M_1^{2},m_3^{2})\,\lambda(M_2^{2},m_5^{2},m_4^{2})}} ,
\end{align}
which are just Eq.~\eqref{eq:costheta-form} with some replacements in variables. The third angle is the azimuthal angle $\theta_{2,4}$, which is defined by moving into $I_1$ and measuring the angle between the planes formed by $(\vec{p}_2, \vec{p}_3)$ and $(\vec{p}_3, \vec{p}_4)$. (One could also define the angle from the perspective of $I_2$ and obtain the same answer.) This angle is the so-called ``Toller'' angle of the multi-Regge literature~\cite{Bali:1967zz,Toller:1969gx,Toller:1969vt,Brower:1974yv}. It can be written as
\begin{equation}\label{eq:costheta24}
    \cos \theta_{2,4} = \frac{n_{2,3} \cdot n_{3,4}}{\sqrt{|n_{2,3}^2|} \sqrt{|n_{3,4}^2|}},
\end{equation}
where (working in $d = 4$ for one moment) we have 
\begin{equation}\label{eq:ns-5pt}
    (n_{2,3})_\mu = \epsilon_{\mu\nu\sigma\omega}q_1^\nu p_2^\sigma p_3^\omega, \qquad (n_{3,4})_\mu = \epsilon_{\mu\nu\sigma\omega}q_1^\nu p_3^\sigma p_4^\omega,
\end{equation}
with $\epsilon$ the four-component totally anti-symmetric tensor and $q_1 = p_1 + p_2$ the momentum of the internal particle $I_1$. Note that, in rest frame of $I_1$, $q_1^\nu$ is purely timelike and the four-vectors Eq.~\eqref{eq:ns-5pt} are proportional to $\vec{p}_2 \cross \vec{p}_3$ and $\vec{p}_3 \cross \vec{p}_4$, respectively. Also note that we can freely swap $q_1$ with $q_2$ (the momentum of $I_2$) in the above formulae, since, in the replacement $q_1 = q_2 - p_3$, the $p_3$ term drops out by anti-symmetry. So this is the sense in which we can work in either the rest frame of $I_1$ or the rest frame of $I_2$. 

It is well known that the contractions of vectors like those in Eq.~\eqref{eq:ns-5pt} can be represented as \textit{determinants}. Defining 
\begin{equation}\label{eq:G24}
G(a;b)=\det\!\begin{pmatrix}
a\!\cdot\!b & a\!\cdot\!q_1 & a\!\cdot\!q_2\\
q_1\!\cdot\!b & q_1\!\cdot\!q_1 & q_1\!\cdot\!q_2\\
q_2\!\cdot\!b & q_2\!\cdot\!q_1 & q_2\!\cdot\!q_2
\end{pmatrix},
\end{equation}
we can re-write Eq.~\eqref{eq:costheta24} as
\begin{equation}\label{eq:cos24-fin}
    \cos \theta_{2,4} = \frac{G(p_2; p_4)}{\sqrt{G(p_2;p_2)G(p_4;p_4)}}.
\end{equation}
Note that, although we started this discussion in $d = 4$, the formulae Eq.~\eqref{eq:G24} and Eq.~\eqref{eq:cos24-fin} are dimension-agnostic. So, we take these as the \textit{definition} of the angle $\theta_{2,4}$ in higher dimension.

Now that we have established the kinematics, we can move on to deriving the partial wave basis $R^{(5)}$. As in the four-point case, we start by enforcing Eq.~\eqref{eq:I1-ang-mom} for $I_1$. However, since we now have an azimuthal angle $\theta_{2,4}$, our formula for $\Delta_{S^{d-2}}$ picks up a correction:
\begin{equation}\label{Eq:Delta-5}
    \Delta_{S^{d-2}} = \partial_{\theta_{2,3}}^2 + (d - 3) \cot \theta_{2,3} \partial_{\theta_{2,3}} + \frac{1}{\sin^2 \theta_{2,3}} \Delta_{S^{d-3}},
\end{equation}
where $\Delta_{S^{d-3}}$ acts on the angle $\theta_{2,4}$:
\begin{equation}
    \Delta_{S^{d-3}} = \partial_{\theta_{2,4}}^2 + (d - 4) \cot \theta_{2,4} \partial_{\theta_{2,4}}.
\end{equation}
Let us consider a solution to this equation which is separable:
\begin{equation}\label{eq:5pt-sep}
    R_{s_1,s_2}^{(5)}(\theta_{2,3}, \theta_{2,4}, \theta_{3,4}) = f(\theta_{2,3}) g(\theta_{2,4}) h(\theta_{3,4}).
\end{equation}
Plugging Eq.~\eqref{eq:5pt-sep} into our eigenvalue equation, and enforcing that
\begin{equation}\label{eq:g-eigen-equation}
    -\Delta_{S^{d-3}} g(\theta_{2,4}) = \mu (\mu + d - 4)g(\theta_{2,4}),
\end{equation}
for some $\mu$, we find that $f$ satisfies
\begin{equation}\label{eq:f-5pt-equation}
    f'' + (d - 3) \cot \theta_{2,3} f' + \left(s_1 (s_1 + d - 3) - \frac{\mu(\mu + d - 4)}{\sin^2 \theta_{2,3}} \right)f = 0.
\end{equation}
Unsurprisingly, this is a familiar equation: the ``associated Gegenbauer equation,'' the arbitrary $d$ generalization of the $d = 4$ associated Legendre equation. The solutions to this equation are thus called the \textit{associated Gegenbauer polynomials}, which we denote as
\begin{equation}\label{eq:A-defs}
    A^d_{s,\mu} (\theta) \equiv (\sin\theta)^\mu C_{s - \mu}^{\left( \mu + \frac{d - 3}{2}\right)}(\cos \theta).
\end{equation}
Thus, $f = A^d_{s_1,\mu}$, where $\mu = 0, 1, \ldots, s_1$. (The orthogonality of this function is demonstrated in Eq.~\eqref{eq:A-ortho}.) For $g$, we see that the eigenvalue equation~\eqref{eq:g-eigen-equation} is precisely the same as that satisfied by the four-point partial wave basis, just in one-dimension lower, and so we find $g = C_\mu^{\left( \frac{d - 4}{2} \right)}$.

So, we are left with determining $h(\theta_{3,4})$. This, of course, can be done by enforcing that $R_{s_1,s_2}^{(5)}$ is an eigenvector of angular momentum in the rest frame of $I_2$. Then, we find exactly the same equation as in Eq.~\eqref{eq:f-5pt-equation} just with $s_1 \to s_2$. And, due to the fact that $\theta_{2,4}$ is the same whether it is measured in the rest frame of $I_1$ or in the rest frame of $I_2$, $g(\theta_{2,4})$ satisfies the azimuthal equation arising from this eigenvalue problem automatically. So, we find that $h = A^d_{s_2,\mu}$, and $\mu$ is now restricted to range from $0$ to the \textit{minimum} of $s_1$ and $s_2$. 

Putting it all together, we find that the five-point partial wave basis elements are given by
\begin{equation}\label{eq:five-pt-basis}
    R^{(5)}_{s_1,s_2,\mu}(\theta_{2,3},\theta_{2,4},\theta_{3,4}) = A^d_{s_1,\mu}(\theta_{2,3}) C_\mu^{(\frac{d-4}{2})}(\theta_{2,4}) A^d_{s_2,\mu}(\theta_{3,4}),
\end{equation}
and thus that the full expansion is
\begin{equation}\label{eq:5-pt-exp}
    R^{(5)} = \sum_{s_1,s_2 = 0}^\infty \sum_{\mu = 0}^{\min(s_1,s_2)} y_{s_1,s_2}^\mu A^d_{s_1,\mu}(\theta_{2,3}) C_\mu^{(\frac{d-4}{2})}(\theta_{2,4}) A^d_{s_2,\mu}(\theta_{3,4}),
\end{equation}
for coupling matrix $y_{s_1,s_2}^\mu$. Note that the vertex quantum number $\mu$ specifies a particular \textit{orthogonal} basis of the internal three-point amplitude. Denoting a basis element in the $b$-basis~\eqref{eq:3pt-expansion} at the vertex joining $I_i$ and $I_{i+1}$ as
\begin{equation}\label{eq:p-basis-3pt}
    M_3^{(b)}(\epsilon_{I_i}, i+2, \epsilon_{I_{i+1}}) = \mathcal{N}_{s_{i+1}}\big(\epsilon_{I_i}\!\cdot\epsilon_{I_{i+1}}\big)^{b}
    \big(\epsilon_{I_i}\!\cdot p_{i+2}\big)^{s_i-b}
    \big(\epsilon_{I_{i+1}}\!\cdot p_{i+2}\big)^{s_{i+1}-b},
\end{equation}
with normalization\footnote{Here $(x)_j = x(x+1)\cdots(x+j-1)$ is the Pochhammer symbol.}
\begin{equation}
    \mathcal N_{s}\equiv\frac{(-1)^{s}\,s!}{2^{s}\,(\nu_0)_{s}},\qquad \nu_0=\frac{d-3}{2}
\end{equation}
and a basis element in the $\mu$-basis $B_3^{(\mu)}(\epsilon_{I_i}, i+2, \epsilon_{I_{i+1}})$, we find the relation
\begin{equation}
    M_3^{(b)} = \sum_{\mu = 0}^b a^{(i)}_{b,\mu} B_3^{(\mu)}
\end{equation}
with expansion coefficients $a^{(i)}_{b,\mu}$ given by
\begin{equation}\label{eq:a-k-mu-form}
    a^{(i)}_{b,\mu} = \mathcal{N}_{s_{i+1}}(-1)^{\mu}\,2^{\mu}\,b!\,\frac{\big[(\nu_0)_\mu\big]^{2}}{(\nu_1)_\mu}
    \prod_{j=i,i+1}\frac{(s_j-\mu)!}{s_j!}\Big(\frac{\sqrt{\lambda_i}}{2M_j}\Big)^{s_j-\mu}
    \sum_{\ell=0}^{\lfloor (b-\mu)/2\rfloor}
    \frac{\sigma_i^{\,b-\mu-2\ell}\,(\rho_i/2)^{2\ell}}{(b-\mu-2\ell)!\;\ell!\;(\nu_1+\mu+1)_\ell},
\end{equation}
where $\nu_1 = \frac{d-4}{2}$, and
\begin{equation}
    \lambda_i \equiv \lambda(M_i^2,M_{i+1}^2,m_{i+2}^2),\qquad
    \sigma_i = \frac{2\,(M_i^2+M_{i+1}^2-m_{i+2}^2)}{\lambda_i},\qquad
    \rho_i = \frac{4M_iM_{i+1}}{\lambda_i}.
\end{equation}
One can view $a^{(i)}_{b,\mu}$ as a lower-triangular (and invertible) matrix, where $a^{(i)}_{b,\mu} = 0$ for all $\mu > b$. As a result, at five points (where $i = 1$) we can write the partial wave element in the $b$-basis in terms of that in the $\mu$-basis as
\begin{equation}\label{eq:five-pt-k}
    \widetilde{R}^{(5)}_{s_1,s_2,b} = \sum_{\mu = 0}^b a^{(1)}_{b,\mu} R^{(5)}_{s_1,s_2,\mu}.
\end{equation}
So, the idea is to use orthogonality to determine $R^{(5)}_{s_1,s_2,\mu}$ in the $\mu$-basis and then rotate to the $b$-basis to understand the partial wave coefficients as couplings of the familiar three-particle structures in terms of epsilons and momenta~\eqref{eq:p-basis-3pt}.

The formula~\eqref{eq:5-pt-exp} matches the result at $d = 4$ in Ref.~\cite{Saha:2026ftv}. In general dimension, the authors propose that the five-point partial wave expansion contains azimuthal dependence of $e^{i\mu \theta_{2,4}}$. The real part of this exponential is, of course, the Chebyshev polynomial, which in turn corresponds to the Gegenbauer polynomial $C_\mu^{(\lambda)}$ with $\lambda = 0$. However, as we have shown in this subsection, the correct azimuthal dependence is a Gegenbauer polynomial with dimension-dependent $\lambda = (d - 4) / 2$, which comes directly from solving the eigenvalue equation Eq.~\eqref{eq:g-eigen-equation}. 

Five- and six-point partial waves in $d = 4$ for massless planar amplitudes in the complex-forward kinematics limit have also been constructed recently in
Ref.~\cite{Jeong:2026xzk}. We comment on the relation to the present construction in
Sec.~\ref{sec:gram}.

Finally, before moving on to the general case in the next section, we note that, by just using the four- and five-point partial wave expansions, there is no positivity result from unitarity one gets beyond that already noted at $n = 4$~\cite{Chandrasekaran:2018qmx}. However, as we will see in the next section where we give the all-$n$ result, there is a new positivity result at every \textit{even} $n$ partial wave expansion, which we will make good use of in the context of string theory amplitudes in Sec.~\ref{sec:string-theory}.

\section{The General Partial Wave Expansion for the Half-Ladder Diagram}\label{sec:halfladder}

In this section, we will define and prove an orthogonal partial wave basis for the $n$-point half-ladder Feynman diagram in general dimension $d$. The basis is parameterized by two sets of quantum numbers: the set of spins of each internal particle $\{ s_i\}$, and the set of internal three-point vertex structures $\{ \mu_i \}$ (shown in Fig.~\ref{fig:n-pt-half-ladder}). By projecting a generic half-ladder diagram onto this basis, one can determine the contribution to the diagram from a particular configuration of internal spins and vertices. We restrict for now to the case when $d \geq n - 1$, i.e., to the situation where all momenta of the problem are independent sans momentum conservation.  

In Sec.~\ref{sec:kinematics}, we first establish the kinematics of the problem, which we take to be a set of angles parameterizing the orientation of the scattering process. In $d = 4$, these angles are the familiar polar and azimuthal angles parameterizing rotations in $3$ spatial dimensions: the polar angles are the angles $\theta_{i,i+1}$ between \textit{intersecting lines} $\vec{p}_i$ and $\vec{p}_{i+1}$ in the center-of-mass frame of particle $I_{i-1}$, and the azimuths $\theta_{i,i+2}$ are the angles between the \textit{intersecting planes} $(\vec{p}_i, \vec{p}_{i+1})$ and $(\vec{p}_{i+1}, \vec{p}_{i+2})$ in the center-of-mass frame of $I_{i-1}$ or $I_i$. These definitions hold in any spacetime dimension. However, when $d > 4$, we unlock ``deeper'' azimuths: for example, when $d \geq 5$, we have $\theta_{i,i+3}$, the angle between the $3$-dimensional intersecting hyperplanes $(\vec{p}_{i}, \vec{p}_{i+1}, \vec{p}_{i+2})$ and $(\vec{p}_{i+1}, \vec{p}_{i+2}, \vec{p}_{i+3})$ in any frame $I_{i-1}, I_i, I_{i+1}$. Given these definitions, it is straightforward to write the angles in a totally Lorentz-invariant manner using \textit{determinants}, which are well-known from linear algebra to compute the angles between intersecting hyperplanes.

In Sec~\ref{sec:six-pt}, we prove the six-point partial wave basis. Though the discussion there is a bit technical, the idea is the same from the previous section: we require that, in the center-of-mass frame of each internal spin-$s_i$ particle $I_i$, the residue $R$ is an eigenstate of angular momentum $s_i ( s_i + d - 3)$. This gives us a set of $n - 3 = 3$ equations
\begin{equation}
    - \nabla_{S^{d-2}}^2 R = s_i (s_i + d - 3) R
\end{equation}
which we solve using standard separation-of-variables techniques. By doing so, we guarantee that the basis is labeled by quantum numbers $\{ s_i \}$ corresponding to the spins of the internal particles. Additionally, this process automatically hands us a particular \textit{orthogonal} basis of the internal three-point vertex labeled by $\mu_i$. We can rotate from this basis to the familiar basis in terms of epsilons and momenta using the $a^{(i)}_{b,\mu}$ matrix defined in the five-point example in Eq.~\eqref{eq:a-k-mu-form}. 

We find that each basis element at six-points is a sum over a product of special functions, with each special function assigned one angular variable $\theta_{i,j}$. There are three special functions that appear in the problem: the Gegenbauer polynomials $C_s^{(\lambda)}$, the ``associated'' Gegenbauer polynomials $A_{s,\mu}^{d}$, and the special $T$-functions $T^d_{s,\mu_1,\mu_2,l}$. These may be thought of as the all-$d$ generalizations, respectively, of the Legendre polynomials, the associated Legendre polynomials, and the Wigner little-$d$ functions, and we write them explicitly in the text. In particular, we give a formula for the $T$ functions as a sum over Gegenbauer polynomials.

It turns out these are all the functions we need for the general case, and so we move on to the general proof in Sec.~\ref{sec:proof}, which is styled as an induction with the six-point taken as the base case. The key observation is that each partial wave basis element at six-points in $d$ dimensions contains within it a five-point partial wave basis element in $d - 1$ dimensions. With a judicious choice of normalization (which we spell out in detail in the subsection), we conjecture that this pattern holds at general $n$, and show that it is true by solving all necessary eigenvalue equations.

While this does prove the partial wave expansion at general $n$ and $d$, it is fundamentally recursive in nature. In Sec.~\ref{sec:writing}, we demonstrate a simple algorithm for writing down the partial wave basis at any $n, d$, where we rely on two upside-down triangles corresponding to the angular variables of the problem (left at $8$-points in Fig.~\ref{fig:8pt-kins}) and its set of quantum numbers (right). This solves the recursion and gives a means of using the expansion practically.

Finally, in Sec.~\ref{sec:gram}, we consider the case when $d < n - 1$, that is, when we have a vanishing Gram determinant. In this situation, the dimension is too small to accommodate the $n - 1$ independent vectors in an $n$-point scattering process. So, we have a reduction in the number of kinematic variables specifying the process, and the partial wave basis changes accordingly. We show that the variables parameterizing the process can be obtained by ``chopping'' the bottom of the upside-down kinematic triangle at a particular level $r_\star = d - 3$--- turning the triangle into a trapezoid! (See Fig.~\ref{fig:8pt-kins-gram} for the example at eight-points.) Starting from the general expansion in $d \geq n - 1$, we show that this ``freezes out'' all angles below the line, such that they play no dynamical role (as expected) in the scattering. For the angles above the line, we carefully consider what happens to each component of the partial wave basis as we approach $d < n - 1$, and we arrive at a final compact expression~\eqref{eq:gen-d-gram}. In particular, we show that this recovers the expressions for the all-$n$ partial wave expansions in $d = 3,4$, where we have, respectively, Chebyshev polynomials and Wigner-$d$ functions. 

We implement the partial wave coefficient extraction in a Python script accompanying this manuscript \texttt{combwaves.py}, which we describe more in App.~\ref{sec:code}. Additionally, we note that, while everything here is done explicitly for the regular half-ladder diagram, where all internal ``rungs'' $p_3^\mu, p_4^\mu, \ldots, p_{n-2}^\mu$ are on the same side of ladder, we may just as well have taken the rungs to be on \textit{either} side of the ladder. That is, we can ``twist'' it however we like, and the expansion remains unchanged. This will prove to be a useful fact in Sec.~\ref{sec:string-theory}, where we apply this formalism to ``string theory'' amplitudes.

\subsection{Kinematics}\label{sec:kinematics}

\begin{figure}[t]
    \centering
    \includegraphics[width=1.0\linewidth]{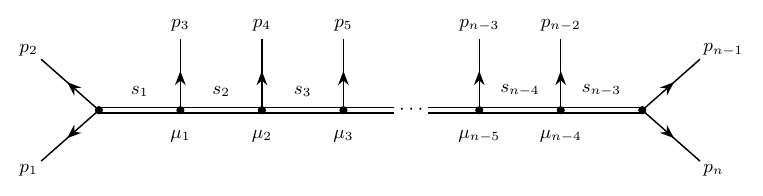}
    \caption{The $n$-point half-ladder diagram. Each external line $p_i$ is a scalar, and each internal line $I_i$ a boson of spin $s_i$ with momentum $q_i = p_1 + p_2 + \ldots + p_{i+1}$. The parameter $0 \leq \mu_i \leq \min(s_i,s_{i+1})$ enumerates the possible internal-external-internal vertex structures, while the external-external-internal vertices are fixed uniquely by Lorentz invariance.}
    \label{fig:n-pt-half-ladder}
\end{figure}

Having reviewed the known expansions at four- and five-points, we are prepared to move on to six-points and higher for the half-ladder in general dimension $d$, as shown in Fig.~\ref{fig:n-pt-half-ladder}. But, before we do this, let us first firmly establish the kinematics of the problem. 

For simplicity, just as at four- and five-points, in the following discussion we will take all external particles to be massive scalars of mass $m_i$, and all internal particles to be massive bosons of spin $s_i$ and mass $M_i$. The kinematics of the problem are then the momenta of the $n$ massive scalars, $p_i^\mu$ with $p_i^2 = -m_i^2$. Just as in the previous section, we continue to work in the all-outgoing convention, and to package our data into the scalar invariants $X_{i,j}$. We will also restrict ourselves temporarily to $d \geq n - 1$, so that there are no additional constraints on kinematic space besides momentum conservation. (Note that this was not a problem for $d \geq 4$ in either $n = 4,5$.)

For a scalar $n$-point process, there are a total of $n(n-3)/2$ of these dynamical $X$'s. Further putting all internal modes on-shell means that $X_{1,i+2} = (p_1 + \ldots + p_{i+1})^2 = q_{I_i}^2 = -M_i^2$, where $q_{I_i}^\mu = p_1 + \ldots + p_{i+1}$ is the momentum of the internal particle $I_i$. There are exactly $n - 3$ of these on-shell conditions; thus, the partial wave expansion depends on 
\begin{equation}
    \frac{(n-2)(n - 3)}{2}
\end{equation}
dynamical variables.

We \textit{could} work with these scalar variables, but, as we saw in the four- and five-point examples, this problem appears to be much better suited to angular variables. This is not surprising, given that the partial wave expansion satisfies angular momentum eigen-equations in each of the internal particles. So, we now switch from scalar variables to angles $\theta_{i,j}$, which we define in terms of the momenta by
\begin{equation}\label{eq:costheta-def}
    \cos\theta_{i,j}
    = -\,\frac{G_{i,j}(A_i;B_j)}{\sqrt{G_{i,j}(A_i;A_i)\,G_{i,j}(B_j;B_j)}},
    \qquad 2\le i<j\le n-1,
\end{equation}
with
\begin{equation}
    A_i=\begin{cases} p_1, & i=2,\\ q_{i-2}, & 3\le i\le n-2,\end{cases}
    \qquad
    B_j=\begin{cases} q_{j-1}, & 3\le j\le n-2,\\ p_{n-1}, & j=n-1,\end{cases}
\end{equation}
where $G$ is the determinant
\begin{equation}
    G_{i,j}(a;b)=\det\begin{pmatrix}
    a\cdot b & a\cdot q_{i-1} & \cdots & a\cdot q_{j-2}\\
    q_{i-1}\cdot b & q_{i-1}\cdot q_{i-1} & \cdots & q_{i-1}\cdot q_{j-2}\\
    \vdots & \vdots & \ddots & \vdots\\
    q_{j-2}\cdot b & q_{j-2}\cdot q_{i-1} & \cdots & q_{j-2}\cdot q_{j-2}
\end{pmatrix}.
\end{equation}
Note that, for the polar angles $\theta_{i,i+1}$, this equation reduces to the general result
\begin{equation}
\cos\theta_{i,i+1}=
-\,
\frac{2M_{i-1}^{2}\big[X_{i,i+2}+M_{i-2}^{2}+M_{i}^{2}\big]
-\big(M_{i-1}^{2}+M_{i-2}^{2}-m_i^{2}\big)\big(M_{i-1}^{2}+M_{i}^{2}-m_{i+1}^{2}\big)}
{\sqrt{\lambda\!\left(M_{i-1}^{2},M_{i-2}^{2},m_i^{2}\right)\,\lambda\!\left(M_{i-1}^{2},M_{i}^{2},m_{i+1}^{2}\right)}},
\end{equation}
where $M_0 = m_1$ and $M_{n-2} = m_n$. We saw examples of this function at four- and five-points in Sec.~\ref{sec:review}. For depth-$1$ azimuthal angles $\theta_{i,i+2}$, one can check that Eq.~\eqref{eq:costheta-def} exactly reproduces the physically-motivated result for $\theta_{2,4}$ in the case of the five-point partial wave expansion~\eqref{eq:cos24-fin}.

Now, at this point it is not at all obvious (besides matching at $n = 4,5$) what these angles are, and why they naturally appear in the problem. So let us now explain. We will begin at six-points in dimension $d \geq 5$. At that multiplicity, there are three polar angles, one for each internal particle: $\theta_{2,3}$, $\theta_{3,4}$, and $\theta_{4,5}$. As we have previously noted, these are defined by sitting in the rest frame of $I_i$ and measuring the angle between incoming and outgoing particles (i.e., they are angles between \textit{lines} that share a \textit{point}). In the six-point problem, there are also two azimuthal angles $\theta_{2,4}$ and $\theta_{3,5}$, which, as we discussed in Sec.~\ref{sec:review}, are defined as angles between \textit{planes} that share a \textit{line}: $(\vec{p}_2, \vec{p}_3)$ and $(\vec{p}_3, \vec{p}_4)$ in either $I_1$ or $I_2$ for the former, and $(\vec{p}_3, \vec{p}_4)$ and $(\vec{p}_4, \vec{p}_5)$ in either $I_2$ or $I_3$ for the latter. In $d = 4$, besides an overall choice of orientation which we will discuss in Sec.~\ref{sec:gram}, this is all the freedom we have, and so this set of angles completely specifies the process. However, starting at $d = 5$, there is another degree of freedom, a deeper azimuth $\theta_{2,5}$. This is the angle (in the rest frame of $I_2$) between the 3D \textit{hyperplanes} $(\vec{p}_2, \vec{p}_3, \vec{p}_4)$ and $(\vec{p}_3, \vec{p}_4, \vec{p}_5)$, which meet on a 2D plane. We know from standard linear algebra that a determinant can be used in any dimension $d$ to represent the angle between two dimension $D$ hyperplanes which meet on a $D - 1$ hyperplane living in that space, and so one can check that $\theta_{2,5}$ as defined just now agrees with what one computes from Eq.~\eqref{eq:costheta-def}.

The general rule is, in sufficiently high dimension, $\theta_{i,j}$ is the angle between the two $(j - i)$-dimensional hyperplanes $(\vec{p}_i, \vec{p}_{i+1}, \ldots, \vec{p}_{j-1})$ and $(\vec{p}_{i+1}, \vec{p}_{i+2}, \ldots, \vec{p}_{j-1}, \vec{p}_j)$, which meet on the $(j - i - 1)$-dimensional plane $(\vec{p}_{i+1}, \ldots, \vec{p}_{j-1})$. The rest frame in which it is defined can be any of the internal particles $I_{i-1}, I_i, \ldots I_{j-2}$ living between $p_i^\mu$ and $p_j^\mu$ on the half-ladder diagram.

Note here that we have chosen our angles such that the indices range from $2$ to $n-1$; that is, we have neglected to include $1$ and $n$, which breaks some of the inherent symmetry of the half-ladder diagram. If one wanted to use, for example, particle $1$ in the place of particle $2$ (so that, e.g., instead of $\theta_{2,3}$ we use $\theta_{1,3}$, which is the angle between $\vec{p}_1$ and $\vec{p}_3$ in the center-of-mass frame of $I_1$), we have the simple relation that $\theta_{1,i} = \pi - \theta_{2,i}$. This follows from the fact that, in the rest frame of $I_1$, $\vec{p}_1$ and $\vec{p}_2$ are anti-parallel vectors. Of course, the same relation holds for particles $n-1$ and $n$ for the analogous reason: $\theta_{i,n-1} = \pi - \theta_{i,n}$. We are \textit{a priori} free to use either convention: while the partial wave expansion we present in the following subsections is designed for $2$ and $n-1$, plugging in these relations yields the same structures for $1$ and $n$, just with modified signs for the normalizations.

\subsection{Six-points}\label{sec:six-pt}

Now that we have established the kinematics, we begin by proving the partial wave expansion for the $n = 6$ half-ladder Feynman diagram. By the discussion in the previous subsection, we parameterize the process by $6$ angles: the polar angles $\theta_{2,3}$, $\theta_{3,4}$, and $\theta_{4,5}$, and the azimuthal angles $\theta_{2,4}$, $\theta_{3,5}$, and $\theta_{2,5}$. We work in spacetime dimension $d \ge 5 = n - 1$ so that all momenta are independent except for the constraint of momentum conservation. (We postpone this discussion until Sec.~\ref{sec:gram}.) 

We will consider the partial wave expansion of a diagram with fixed internal spins $s_1$, $s_2$, and $s_3$ and ``vertex'' quantum numbers $\mu_1$ and $\mu_2$ belonging to each of the internal-external-internal three-point vertices. (This corresponds to $n = 6$ in Fig.~\ref{fig:n-pt-half-ladder}.) To ease notation, we will suppress these labels and denote the object of our interest as $R^{(6)} [ \theta_{i,j} ]$. 

To solve the problem, we note that --- just as at five-points --- $R^{(6)}$ must be an eigenvector of angular momentum in each of the rest frames of the internal particles. Starting with $I_1$, this means that $R^{(6)}$ must satisfy the same equation as at four and five-points:
\begin{equation}
\label{eq:s-1-eigen}
    \hat{L}_{I_1}^2 R^{(6)} = - \Delta_{S^{d-2}} R^{(6)} = s_1 (s_1 + d - 3) R^{(6)},
\end{equation}
where the $\Delta_{S^{d-2}}$ operator is the same as at five-points:
\begin{equation}
\label{eq:def-Delta}
    \Delta_{S^{d-2}} = \partial_{\theta_{2,3}}^2 + (d - 3) \cot \theta_{2,3} \partial_{\theta_{2,3}} + \frac{1}{\sin^2 \theta_{2,3}} \Delta_{S^{d-3}}.
\end{equation}
Here, we have written out $\Delta_{S^{d-2}}$ in the standard way, separating polar from azimuthal: $\theta_{2,3}$ is the polar angle of the problem, while $\Delta_{S^{d-3}}$ is an operator which acts solely on dependence on two of the azimuthal angles $\theta_{2,4}$ and $\theta_{2,5}$. Thus, to solve Eq.~\eqref{eq:s-1-eigen}, we first split $R^{(6)}$ into:
\begin{equation}
    R^{(6)} = f(\theta_{2,3}) g(\theta_{2,4}, \theta_{2,5}) h(\theta_{4,5}, \theta_{3,4}, \theta_{3,5}).
\end{equation}
Plugging this ansatz into Eq.~\eqref{eq:s-1-eigen}, we can cancel out $h$. Then, requiring that
\begin{equation}
\label{eq:Delta-one-d-down}
    \hat{J}_1 g = -\Delta_{S^{d-3}} g = \mu_1 (\mu_1 + d - 4) g,
\end{equation}
for integer $0 \leq \mu_1 \le s_1$, we find that $f$ must satisfy
\begin{equation}
    f'' + (d - 3) \cot \theta_{2,3} f' + \left(s_1 (s_1 + d - 3) - \frac{\mu_1(\mu_1 + d - 4)}{\sin^2 \theta_{2,3}} \right)f = 0,
\end{equation}
the same equation~\eqref{eq:f-5pt-equation} which appeared at five-point. Thus, the solution is $f = A^d_{s_1,\mu_1}$ with $A$ given as in Eq.~\eqref{eq:A-defs}. It is clear that we can do the same procedure for $g$: separate $g(\theta_{2,4}, \theta_{2,5}) = g_1(\theta_{2,4}) g_2(\theta_{2,5})$. Now, in Eq.~\eqref{eq:Delta-one-d-down}, $\theta_{2,4}$ plays the role of the polar angle and $\theta_{2,5}$ the azimuthal angle. Hence, setting
\begin{equation}
\label{eq:g-2-l-eq}
    \hat{J}' g_2 = -\Delta_{S^{d-4}} g_2 = l ( l + d - 5) g_2,
\end{equation}
now for integer $0 \leq l \leq \mu_1$, we naturally find $g_1 = A^{d-1}_{\mu_1, l}$. Finally, $g_2$ depends only on one angle, and thus it satisfies
\begin{equation}
    g_2'' + (d - 5) \cot \theta_{2,5} g_2' + l (l + d - 5) g_2 = 0.
\end{equation}
This gives $g_2 = A^{d-2}_{l,0} = C_l^{\left( \frac{d - 5}{2}\right)}$. 

We can apply the same procedure to $I_3$, which gives the analogous results for the dependence of $R^{(6)}$ on $\theta_{4,5}$, $\theta_{3,5}$, and $\theta_{2,5}$. (In particular, for the dependence in $\theta_{2,5}$, one sees that the eigenvalue problem of $I_3$ agrees with that of $I_1$, so long as we take $l \le \min (\mu_1,\mu_2)$.) 

Thus we have determined exactly how $R^{(6)}$ must depend on all angles save for the middle polar angle $\theta_{3,4}$. To finish the problem, we must therefore invoke the requirement that, in the rest frame of $I_2$, $R^{(6)}$ is an eigenvector of angular momentum with value $s_2(s_2 + d - 3)$. This means that
\begin{equation}
\label{eq:mid-eq}
    \hat{L}_{I_2}^2 R^{(6)} = s_2 (s_2 + d - 3) R^{(6)}.
\end{equation}
The operator $\hat{L}_{I_2}^2$ is slightly more complicated in this case, since it acts on \textit{four} angles rather than on three. It may be written as
\begin{equation}\label{eq:LI2}
    \hat{L}_{I_2}^2 = -\partial^2_{\theta_{3,4}} - (d - 3) \cot \theta_{3,4} \partial_{\theta_{3,4}} + \frac{\hat{J}_1 + \hat{J}_2 - (1 + \cos^2 \theta_{3,4}) \hat{J}' - 2 \cos \theta_{3,4} \hat{\Xi}}{\sin^2 \theta_{3,4}},
\end{equation}
where $\hat{J}_i$, $\hat{J}'$, and $\hat{\Xi}$ are operators acting exclusively on the azimuthal angles. $\hat{J}_1$ and $\hat{J}'$ are the same operators listed in Eq.~\eqref{eq:Delta-one-d-down} and Eq.~\eqref{eq:g-2-l-eq}, respectively, and $\hat{J}_2$ is $\hat{J}_1$ but associated to $I_3$ instead of $I_1$. $\hat{\Xi}$ is the genuinely new operator, given by
\begin{equation}\label{eq:Xi-def}
\begin{split}
\hat{\Xi} = {}& -\cos\theta_{2,5}\,\partial_{\theta_{2,4}}\partial_{\theta_{3,5}}
    +\sin\theta_{2,5}\left[\cot\theta_{3,5}\,\partial_{\theta_{2,4}}
    +\cot\theta_{2,4}\,\partial_{\theta_{3,5}}\right]\partial_{\theta_{2,5}} \\
    & +\cos\theta_{2,5}\cot\theta_{2,4}\cot\theta_{3,5}\,\partial^2_{\theta_{2,5}}
    +(d-5)\,\frac{\cot\theta_{2,4}\cot\theta_{3,5}}{\sin\theta_{2,5}}\,\partial_{\theta_{2,5}}.
\end{split}
\end{equation}
With this information, to solve this eigen-equation we make the following ansatz: we separate $R^{(6)}$ into
\begin{equation}\label{eq:6pt-ansatz}
    R^{(6)} = A^d_{s_1,\mu_1}(\theta_{2,3}) F^d_{s_2,\mu_1,\mu_2}(\theta_{3,4}, \theta_{2,4}, \theta_{2,5}, \theta_{3,5}) A^d_{s_3,\mu_2}(\theta_{4,5}),
\end{equation}
where we take
\begin{equation}
\label{eq:six-pt-ansatz}
    F^d_{s_2,\mu_1,\mu_2} = \sum_{l = 0}^{\min{(\mu_1,\mu_2)}} (-1)^l \widetilde{T}^d_{s_2,\mu_1,\mu_2,l} ( \theta_{3,4}) A^{d-1}_{\mu_1,l}(\theta_{2,4}) A^{d-1}_{\mu_2,l}(\theta_{3,5}) C^{\left( \frac{d - 5}{2}\right)}_l(\cos \theta_{2,5}),
\end{equation}
for some function $\widetilde{T}$ we need to solve for. Note that this ansatz automatically satisfies the eigen-equations of $I_1$ and $I_3$. Denoting by $t_l$ the product of $AAC$ in the above ansatz, one can check that the action of $\hat{\Xi}$ on $t_l$ is to move $l$ up and down by one unit:
\begin{equation}
\label{eq:Xi-action}
    \hat{\Xi} t_l = a_l(\nu_0) t_{l+1} + b_l(\nu_0) t_{l-1},
\end{equation}
with coefficients $a_{l_k}(\nu_r)$ and $b_{l_k}(\nu_r)$ given in general by
\begin{equation}
\label{eq:al-bl}
\begin{aligned}
    a_{l_k}(\nu_r) &= -\frac{2\,(l_k+1)(\nu_{r+1} + l_k)^2}{\nu_{r+2} + l_k},\\[4pt]
    b_{l_k}(\nu_r) &= -\frac{(2\nu_{r+3} + l_k)(\mu_k - l_k + 1)(\mu_{k+1} - l_k + 1)(2\nu_{r+2} + l_k + \mu_k)(2\nu_{r+2} + l_k + \mu_{k + 1})}{8\,(\nu_{r+3}+ l_k)^2 (\nu_{r+2} + l_k)},
\end{aligned}
\end{equation}
where
\begin{equation}
    \nu_r = \frac{d - 3 - r}{2}.
\end{equation}
Thus, plugging Eq.~\eqref{eq:six-pt-ansatz} into the eigenvalue equation Eq.~\eqref{eq:mid-eq}, we find that $\widetilde{T}_l(x)$ (suppressing all other indices) must satisfy:
\begin{equation}
\label{eq:T-eq}
    -(1 - x^2) \widetilde{T}_l'' + (d - 2) x \widetilde{T}_l' + \frac{E^{d-1}_{\mu_1} + E^{d-1}_{\mu_2} - (1 + x^2) E^{d-2}_l}{1 - x^2} \widetilde{T}_l + \frac{2x}{1 - x^2}\left(a_{l-1}(\nu_0) \widetilde{T}_{l - 1} + b_{l + 1}(\nu_0)\widetilde{T}_{l + 1} \right) = E^d_{s_2} \widetilde{T}_l,
\end{equation}
where $E^d_s \equiv s ( s + d - 3)$ and $x \equiv \cos \theta_{3,4}$. Solving this set of coupled ODEs then yields a solution for $\widetilde{T}_l$, in the form of a sum over Gegenbauer polynomials. We find, for a dimension $D = d - r$ for natural number $r$, that
\begin{equation}\label{eq:T-def}
    \widetilde{T}^{d-r}_{s,\mu_1,\mu_2,l}(x)=w_l(\nu_r)\,(1-x^2)^{|\Delta \mu|/2}\;
    \frac{\displaystyle\sum_{j=0}^{K}\gamma_j^{(r)}\,
    C^{\left(\nu_r+l+|\Delta \mu|+j\right)}_{\,s-|\Delta \mu|-l-2j}(x)}
    {\displaystyle\sum_{j=0}^{K}\gamma_j^{(r)}\,
    C^{\left(\nu_r+l+|\Delta \mu|+j\right)}_{\,s-|\Delta \mu|-l-2j}(1)}\,,
\end{equation}
where $\Delta \mu  = \mu_1 - \mu_2$ and $K = \min( \min(\mu_1,\mu_2) - l, \lfloor \frac{s - |\Delta \mu| - l}{2}\rfloor)$. The denominator is just a normalization condition. Note that, when $l$ is extremal, that is, when $l = \min(\mu_1,\mu_2)$, the sum above collapses into a single Gegenbauer. Further, with $\gamma^{(r)}_0 = 1$, the coefficients of the expansion $\gamma^{(r)}_j$ are given by
\begin{equation}
\begin{split}
    \gamma_j^{(r)} = \frac{(\alpha)_j}{(\alpha-|\Delta \mu|)_j}
    \sum_{i=0}^{\min(j,\,K'-j)} &(-1)^{j+i}\,i!\,
    \binom{K'}{j}\binom{j}{i}\binom{K'-j}{i} \\
    &\times\frac{(\beta+|\Delta \mu|+K'-1)_j\,(\beta+j+i-1)_{j-i}\,(\beta+|\Delta \mu|+K'+j-1)_i}
    {(n_0-j+1)_j\,(n_0+|\Delta \mu|-j+1)_j\,(n_0-j-i+1)_i},
\end{split}
\end{equation}
where
\[
\alpha=\nu_r+l+|\Delta \mu|,\qquad
n_0=s-|\Delta \mu|-l,\qquad
K'=\min(\mu_1,\mu_2)-l,\qquad
\beta=2\nu_r+2l .
\]
Finally, the $w_l$'s out front come out quite simply to be related to the coefficients of the Gegenbauer classical addition theorem $b_{n,m}(\mu)$: with $w_0(\nu_r) = 1$, they are
\begin{equation}
\label{eq:rat-w-6}
    w_\ell(\nu_r)
    =\frac{\binom{p}{\ell}}{\binom{q}{\ell}}\,
    \frac{b_{p\ell}(\nu_{r+1})}{b_{p0}(\nu_{r+1})}
    =4^{\ell}\,\frac{(q-\ell)!}{q!}\;\frac{\nu_{r+2}+\ell}{\nu_{r+2}}\;
    \frac{\big[\big(\nu_{r+1}\big)_\ell\big]^2}{(2\nu_{r+1} +p)_\ell},
\end{equation}
where $p = \min(\mu_1, \mu_2)$ and $q=\max(\mu_1,\mu_2)$. These $b_{n,m}(\nu)$ are defined via the identity
\begin{equation}\label{eq:bnm-thm}
\begin{split}
C^{(\lambda)}_n\!\big(\cos\theta_1\cos\theta_2+\sin\theta_1\sin\theta_2\cos\phi\big)
=\sum_{m=0}^{n} b_{n,m}(\lambda)\,
&(\sin\theta_1)^m C^{(\lambda+m)}_{n-m}(\cos\theta_1)\,
(\sin\theta_2)^m C^{(\lambda+m)}_{n-m}(\cos\theta_2) \\
&\times\, C^{(\lambda-\frac12)}_m(\cos\phi),
\end{split}
\end{equation}
and are explicitly given by
\begin{equation}
b_{n,m}(\lambda)
=\frac{2^{2m}\,(n-m)!\,\big[(\lambda)_m\big]^2\,(2\lambda+2m-1)}{(2\lambda-1)_{n+m+1}},
\qquad 0\le m\le n,
\label{eq:bnm}
\end{equation}
so that $b_{n,0}(\lambda)=n! / (2\lambda)_n$ and $b_{0,0}=1$. As one can see, $b_{n,m}$ is a function of the order $\lambda$ of the Gegenbauer polynomial in Eq.~\eqref{eq:bnm-thm}. $\widetilde{T}_l$ itself turns out to be the all-$d$ generalization of the Wigner-$d$ matrix in $d = 4$; in Sec.~\ref{sec:gram}, we elaborate further on this relationship.

Roughly, the reason these coefficients appear is the following. Suppose the internal-external-internal vertex were trivial, so that the emitted particle did nothing to the diagram. Then the five-point half-ladder would be the same as the four-point one: a single Gegenbauer polynomial $C^{(\lambda)}_s(\cos\Theta)$,
where $\Theta$ is the angle between $\vec p_2$ and $\vec p_4$ in the rest frame of the internal line. We write $\Theta$ in terms of the angular variables of the six-point problem as $\cos\Theta=-\cos\theta_{2,3}\cos\theta_{3,4}
+\sin\theta_{2,3}\sin\theta_{3,4}\cos\theta_{2,4}$, which is precisely the argument in Eq.~\eqref{eq:bnm-thm}. Rewriting $C^{(\lambda)}_s(\cos\Theta)$ in the five-point basis~\eqref{eq:five-pt-basis} is therefore exactly the addition
theorem, with $b_{s,\mu}(\lambda)$ as the expansion coefficients; at higher-points, the same step is repeated once per internal vertex.

Thus, putting it all together, we have proved a totally explicit partial wave expansion for the six-point half-ladder in any spacetime dimension: for a given set of $\{s_i\}$, $\{\mu_i\}$, it reads
\begin{equation}\label{eq:six-pt-partial-wave}
\begin{aligned}
    R^{(6)}=N^{(6)}\,A^d_{s_1,\mu_1}(\theta_{2,3})\,A^d_{s_3,\mu_2}(\theta_{4,5})
    \sum_{l=0}^{\min(\mu_1,\mu_2)}&(-1)^l\sqrt{b_{\mu_1,l}(\nu_1)\,b_{\mu_2,l}(\nu_1)}\;
    T^{d}_{s_2,\mu_1,\mu_2,l}(\theta_{3,4})\\
    &\times A^{d-1}_{\mu_1,l}(\theta_{2,4})\,A^{d-1}_{\mu_2,l}(\theta_{3,5})\,
    C^{\left(\nu_2\right)}_l(\cos\theta_{2,5}),
\end{aligned}
\end{equation}
up to an overall normalization $N^{(6)}$ which we will deal with in the next subsection. The rescaled $T_l$ is related to the $\widetilde{T}_l$ by\footnote{We will see why this particular normalization choice is useful when we prove the generic expansion in the next subsection.}
\begin{equation}\label{eq:rescaled-T}
T^{\,d-r}_{s,\mu_1,\mu_2,l}
\equiv\frac{\widetilde T^{\,d-r}_{s,\mu_1,\mu_2,l}}{\sqrt{b_{\mu_1,l}(\nu_{r+1})\,b_{\mu_2,l}(\nu_{r+1})}}
\sqrt{\frac{(s-q+1)_{q-p}\,(s+p+2\nu_{r})_{q-p}}{c_{q,p}(\nu_r)}},
\end{equation}
for $p=\min(\mu_1,\mu_2),\ q=\max(\mu_1,\mu_2)$, and $c_{q,p}$ given by
\begin{equation}
    c_{q,p}(\nu_r)=\big[(q-p)!\big]^{2}\,\frac{p!}{q!}\;4^{\,q-p}\;
    \frac{\big(\nu_{r+1}+p\big)_{q-p}\,\big(\nu_{r-1}+p\big)_{q-p}}
    {(2\nu_{r+1}+p)_{q-p}}.
\end{equation}
For a generic six-point half-ladder residue, we use the orthogonality of the partial wave basis to compute the coupling matrix $y_{\vec{s},\vec{\mu}}$. We give the explicit formulae in App.~\ref{sec:ortho}; in particular, the answer at $n = 6$ is given in Eqs.~\eqref{eq:ortho-con} and~\eqref{eq:six-pt-ortho}.

All special functions appearing in the expansion are (sums of) Gegenbauer polynomials, and all other components necessary to practically implement this expansion are given throughout this subsection. As it turns out, the $n = 6$ case is sufficiently general to introduce the full cast-of-characters at any multiplicity, and so now we turn to a full proof of the partial-wave expansion.

\begin{figure}[t]
    \centering
    \includegraphics[width=0.5\linewidth]{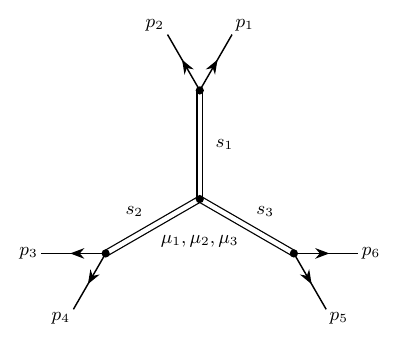}
    \caption{The ``Mercedes Benz'' diagram at six-points. Each external vertex is known uniquely from Lorentz invariance, while the middle, purely-internal vertex is labeled by three quantum numbers $\mu_1, \mu_2, \mu_3$.}
    \label{fig:benzy}
\end{figure}

\medskip
\noindent \textbf{Mercedes Benz Diagram.} At six-points, unlike at four- and five-points, there is a new diagram topology that can appear, called the ``snowflake'' or the ``Mercedes Benz.'' This diagram --- shown in Fig.~\ref{fig:benzy} --- consists of three external-external-internal vertices (known uniquely from Lorentz invariance) and one new purely internal-internal-internal vertex in the middle. This vertex can be parameterized by three quantum numbers $a,b,c$, each corresponding to the contraction between a pair of the three particles' polarization tensors:
\begin{equation}
\mathcal A_3 \;=\; \sum_{a,b,c\ge0} g_{abc}\,
(\epsilon_1\!\cdot\!\epsilon_2)^a (\epsilon_2\!\cdot\!\epsilon_3)^b (\epsilon_3\!\cdot\!\epsilon_1)^c\,
(\epsilon_1\!\cdot\! q_2)^{s_1-a-c} (\epsilon_2\!\cdot\! q_3)^{s_2-a-b} (\epsilon_3\!\cdot\! q_1)^{s_3-b-c},
\end{equation}
where $a + c\leq s_1$, $a + b \leq s_2$, and $b + c \leq s_3$. So now our basis element will have six quantum numbers, as opposed to five in the six-point half-ladder.

So long as we are in $d \ge 5$, the scattering process is still parameterized by six angular variables. In an analogy to the half-ladder, we take the polar angle $\theta_i$ for $i = 1, 2, 3$ to be the angle between $\vec{p}_{2i-1}$ and $\vec{q}_{i+1} = -\vec{q}_{i+2}$ in the rest frame of $I_i$, where $q_i = p_{2i-1} + p_{2i}$ is the momentum of $I_i$ ($p_i$'s are to be understood mod $6$, and $q_i$'s mod $3$.) The three azimuths are then all depth-$1$ azimuths: we take $\phi_{i,j}$ to be the angle between $(\vec{p}_{2i-1}, \vec{q}_{i+1})$ and $(\vec{q}_{i+1}, \vec{p}_{2j-1})$ in the rest frame of $I_i$, with $i \neq j$. (This is in contrast to the situation with the half-ladder, where one azimuth is at depth-$2$.) The independent angles are thus $\phi_{1,2}$, $\phi_{2,3}$, and $\phi_{3,1}$. With this parameterization, we observe that the partial wave basis takes the following form:
\begin{equation}
    R_{\mathrm{MB}}^{(6)} = A_{s_1,\mu_1}^d (\theta_1) A_{s_2,\mu_2}^d(\theta_2) A_{s_3, \mu_3}^d (\theta_3) K^{d}_{\mu_1, \mu_2, \mu_3} (\cos \phi_{1,2}, \cos \phi_{2,3}, \cos \phi_{3,1}).
\end{equation}
Each polar angle comes with an associated Gegenbauer, which is exactly what one would expect from our explorations of the half-ladder diagram. We find that the $\mu$'s themselves range over all values such that:
\begin{equation}
0\le\mu_i\le s_i,\qquad |\mu_i-\mu_j|\le\mu_k\le\mu_i+\mu_j,\qquad \mu_1+\mu_2+\mu_3\ \ \text{even}.
\end{equation}
The new function is $K$, which depends only on these three azimuthal angles. Some examples of this function are given here:
\begin{align}
K_{110} &= x, \qquad
K_{220} = x^2-\frac1D, \qquad
K_{330} = x^3-\frac{3x}{D+2}, \qquad
K_{440} = x^4-\frac{6x^2}{D+4}+\frac{3}{(D+2)(D+4)}, \nonumber\\[4pt]
K_{211} &= xz-\frac{y}{D}, \qquad\qquad
K_{321} = x^2z-\frac{2xy}{D+2}-\frac{z}{D+2}, \nonumber\\[4pt]
K_{222} &= xyz-\frac{x^2+y^2+z^2}{D}+\frac{2}{D^2}, \\[4pt]
K_{422} &= x^2z^2-\frac{x^2+z^2+4xyz}{D+4}+\frac{2y^2+1}{(D+2)(D+4)}, \nonumber\\[4pt]
K_{332} &= x^2yz-\frac{x^3}{D}-\frac{2x(y^2+z^2)}{D+2}-\frac{(D-2)\,yz}{(D+2)^2}+\frac{(5D+6)\,x}{D(D+2)^2}, \nonumber\\[4pt]
K_{334} &= xy^2z^2-\frac{4x^2yz}{D+4}-\frac{2yz(y^2+z^2)}{D+2}-\frac{(D-4)\,x(y^2+z^2)}{(D+2)(D+4)}
+\frac{2x^3+12\,yz}{(D+2)(D+4)}+\frac{(D-10)\,x}{(D+2)^2(D+4)}, \nonumber
\end{align}
with $D = d - 2$ and $x = \cos\phi_{1,2}$, $y = \cos\phi_{2,3}$ and $z = \cos \phi_{3,1}$. You'll note that, when $\mu_1 = \mu_2 = \mu$ and $\mu_3 = 0$ for example, we find $K_{\mu\mu0}(x,y,z) \propto C^{\left( \frac{d -4}{2} 
\right)}_\mu(x)$, exactly what one would expect from the half-ladder diagram. However, it is clear that $K$ does not, in general, separate: when all $\mu$'s are nonzero, the $K$'s mix $x,y,z$ together. This is also true of the half-ladder diagram, where the function $F^d$~\eqref{eq:6pt-ansatz} can be thought of as the analog of $K^d$.

Furthermore, as it turns out, we find that these $K^d$ functions are \textit{orthogonal} under measure
\begin{equation}
    d\mu \propto (1 - x^2 - y^2 - z^2 + 2xyz)^{\frac{D-4}{2}} dx dy dz.
\end{equation}
As such, the basis of internal three-point amplitudes defined by $\mu_1, \mu_2, \mu_3$ is itself orthogonal, in perfect analogy to the situation with the single vertex quantum number $\mu$ for the half-ladder.

These are just some preliminary facts about the object; we leave more detailed work to future investigations.

\subsection{All multiplicity and dimension}\label{sec:proof}

Let us now move to a general construction of the half-ladder partial wave expansion in any spacetime dimension and at any multiplicity. This will be a recursive proof, with $n = 6$ in Eq.~\eqref{eq:six-pt-partial-wave} providing the base case. As in the previous subsection, we assume here that $d \ge n - 1$. We also will only consider internal lines $I_i$ which are symmetric traceless tensor (STT) irreps of the massive little group $SO(d-1)$; the case for more interesting irreps should be a straightforward generalization, though we do not pursue it here.\footnote{In the previous section, we implicitly assumed this, although in the $n = 6$ half-ladder it is indeed possible for a $2$-form to propagate as $I_2$ in sufficiently high dimension.}

We begin by taking a closer look at Eq.~\eqref{eq:six-pt-partial-wave}, where we recognize in the $l$-sum the presence of a five-point partial wave basis element in precisely one lower dimension:
\begin{equation}
    R^{(5, d-1)}_{(\mu_1,\mu_2);l}(\phi_{2,3},\phi_{3,4}, \phi_{2,4}) = A^{d-1}_{\mu_1,l}(\phi_{2,3}) A^{d-1}_{\mu_2,l}(\phi_{3,4}) C^{\left( \nu_{2}\right)}_l(\cos \phi_{2,4}),
\end{equation}
where we have $\phi_{i,j} = \theta_{i,j+1}$ (and, also, we are no longer suppressing indices). Thus, the six-point partial wave basis element can be written as
\begin{equation}\label{eq:6pt-w-5pt-inside}
\begin{aligned}
    R^{(6,d)}_{(s_1,s_2,s_3);(\mu_1,\mu_2)} = N^{(6)}\, A^d_{s_1,\mu_1}(\theta_{2,3})\, A^d_{s_3,\mu_2}(\theta_{4,5})
    \sum_{l = 0}^{\min(\mu_1,\mu_2)} &(-1)^l\sqrt{b_{\mu_1,l}(\nu_1)\,b_{\mu_2,l}(\nu_1)}\\
    &\times T^d_{s_2,\mu_1,\mu_2,l}(\theta_{3,4})\, R^{(5,d-1)}_{(\mu_1,\mu_2);l}.
\end{aligned}
\end{equation}
Additionally, for an inductive proof we must find a definitive way to fix the normalization $N^{(6)}$. A particularly convenient way to do this is to require that:
\begin{equation}
\label{eq:6-Xi-cond}
    \hat{\Xi}_k R^{(6,d-1)}_{\vec{\mu};\vec{l}} = a_{l_{k-1}}(\nu_0) R^{(6,d-1)}_{\vec{\mu};\vec{l} + e_{k-1}} + b_{l_{k-1}}(\nu_0) R^{(6,d-1)}_{\vec{\mu};\vec{l} - e_{k-1}},
\end{equation}
for both $k= 2,3$, and where the vector $e_{i} = l_i \hat{l_i}$. The operator $\hat{\Xi}_k$ is Eq.~\eqref{eq:Xi-def} associated with the eigen-equation of internal particle $I_k$ instead of with $I_2$. Note that $R^{(5)}$ already satisfies this condition via Eq.~\eqref{eq:Xi-action}. One can show this is possible by selecting:
\begin{equation}
    N^{(6)}=(-1)^{\min(\mu_1,\mu_2)}\,
    \sqrt{\frac{b_{\mu_1,0}(\nu_1)\,b_{\mu_2,0}(\nu_1)}{b_{s_2,\mu_1}(\nu_0)\,b_{s_2,\mu_2}(\nu_0)}},
\end{equation}
where we have moved back up to $d$ dimensions so that $N^{(6)}$ corresponds to the normalization of Eq.~\eqref{eq:6pt-w-5pt-inside}. So, we will take $R^{(6)}$ to have this canonical normalization.

Having established the base case, let us now proceed with the inductive step. We first want to note that a given $\theta_{i,j}$ of the $(n+1)$-point problem with $j-i\ge 2$ has the same weight in the orthogonality measure, and enters the eigenvalue equations in the same way, as the angle $\theta_{i,j-1}$ of the $n$-point problem in one dimension lower. For example, the azimuthal angle $\theta_{2,4}$ of the $(n+1,d)$ problem plays the role of the polar angle $\theta_{2,3}$ of the $(n,d-1)$ problem, as one can see at $n = 4$ in Eq.~\eqref{eq:five-pt-basis}. We just showed that this holds at $n = 6$, and in this case it is really manifest in Eq.~\eqref{eq:LI2}, whose azimuthal operators $\hat J_1$, $\hat J_2$, and $\hat J'$ are precisely those of the five-point problem in $d-1$ dimensions.

The $(n+1)$-point partial wave basis element $R^{(n+1,d)}_{(s_1, \ldots, s_{n-2});(\mu_1,\ldots, \mu_{n-3})}$ must be an eigenvector of angular momentum in each of the internal particles $I_1$, $I_2$, $\ldots$, $I_{n-2}$. We now venture an ansatz which is the natural extension of Eq.~\eqref{eq:6pt-w-5pt-inside}: we write
\begin{equation}
    R^{(n+1,d)}_{\vec{s};\vec{\mu}} = N^{(n+1)} A^d_{s_1,\mu_1}(\theta_{2,3}) A^d_{s_{n-2},\mu_{n-3}}(\theta_{n-1,n}) \sum_{l_i = 0}^{\min(\mu_i,\mu_{i+1})} Y_{\vec{l}}(\nu_1)\left( \prod_{k = 2}^{n - 3} T^d_{s_k, \mu_{k-1}, \mu_k, l_{k-1}}(\theta_{k+1,k+2}) \right) R^{(n,d-1)}_{\vec{\mu};\vec{l}},
\end{equation}
with shorthand $\vec{s} = (s_1, \ldots, s_{n-2})$ and $\vec{\mu} = (\mu_1,\ldots, \mu_{n-3})$, and $N^{(n+1)}$ is the normalization which we will return to momentarily. We take $R^{(n,d-1)}_{\vec{\mu};\vec{l}}$ to satisfy the condition Eq.~\eqref{eq:6-Xi-cond} for $k = 2, \ldots, n-3$, just as it did at $n = 6$.

This form automatically satisfies the eigenvalue equations of the internal particles on the far-left and far-right, $I_1$ and $I_{n - 2}$. For any other internal particle $I_k$, a set of ODEs exactly as we had with $I_2$ in Eq.~\eqref{eq:T-eq}:
\begin{equation}
\begin{split}
    -(1-x^2)\,f_{l_{k-1}}'' + (d-2)\,x\,f_{l_{k-1}}'
    &+ \frac{E^{d-1}_{\mu_{k-1}} + E^{d-1}_{\mu_k} - (1+x^2)\,E^{d-2}_{l_{k-1}}}{1-x^2}\, f_{l_{k-1}} \\
    &- \frac{2x}{1-x^2}\, \Big( a_{l_{k-1}-1}(\nu_0)\, f_{l_{k-1}-1} + b_{l_{k-1}+1}(\nu_0)\, f_{l_{k-1}+1} \Big)
    = E^{d}_{s_k}\, f_{l_{k-1}},
\end{split}
\end{equation}
where $x \equiv \cos \theta_{k+1,k+2}$ and
\begin{equation}
    f_{l_{k-1}} \equiv Y_{\vec{l}}(\nu_1) T^d_{s_k, \mu_{k - 1},\mu_k, l_{k-1}}.
\end{equation}
Thus, the solution is simply the product of the solutions to Eq.~\eqref{eq:T-eq}, yielding a coefficient matrix of
\begin{equation}
    Y_{\vec l}(\nu_1)=\prod_{a=1}^{n-4}(-1)^{l_a}\sqrt{b_{\mu_a,l_a}(\nu_1)\,b_{\mu_{a+1},l_a}(\nu_1)}.
\end{equation}
Finally, to close the recursion, we must fix the overall normalization $N^{(n+1)}$. Given that we assumed it was a property of the $n$-point partial wave basis, we need our new $(n+1)$-point element to satisfy Eq.~\eqref{eq:6-Xi-cond} (with $6$ replaced with $n + 1$) for $k = 2, \ldots, n-2$. This naturally gives us,
\begin{equation}
    N^{(n+1)}=(-1)^{\sum_{i=1}^{n-4}\min(\mu_i,\mu_{i+1})}\prod_{j=2}^{n-3}
    \sqrt{\frac{b_{\mu_{j-1},0}(\nu_1)\,b_{\mu_j,0}(\nu_1)}{b_{s_j,\mu_{j-1}}(\nu_0)\,b_{s_j,\mu_j}(\nu_0)}},
\end{equation}
completing the recursive proof.

App.~\ref{sec:ortho} explicitly gives formulae relating to the orthogonality of the general $n$-point half-ladder, providing for us a practical means to expand any given half-ladder residue in this partial wave basis.

Additionally, as we did at five-points~\eqref{eq:five-pt-k}, we can move the $n$-point partial-wave element in the $\mu$-basis $R^{(n,d)}_{\vec s;\vec\mu}$ into that written in the $b$-basis~\eqref{eq:3pt-expansion} of the internal three-particle vertex $\widetilde{R}^{(n,d)}_{\vec s;\vec b}$ as:
\begin{equation}\label{eq:k-to-mu-npt}
    \widetilde{R}^{(n,d)}_{\vec s;\vec b} = \sum_{\mu_1 = 0}^{b_1}\cdots\sum_{\mu_{n-4} = 0}^{b_{n-4}}
    \Big[\prod_{i=1}^{n-4} a^{(i)}_{b_i,\mu_i}\Big]\, R^{(n,d)}_{\vec s;\vec\mu},
\end{equation}
with $a^{(i)}_{b,\mu}$ the matrix~\eqref{eq:a-k-mu-form} of the vertex joining $I_i$ and $I_{i+1}$. Since gluing is linear in each vertex, the change of basis is simply the tensor product of the five-point matrices. Therefore, we can proceed in exactly the same manner as at five-points: extract the partial-wave coefficients in the $\mu$-basis due to its inherent orthogonality, and then rotate back to the $b$-basis to get the picture in terms of the familiar dot products of epsilons and momenta that make up the structure of the internal three-point amplitudes.

\subsection{A graphical rule for the partial wave basis}\label{sec:writing}

In the previous subsection, we found and proved a general expression for the partial-wave expansion of the half-ladder at any $n$ and any $d \ge n-1$. However, the final expression was given in the form of a recursion. Now we would like to lay out an algorithmic means of constructing the basis directly from the special functions $T$, $A$, and $C$.

\begin{figure}[t]
    \centering
    \includegraphics[width=1.0\linewidth]{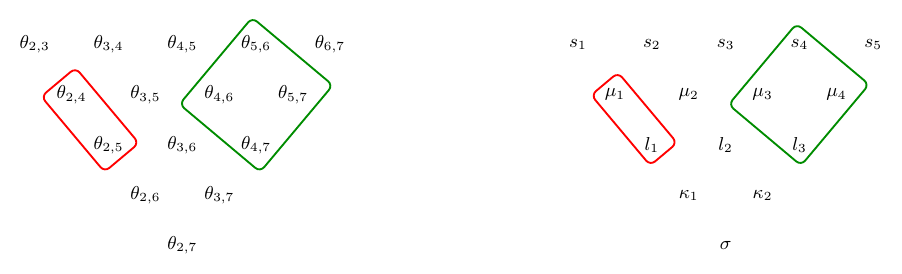}
    \caption{The kinematics (left) and the quantum numbers (right) for the half-ladder partial wave basis at $n = 8$ points. Following the rules laid out in Sec.~\ref{sec:writing}, the rectangle in red yields a contribution of $A^{d-1}_{\mu_1,l_1}(\theta_{2,4})$, and the diamond in green gives $T^d_{s_4,\mu_3,\mu_4,l_3}(\theta_{5,6})$.}
    \label{fig:8pt-kins}
\end{figure}

First, we note again that the number of variables at $n$-points is precisely:
\begin{equation}
    \frac{n(n-3)}{2} - (n - 3) = \frac{(n-2)(n-3)}{2},
\end{equation}
which are the \textit{triangle numbers}. Thus we may organize the kinematic data $\theta_{i,j}$ by an upside-down triangle, as shown in the left-hand side of Fig.~\ref{fig:8pt-kins} for $n = 8$. The highest layer are the polar angles, and the rest are azimuthal angles. To each row of the triangle, we associate a set of quantum numbers: depth $0$ (the polar angles) gets the spins $s_i$, depth $1$ gets the vertex quantum numbers $\mu_i$, and so on and so forth down to the tip of the triangle. (We show this on the right-hand side of Fig.~\ref{fig:8pt-kins} for $n = 8$.) For the purposes of this discussion, let us write any quantum number as $l^{(r)}_p$, where $r$ is the depth (starting at $0$) and $p$ is the location of $l^{(r)}_p$ from left-to-right starting at $p = 1$. For example, we have $s_i = l^{(0)}_i$ and $\mu_i = l^{(1)}_i$. 

Let us begin the algorithm by determining the functional dependence on these angles, via four separate cases. In the case where an angle is located on the \textit{left outermost} edge of the triangle at depth $r \neq n - 4$ (that is, not at the tip), it enters as
\begin{equation}
    A^{d - r}_{l^{(r)}_1, l^{(r+1)}_1}(\theta_{2, 3 + r}).
\end{equation}
For the \textit{right outermost} edge, it reads
\begin{equation}
    A^{d - r}_{l^{(r)}_{n - 3 - r}, l^{(r+1)}_{n - 4 - r}}(\theta_{n - 2 - r, n - 1}).
\end{equation}
Both of these cases can be understood graphically by drawing rectangles on the upside-down triangles, as drawn in red in Fig.~\ref{fig:8pt-kins}. When $r = n - 4$, that is, for the bottommost angle $\theta_{2,n-1}$, we have
\begin{equation}
    C^{\left(\nu_{n-4}\right)}_{l^{n-4}_1}(\cos\theta_{2,n-1}).
\end{equation}
Finally, for all other cases, we draw diamonds (shown in green in Fig.~\ref{fig:8pt-kins}), which illustrate the rule 
\begin{equation}
    T^{\,d-r}_{\,l^{(r)}_p,\;l^{(r+1)}_{p-1},\;l^{(r+1)}_{p},\;l^{(r+2)}_{p-1}}(\theta_{p+1,\,p+2+r}).
\end{equation}
We do this for all angles in the triangle, and then multiply them together. For fixed $\vec{s}, \vec{\mu}$, as we saw in Sec.~\ref{sec:proof} we then need to sum over all quantum numbers $l^{(r)}_p$ for $r \ge 2$ with some  matrix $Y_{\{ l^{(r)}_p \}}$ inserted. The range for a given $l^{(r)}_p$ can be read off the upside-down triangle as the minimum of the two quantum numbers directly above it:
\begin{equation}
    0 \leq l^{(r)}_p \leq \min(l^{(r-1)}_p, l^{(r-1)}_{p+1}).
\end{equation}
The coefficient matrix $Y_{\{ \vec{l}^{(r)} \}}$ (including the overall $n$-point normalization factor) can then be worked out straightforwardly to be
\begin{equation}\label{eq:full-Y-form}
\begin{aligned}
    Y_{\{\vec l^{(r)}\}}=N^{(n)}\prod_{r=2}^{n-4}\Bigg[\;
    &(-1)^{\sum_{p=1}^{n-3-r} l^{(r)}_p\;+\;\sum_{p=1}^{n-4-r}\min\big(l^{(r)}_p,\,l^{(r)}_{p+1}\big)}\;
    \sqrt{b_{l^{(r-1)}_1,\,l^{(r)}_1}(\nu_{r-1})\;b_{l^{(r-1)}_{n-2-r},\,l^{(r)}_{n-3-r}}(\nu_{r-1})}\\
    &\times\prod_{p=1}^{n-3-r}\Big[b_{l^{(r)}_p,\,0}(\nu_r)\Big]^{\,1-\frac12\left(\delta_{p,1}+\delta_{p,\,n-3-r}\right)}\;\Bigg],
\end{aligned}
\end{equation}
with
\begin{equation}
    N^{(n)}=(-1)^{\sum_{i=1}^{n-5}\min(\mu_i,\mu_{i+1})}\;\prod_{j=2}^{n-4}
    \sqrt{\frac{b_{\mu_{j-1},0}(\nu_1)\,b_{\mu_j,0}(\nu_1)}{b_{s_j,\mu_{j-1}}(\nu_0)\,b_{s_j,\mu_j}(\nu_0)}} .
\end{equation}
Thus, following this procedure we can construct a general partial wave element for the half-ladder diagram for any multiplicity $n$ and dimension $d \geq n - 1$.

There is one small subtlety which exists at $d = n - 1$, just at the bottom tip of the kinematic triangle. In this situation, we observe that the order of the corresponding Gegenbauer $C_\sigma^{(\lambda)}$ vanishes:
\begin{equation}
    \lambda = \frac{d - n + 1}{2} = 0.
\end{equation}
For general $\sigma \neq 0$, the limit as $\lambda \to 0$ is given by
\begin{equation}
    C_\sigma^{(\lambda)}(\cos\theta) \to \frac{2\lambda}{\sigma} \cos(\sigma \theta).
\end{equation}
So, it vanishes for $\sigma \geq 1$, and we take it to be $1$ at $\sigma = 0$. This seems to be a problem. However, the $b$ factors in the coefficient matrix $Y$ precisely compensate for this behavior: as $\lambda \to 0$ for $\sigma \neq 0$, they go as
\begin{equation}
    b_{\kappa,\sigma}(\lambda + 1/2) \propto \frac{1}{\lambda},
\end{equation}
whereas when $\sigma = 0$, they go to $b_{\kappa,0}(\lambda + 1/2) \to 1$. So, the full expansion is finite at $d = n - 1$, but one must take care to define it as a limit where $d \to n - 1$ from above.

\medskip
\noindent \textbf{Example.} At $n = 8$, we can follow the algorithm using the diagrams in Fig.~\ref{fig:8pt-kins}: doing this, we find
\begin{equation}
    R^{(8,d)}_{\vec{s};\vec{\mu}} = A^d_{s_1,\mu_1}(\theta_{2,3}) \Phi^{s_2,s_3,s_4}_{\mu_1, \mu_2,\mu_3,\mu_4} A^d_{s_5,\mu_4}(\theta_{6,7}),
\end{equation}
where
\begin{equation}
\begin{aligned}\label{eq:8pt-Phi}
    \Phi=\sum_{\vec l,\vec\kappa,\sigma} Y_{\vec l,\vec\kappa,\sigma}\;
    &T^{d}_{s_2,\mu_1,\mu_2,l_1}(\theta_{3,4})\,
     T^{d}_{s_3,\mu_2,\mu_3,l_2}(\theta_{4,5})\,
     T^{d}_{s_4,\mu_3,\mu_4,l_3}(\theta_{5,6})\\
    &\times A^{d-1}_{\mu_1,l_1}(\theta_{2,4})\,
     T^{d-1}_{\mu_2,l_1,l_2,\kappa_1}(\theta_{3,5})\,
     T^{d-1}_{\mu_3,l_2,l_3,\kappa_2}(\theta_{4,6})\,
     A^{d-1}_{\mu_4,l_3}(\theta_{5,7})\\
    &\times A^{d-2}_{l_1,\kappa_1}(\theta_{2,5})\,
     T^{d-2}_{l_2,\kappa_1,\kappa_2,\sigma}(\theta_{3,6})\,
     A^{d-2}_{l_3,\kappa_2}(\theta_{4,7})\,
     A^{d-3}_{\kappa_1,\sigma}(\theta_{2,6})\,
     A^{d-3}_{\kappa_2,\sigma}(\theta_{3,7})\,
     C^{\left(\frac{d-7}{2}\right)}_{\sigma}(\cos\theta_{2,7}).
\end{aligned}
\end{equation}
The coefficient matrix $Y_{\vec l,\vec\kappa,\sigma}$ is given by
\begin{equation}\label{eq:Y-8-pt}
\begin{aligned}
    Y_{\vec l,\vec\kappa,\sigma}=N^{(8)}
    &\,(-1)^{l_1+l_2+l_3+\min(l_1,l_2)+\min(l_2,l_3)}
    \sqrt{b_{\mu_1,l_1}(\nu_1)\,b_{\mu_4,l_3}(\nu_1)}\,\sqrt{b_{l_1,0}(\nu_2)\,b_{l_3,0}(\nu_2)}\;b_{l_2,0}(\nu_2)\\
    &\times(-1)^{\kappa_1+\kappa_2+\min(\kappa_1,\kappa_2)}
    \sqrt{b_{l_1,\kappa_1}(\nu_2)\,b_{l_3,\kappa_2}(\nu_2)}\,\sqrt{b_{\kappa_1,0}(\nu_3)\,b_{\kappa_2,0}(\nu_3)}\\
    &\times(-1)^{\sigma}\sqrt{b_{\kappa_1,\sigma}(\nu_3)\,b_{\kappa_2,\sigma}(\nu_3)},
\end{aligned}
\end{equation}
with
\begin{equation}
    N^{(8)}=(-1)^{\min(\mu_1,\mu_2)+\min(\mu_2,\mu_3)+\min(\mu_3,\mu_4)}\,
    \frac{\sqrt{b_{\mu_1,0}(\nu_1)\,b_{\mu_4,0}(\nu_1)}\;b_{\mu_2,0}(\nu_1)\,b_{\mu_3,0}(\nu_1)}
    {\sqrt{b_{s_2,\mu_1}(\nu_0)\,b_{s_2,\mu_2}(\nu_0)\,b_{s_3,\mu_2}(\nu_0)\,b_{s_3,\mu_3}(\nu_0)\,b_{s_4,\mu_3}(\nu_0)\,b_{s_4,\mu_4}(\nu_0)}}.
\end{equation}
Note that Eq.~\eqref{eq:Y-8-pt} has a clear recursive structure.

\subsection{Vanishing Gram determinant}\label{sec:gram}

In all of the previous discussion, we worked under the assumption that $d \geq n -1$. This was so that all momenta $p_i^\mu$ could be treated as independent, except for the fact that they are related by momentum conservation~\cite{Byers:1964ryc, Asribekov:1962tgp}. Now, in this subsection we will show how it is possible to take the general algorithm laid forth in Sec.~\ref{sec:writing} and modify it to obtain a partial wave expression even when $d < n - 1$.

\begin{figure}[t]
    \centering
    \includegraphics[width=0.7\linewidth]{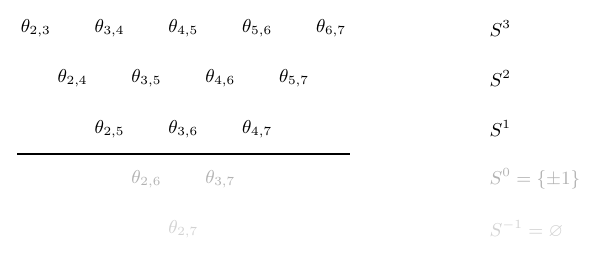}
    \caption{The kinematic variables for the $8$-point partial wave expansion in dimension $d = 5$. All variables above the black line are dynamical in the expansion, while those below are ``frozen out'' by the fact that the Gram determinant vanishes. To the right of each row in the triangle, we give the sphere on which the angles in the row live, starting with $S^3 = S^{d-2}$.}
    \label{fig:8pt-kins-gram}
\end{figure}

First, let us explain what exactly we mean by ``vanishing Gram determinant.'' For a set of $n$ momenta in $d$ dimensions, $p_1, p_2, \ldots, p_n$ satisfying momentum conservation, we form a $d \times (n-1)$ matrix $P$ whose components are
\begin{equation}
    P_{i,j} = p^i_j.
\end{equation}
We do not include $p_n$ in the matrix since it is given in terms of all the other momenta by
\begin{equation}
    p_n^\mu = -\sum_{i = 1}^{n-1} p_i^\mu.
\end{equation}
Then the \textit{Gram matrix} is
\begin{equation}
    M = P^T \eta P,
\end{equation}
with $\eta$ the mostly-plus metric tensor. This means, in terms of dot products of momenta, that
\begin{equation}
    M_{i,j} = p_i \cdot p_j.
\end{equation}
If $d = n-1$, the determinant of $M$ comes out as
\begin{equation}\label{eq:negative-M}
    \det M = (\det \eta) (\det P)^2 = - (\det P)^2 \leq 0,
\end{equation}
For general $d,n$ however, $\det M$ can be positive, negative, or zero, since $\eta$ is a pseudo-Riemannian metric. 

If $d < n - 1$, the momenta develop dependency conditions in addition to that coming from momentum conservation. For example, let's say we are in $d = 3$ and are scattering $n = 5$ particles. The $3$-dimensional space is spanned by $3$ linearly-independent vectors, so we can have at most $4$ generic momentum-conserving momenta. (This is the point $d = n - 1$ for which Eq.~\eqref{eq:negative-M} holds.) For $5$ momenta then, there must be an additional constraint relating them together, one which does not derive from momentum conservation but rather from the fact that the dimension of the space cannot accommodate so many independent vectors. So, when $d < n - 1$, we find that the ``Gram determinant'' vanishes:
\begin{equation}
    \det M = \det (p_i \cdot p_j) = 0,
\end{equation}
since the rows/ columns become linearly dependent. So, in this case there are a reduced number of kinematic variables needed to fully specify the scattering process. 

It is clear that the discussion here mirrors that of the Hankel matrix in App.~\ref{sec:hankel}. In fact, the techniques therein, when translated to this situation, can be used to find the precise dependencies between momenta created by scattering in $d < n - 1$, though this will be unnecessary for the following discussion.

Now that we have reviewed the mechanism by which the kinematic space shrinks for $d < n - 1$, let us understand how it affects the angular parameterization. For a concrete example, we will focus on computing the $n = 8$ point partial wave expansion in $d = 5$ dimensions. The reduced set of kinematics can be worked out by again examining the $n = 8$ upside-down triangle of angular variables, and labeling each row by which dimensional sphere the angles live on. (See Fig~\ref{fig:8pt-kins-gram}.) In this case, the top ($r = 0$) row lives on $S^3$, the $r = 1$ row lives on sphere $S^2$, and the $r = 2$ row lives on the circle $S^1$. This means that the fourth row $r = 3$ lives on $S^{0} = \{ \pm 1\}$, forcing both of its angles $\theta_{2,6}, \theta_{3,7} = 0, \pi$. Anything below (in this case, just $\theta_{2,7}$) is completely frozen out. So, to work out the dynamical variables of the restricted problem in general, we draw the $n$-point upside-down triangle, label each row by $S^{d - 2 - r}$, and draw a line directly below the row corresponding to $S^1$. In our eight-point example, this is all illustrated in Fig.~\ref{fig:8pt-kins-gram}. In fact, this ``chopping'' procedure is a nice way to visualize the reduction in number of variables due to a vanishing Gram determinant when $d < n - 1$: they go from ``triangle'' numbers to the ``trapezoid'' numbers
\begin{equation}
    \sum_{r = 0}^{d - 3} (n - 3 - r) = \frac{(d-2)(2n - d - 3)}{2} = \frac{(n - 3)(n - 2)}{2} - \frac{(n - d - 1)(n - d)}{2},
\end{equation}
which are the difference of two triangle numbers.

Now that we have worked out the kinematics, we need a prescription for modifying the partial wave expansion. We will keep with the $d = 5, n = 8$ example, which entails modifying Eq.~\eqref{eq:8pt-Phi}. The first thing that happens is the sum over $\sigma$ collapses to just $\sigma = 0$. One can see this by looking at the factors $A^{2}_{\kappa_1,\sigma}(\theta_{2,6})$ and $A^{2}_{\kappa_2,\sigma}(\theta_{3,7})$, which have $(\sin \theta)^\sigma$ prefactors as in Eq.~\eqref{eq:A-defs}. For $\theta = 0, \pi$, these prefactors are zero unless $\sigma = 0$. So, 
\begin{equation}\label{eq:A-2-rel}
    A^{2}_{\kappa,0}(\theta) \to C_{\kappa}^{(-1/2)}(\cos \theta),
\end{equation}
and the deepest row $C_0^{(-1)}(\cos \theta_{2,7}) = 1$. Since $\cos \theta = \pm 1$ in the case above, Eq.~\eqref{eq:A-2-rel} further degenerates into
\begin{equation}\label{eq:C-0.5}
    C^{(-1/2)}_{s}(\pm 1)=
    \begin{cases}
    1, & s=0,\\[2pt]
    \mp\,1, & s=1,\\[2pt]
    0, & s\ge 2,
\end{cases}
\end{equation}
which restricts $\kappa_i = 0,1$ in the sum Eq.~\eqref{eq:8pt-Phi}. This means that the second deepest row in Fig.~\ref{fig:8pt-kins-gram} contributes
\begin{equation}
    (- \epsilon_1)^{\kappa_1} (-\epsilon_2)^{\kappa_2},
\end{equation}
where $\epsilon_1 = \cos\theta_{2,6}$ and $\epsilon_2 = \cos\theta_{3,7}$
for $\kappa_i = 0, 1$. For these values of $\kappa_i$, we additionally find that the third-from-last row $r = 2$ (the deepest dynamical row) has
\begin{equation}
\lim_{d\to5} A^{d-2}_{l,\kappa}(\theta)\;\propto\;
\begin{cases}
\cos(l\theta), & \kappa=0,\\[2pt]
\sin(l\theta), & \kappa=1,
\end{cases}
\qquad
\lim_{d\to5} T^{d-2}_{l,\kappa_1,\kappa_2,0}(\theta)\;\propto\;
\begin{cases}
\cos(l\theta), & \kappa_1=\kappa_2,\\[2pt]
\sin(l\theta), & \kappa_1\neq\kappa_2.
\end{cases}
\end{equation}
So, we derive thus far that
\begin{equation}
\begin{aligned}
\Phi^{(8, 5)}=\sum_{\vec l}\sum_{\kappa_1,\kappa_2\in\{0,1\}}
\widetilde{Y}^{(5)}_{\vec l,\vec\kappa}\,(-\epsilon_1)^{\kappa_1}(-\epsilon_2)^{\kappa_2}\;
&T^{5}_{s_2,\mu_1,\mu_2,l_1}(\theta_{3,4})\,T^{5}_{s_3,\mu_2,\mu_3,l_2}(\theta_{4,5})\,T^{5}_{s_4,\mu_3,\mu_4,l_3}(\theta_{5,6})\\
\times\;&A^{4}_{\mu_1,l_1}(\theta_{2,4})\,T^{4}_{\mu_2,l_1,l_2,\kappa_1}(\theta_{3,5})\,T^{4}_{\mu_3,l_2,l_3,\kappa_2}(\theta_{4,6})\,A^{4}_{\mu_4,l_3}(\theta_{5,7})\\
\times\;&f_{\kappa_1}\!\big(l_1\theta_{2,5}\big)\,f_{\kappa_1+\kappa_2}\!\big(l_2\theta_{3,6}\big)\,f_{\kappa_2}\!\big(l_3\theta_{4,7}\big),
\end{aligned}
\end{equation}
where $f_{\mathrm{odd}} = \sin$ and $f_{\mathrm{even}}= \cos$.

Now, we also need to figure out what happens to the row $r = 1$, which contains $d = 4$ $T$ special functions and associated Gegenbauers. One can derive that, for the values under consideration $\kappa_i = 0,1$, the former are simply combinations of the familiar Wigner $d$-matrices:\footnote{The Wigner $D$-matrix is the rotation operator in three spatial dimensions written in its standard basis $D_{\mu_1,\mu_2}^j(\alpha, \beta, \gamma) \equiv \langle j \mu_1 \lvert \mathcal{R} \rvert j \mu_2 \rangle$ for $\alpha,\beta,\gamma$ the Euler angles. The Wigner $d$-matrix is then just $d_{\mu_1,\mu_2}^j(\beta) \equiv D_{\mu_1,\mu_2}^j(0,\beta,0)$.}
\begin{equation}\label{eq:wigner-d}
    T^{4}_{\mu_2,\,l_1,\,l_2,\,\kappa}(\theta)\;\propto \;\Big[\,d^{\,\mu_2}_{l_1,\,l_2}(\theta)+(-1)^{\kappa+l_2}\,d^{\,\mu_2}_{l_1,\,-l_2}(\theta)\Big],\qquad \kappa=0,1.
\end{equation}
Additionally, $A$ in $d = 4$ becomes an associated Legendre polynomial:
\begin{equation}\label{eq:legendre}
    A^4_{\mu,l}(\theta) \propto P_\mu^l(\cos\theta) \propto (\sin\theta)^l C_{\mu-l}^{\left(l + \frac{1}{2}\right)}(\cos\theta)
\end{equation}
So, putting this all together, we find that the $n = 8$ partial wave expansion in $d = 5$ is given by
\begin{equation}
\begin{aligned}\label{eq:Phi-8-5-expand}
\Phi^{(8,5)}_{s_2,s_3,s_4;\,\mu_1,\dots,\mu_4}
=\sum_{l_1,l_2,l_3}\ \sum_{\kappa_1,\kappa_2\in\{0,1\}}
&\widetilde{Y}^{(5)}_{\vec l,\vec\kappa}\,(-\epsilon_1)^{\kappa_1}(-\epsilon_2)^{\kappa_2}\;
T^{5}_{s_2,\mu_1,\mu_2,l_1}(\theta_{3,4})\,T^{5}_{s_3,\mu_2,\mu_3,l_2}(\theta_{4,5})\,T^{5}_{s_4,\mu_3,\mu_4,l_3}(\theta_{5,6})\\
\times\;&P_{\mu_1}^{l_1}(\theta_{2,4})\Big[d^{\,\mu_2}_{l_1,l_2}+(-1)^{\kappa_1+l_2}d^{\,\mu_2}_{l_1,-l_2}\Big](\theta_{3,5})
\Big[d^{\,\mu_3}_{l_2,l_3}+(-1)^{\kappa_2+l_3}d^{\,\mu_3}_{l_2,-l_3}\Big](\theta_{4,6})\,\\
\times\;&P^{l_3}_{\mu_4}(\theta_{5,7})f_{\kappa_1}\!\big(l_1\theta_{2,5}\big)\,f_{\kappa_1+\kappa_2}\!\big(l_2\theta_{3,6}\big)\,f_{\kappa_2}\!\big(l_3\theta_{4,7}\big),
\end{aligned}
\end{equation}
with a modified coefficient matrix $\widetilde{Y}_{\vec{l};\vec{\kappa}}$ which we will discuss momentarily. 

One can understand the meaning of these $\epsilon_i = \pm 1$ by imagining $d = 4$ at, say, six-points. There, we have one $\epsilon_1$ orientation. We move into the rest frame of the middle propagator $I_2$. In this frame, the momenta of $3$, $4$, $I_3$, and $I_1$ lie in a plane, and the pairs $1$, $2$ and $5$, $6$ generically come out of the plane by their azimuths. Let us consider the azimuth of $\vec{p}_2$ and the azimuth of $\vec{p}_5$. Up to a reflection through the plane, these angles relative to this plane (through our Eq.~\eqref{eq:costheta-def}) are fixed by knowledge of $\theta_{2,4},\theta_{2,3}$ and $\theta_{3,5}, \theta_{4,5}$. The one thing that isn't fixed is their orientation relative to each other. So, $\epsilon_1 = 1$ means that $\vec{p}_2$ and $\vec{p}_5$ are on the \textit{opposite} sides of this plane, whereas $\epsilon_1 = -1$ means they are on same side. Note that this also agrees with the characterization of the angle $\theta_{2,5}$ as that between $(\vec{p}_2, \vec{p}_3, \vec{p}_4)$ and $(\vec{p}_3, \vec{p}_4, \vec{p}_5) $ as discussed in Sec.~\ref{sec:kinematics}.

While correct, the form~\eqref{eq:Phi-8-5-expand} is a bit unsightly. However, it can be cleaned up significantly. To illustrate the mechanism, let us first make some remarks on the cases at $d = 3,4$. At $d = 3$, the massive little group is $SO(2)$, whose irreps are labeled by ``spins'' $s_i$ with no vertex quantum number. These $s$'s are really equivalent to the familiar helicities of the $d = 4$ problem rather than to true spins--- and helicities, unlike spins, can be \textit{negative}. However, nowhere in the previous discussions do we ever sum a quantum number over negative values, so there appears to be an issue with our formulae. Let us investigate it further by examining a concrete example, at $n = 5$ in $d = 3$. In this case, we have the dynamical angles $\theta_{2,3}$ and $\theta_{3,4}$ and one orientation $\cos \theta_{2,4} = \epsilon = \pm 1$. Following the outline for $d = 5, n = 8$ now for this situation, we find that the full expansion is given by
\begin{equation}
    R^{(5,3)}\big(\theta_{2,3},\theta_{3,4},\epsilon\big)
=\sum_{s_1,s_2=0}^\infty \sum_{\mu= 0, 1}\;y^\mu_{s_1,s_2}\,(-\epsilon)^\mu f_\mu(s_1 \theta_{2,3}) f_\mu(s_2 \theta_{3,4}).
\end{equation}
Defining an orientation $\sigma = \pm 1$ via the fact that
\begin{equation}
    \cos (s_1 \theta_{2,3} + \sigma \epsilon s_2 \theta_{3,4}) = R^{(5,3)}_{s_1,s_2,0} + \sigma R^{(5,3)}_{s_1,s_2,1},
\end{equation}
we switch bases and find that
\begin{equation}
R^{(5,3)}\big(\theta_{2,3},\theta_{3,4},\epsilon\big)
=\sum_{s_1,s_2=0}^\infty \sum_{\sigma = \pm 1}\;y^\sigma_{s_1,s_2}\,
\cos\!\big(s_1\theta_{2,3}+\sigma \epsilon\,s_2\,\theta_{3,4}\big),
\end{equation}
with $y^\sigma_{s_1,s_2}$ the partial wave coefficients and $R^{(5,3)}_{s_1,s_2,\mu} = (-\epsilon)^\mu f_\mu(s_1\theta_{2,3}) f_\mu (s_2 \theta_{3,4})$. The label $\sigma$ is precisely the relative sign of $s_1$ and $s_2$. Thus, if we define $\lambda_1 = s_1$ and $\lambda_2 = \sigma s_2$, then the sum can be rewritten into
\begin{equation}
    R^{(5,3)}\big(\theta_{2,3},\theta_{3,4},\epsilon\big)
    =\frac{1}{2} \sum_{\lambda_1,\lambda_2\in\mathbb Z}y_{\lambda_1\lambda_2}\,
    \cos\!\big(\lambda_1\theta_{2,3}+\epsilon\,\lambda_2\,\theta_{3,4}\big),
\end{equation}
where we have doubled the sum to include negative values of $\lambda_1$ using the fact that $\cos$ is an even function. So, we find that the signed helicity nature of these labels $s_i$ in $d = 3$ is captured by our formalism. This structure holds analogously at $d = 3$ for all $n \geq 5$, where we can derive that
\begin{equation}
    R^{(n, 3)} = \frac{1}{2}\sum_{\lambda_i \in \mathbb Z} y_{\vec{\lambda}} \cos \left( \sum_{p = 1}^{n-3} \lambda_p \phi_p \right)
\end{equation}
up to an overall normalization, where, as at $n = 5$, we have signed $\lambda_p = \sigma_p s_p$, and the $\phi$'s are \textit{oriented} polar angles given by
\begin{equation}\label{eq:orient-azi}
    \phi_p = \epsilon_1 \epsilon_2 \cdots \epsilon_{p-1} \theta_{p+1,p+2}, \qquad p = 1, \ldots, n - 3.
\end{equation}
In $d = 4$, one can show that the exact same situation occurs: the massive little group is $SO(3)$, whose irreps are labeled by an integer spin $s = 0, 1, \ldots$ and integer helicity $-s\leq \mu \leq s$.\footnote{Of course, one can also have half-integer spins, but these are not relevant to our partial-wave discussion as constructed in this paper.} Although the sum of $\mu$ is only over positive integers in our general prescription, the quantum numbers at one layer deeper $l_i = \{ 0,1\}$ convert in exactly the same way as they did in $d = 3$ into a relative sign between the $\mu$'s. For general multiplicity $n \geq 6$, we find the simple structure that, up to an overall normalization,
\begin{equation}\label{eq:d4-helicity}
    R^{(n,4)} = \frac{1}{2}\sum_{s_i = 0}^\infty \sum_{\lambda_i = -\min(s_i,s_{i+1})}^{\min(s_i, s_{i+1})} y_{\vec{s}, \vec{\lambda}} P^{\lambda_1}_{s_1}(\theta_{2,3})\left( \prod_{p = 2}^{n - 4} d_{\lambda_{p-1}, \lambda_{p}}^{s_p}(\theta_{p+1, p+2})\right) P^{\lambda_{n-4}}_{s_{n-3}}(\theta_{n-2,n-1})\cos\left( \sum_{p = 1}^{n-4} \lambda_p \phi_p \right),
\end{equation}
where we have signed helicities $\lambda_p = \sigma_p \mu_p$, and the oriented azimuths $\phi_p$ are now given by
\begin{equation}
    \phi_p = \epsilon_1 \epsilon_2 \cdots \epsilon_{p-1} \theta_{p+1, p+3}, \qquad p = 1, 2 \ldots, n-4.
\end{equation}
In $d = 4$, the partial-wave expansion at general $n$ has a long prehistory. In the multi-Regge program of the late 60s and 70s, amplitudes were expanded in a manner equivalent to the above formulae, in Wigner $d$-matrices using the ``Toller angles'' (our azimuthal $\phi_p$'s above)~\cite{Bali:1967zz,Toller:1969gx,Jones:1971eu,Goddard:1971fq,White:1973ola,Brower:1974yv}. More recently, Ref.~\cite{Jeong:2026xzk} constructed four-, five-, and six-point partial waves in $d = 4$ for massless planar amplitudes, in the complex-forward kinematics underlying the multi-positivity bounds in Ref.~\cite{Cheung:2025nhw}. That construction
resolves the spin of the single state exchanged in one factorization channel, with the remaining invariants expressed as lower-point factors through Wigner $D$-functions. It is complementary to the expansion presented here, which puts all $n-3$ internal lines on-shell simultaneously, keeps the vertex labels $\mu_i$ explicit, and holds in any $d$; at $d = 4$ the two share the Wigner-$d$ and helicity structure of Eq.~\eqref{eq:d4-helicity}.

As it turns out, this structure we observe at $d = 3,4$ holds at general $d$. Let us label the depth at which the angles are defined on $S^1$ (that is, the last dynamical row) as $r_\star = d - 3$. (For the $n = 8$, $d = 5$ running example, we have $r_\star = 2$.) Then, the sum over indices $l^{(r_\star + 1)}_p \in \{ 0,1 \}$ is precisely proportional to
\begin{equation}\label{eq:gen-d-gram}
    P^{\lambda_1}_{l^{(r_\star-1)}_1}(\theta_{2,d-1})\left( \prod_{p = 2}^{n - d} d_{\lambda_{p-1}, \lambda_{p}}^{l^{(r_\star-1)}_p}(\theta_{p+1, p-2+d})\right) P^{\lambda_{n-d}}_{l^{(r_\star - 1)}_{n-d+1}}(\theta_{n-d+2,n-1})\cos\left( \sum_{p = 1}^{n-d} \lambda_p \phi_p \right),
\end{equation}
analogous to the $d = 4$ general-$n$ basis element~\eqref{eq:d4-helicity}, where now the signed quantum numbers $\lambda_p = \sigma_p l^{(r_\star)}_p$ are summed from $-\min(l^{(r_\star -1)}_{p}, l^{(r_\star -1)}_{p+1})$ to $\min(l^{(r_\star -1)}_{p}, l^{(r_\star -1)}_{p+1})$. (This is because in these restricted kinematics, $l^{(r_\star)}_p$ is truly an eigenvalue of $SO(2)$ and therefore a bona fide helicity rather than a spin.) Further, we have the oriented azimuths $\phi$, this time given by
\begin{equation}
    \phi_p = \epsilon_1 \epsilon_2 \cdots \epsilon_{p-1} \theta_{p+1,p-1+d}, \qquad p = 1, 2, \ldots, n - d
\end{equation}
So, at this point we are prepared to lay out the modification of the algorithm of Sec.~\ref{sec:writing} that determines the partial wave basis when $d < n - 1$. First, we see that everything at depth $r \leq r_\star - 2$ is exactly the same as in Sec.~\ref{sec:writing}. At depth $r = r_\star - 1$ and $r_\star$, the new rule is to use exactly the expression in Eq.~\eqref{eq:gen-d-gram}. The sums over internal quantum numbers $l_p^{(r)}$ for $ r \geq 2$ in the basis element are unchanged except for the case when $r = r_\star$, where the oriented quantum number $\lambda_p = \sigma_{p} l_p^{(r_\star)}$ now runs over \textit{negative} values $-\min(l^{(r_\star -1)}_{p}, l^{(r_\star -1)}_{p+1}) \leq \lambda_p \leq \min(l^{(r_\star -1)}_{p}, l^{(r_\star -1)}_{p+1})$ (and we must multiply by a $1/2$ to account for doubling the range of the quantum number at $p = 1$). Further, the sum over $l^{(r_\star + 1)}_p \in \{ 0,1 \}$ is already carried out in Eq.~\eqref{eq:gen-d-gram}.

The last thing to determine is the coefficient matrix $\widetilde{Y}$. In dimension $d_\star\ge5$, one simply takes the limit $d\to d_\star$ of the general expansion of Sec.~\ref{sec:writing}, where everything is understood to be analytically continued in $d$. The limit turns out to be finite, although the individual factors are not. In the row at depth $r_\star$ (where $\nu_{r_\star} \to 0$, $D = d - r_\star \to 3$) the corresponding special functions vanish: $A^D_{l,0}=C^{(\nu_{r_\star})}_l\to\tfrac{2\nu_{r_\star}}{l}\cos l\theta$ for $l\ge1$, and in the row above ($ D\to4$) the special functions $T^D$ vanish like $(D-4)^{(g(l)+g(l'))/2}$, with $g(l) = 1 - \delta_{l,0}$. So, at first glance, it appears as if the partial wave expansion goes to zero in this limit. However, at the same time, we have that the coefficients $b_{l,0}(\nu)$ and $b_{\mu,l}(\nu)$ attached to these rows diverge, as $\nu=(D-3)/2\to0$ and $\nu=(D-2)/2\to\tfrac12$, respectively. It turns out that these divergences and zeros exactly cancel against each other, leaving behind a finite answer for $d_\star\ge5$. All the rows below $r_\star + 1$ drop out in this limit as well. Thus, the new coefficient matrix can then be simply read off from this limit.

For $d_\star = 3,4$, however, this is no longer the situation. In these cases, the relevant $b$'s attached to the $r = r_\star, r_\star - 1$ row(s) are now $b_{s,\mu}(\nu)$ appearing in the \textit{denominator} of the normalization. Since these diverge as $\nu \to 0, 1/2$, the general-form partial wave expansion behaves as
\begin{equation}\label{eq:limit-factor}
    (d - d_\star)^p
\end{equation}
for some $p$ dependent on the dimension: for $d = 4$, we have
\begin{equation}
    p = \sum_{q = 1}^{n-4} g(\mu_q), \qquad g(\mu_q) = 1 - \delta_{\mu_q,0},
\end{equation}
and for $d = 3$, where $\mu_q \in \{ 0,1 \}$,
\begin{equation}
    p = \sum_{q : s_q \geq 1} ( 1 - \mu_{q-1} - \mu_q), \qquad \mu_0 = \mu_{n-3} = 0.
\end{equation}
So, if we first divide the non-restricted partial wave expansion by this factor and then take the limit $d \to d_\star$ for $d_\star = 3,4$, we are left with a finite result. Hence, we can extract with this limit the matrix $\widetilde{Y}$ at $d_\star = 3,4$.

Finally, for orthogonality, we must make a modification to the definition of the inner product, which now truncates at the depth $r_\star$. We give the formula for $d < n - 1$ in Eq.~\eqref{eq:restrict-inner}. With this definition, orthogonality follows directly in the exact same way from the preceding discussion in App.~\ref{sec:ortho}.

\section{Application: String Theory Amplitudes}\label{sec:string-theory}

\subsection{The $n$-point amplitude and its residues}\label{sec:string-amps}

\begin{figure}[t]
    \centering
    \includegraphics[width=0.5\linewidth]{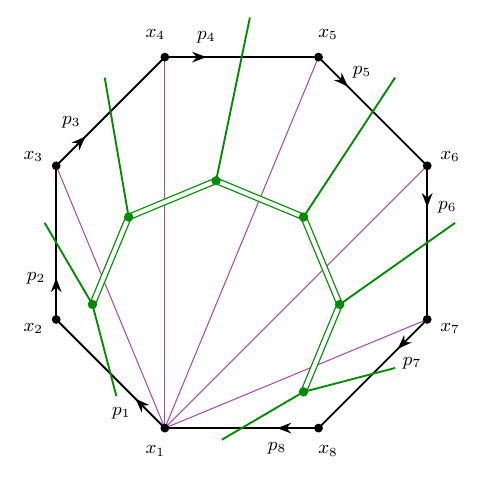}
    \caption{The ray-like triangulation (anchored at $1$) of the octagon. Each edge of the octagon is labeled by an external momentum $p_i$, with dual variables $x_i$ labeling the vertices. The chords (in purple) correspond to the planar invariants $X_{1,j}$ that appear in the denominator of the diagram. This triangulation is precisely dual to the $n = 8$ half-ladder diagram, which one can see overlaid in green on the octagon.}
    \label{fig:ray-tri}
\end{figure}

As a natural application of the results of the previous sections, we turn to ``string theory amplitudes''. In particular, we wish to consider here amplitudes of the form
\begin{equation}\label{eq:string-amp}
    \mathcal{A}_n (X_{i,j}) = \int_0^{\infty} \prod_{\mathcal{C} \in \mathcal{T}} \frac{d y_{\mathcal{C}}}{y_\mathcal{C}} \prod_{i < j} u_{i,j}^{\alpha' X_{i,j} + \alpha_0},
\end{equation}
where each $y_\mathcal{C}$ is a positive coordinate associated to a chord $\mathcal{C}$ in some triangulation $\mathcal{T}$ of the $n$-gon, and the $u_{i,j}$ satisfy the relations
\begin{equation}\label{eq:u-eq}
    u_{i,j} + \prod_{(a,b) \cap (i,j)} u_{a,b} = 1
\end{equation}
known as the \textit{u-equations}~\cite{Koba:1969rw,Koba:1969kh,Arkani-Hamed:2019mrd,Arkani-Hamed:2019plo}, which can be solved by parameterizing in terms of the $n - 3$ $y$-variables. Note that we have shifted the planar Mandelstam invariants $\alpha' X_{i,j}$ by the Regge intercept $\alpha_0$~\cite{Veneziano:1968yb,Virasoro:1969pd,Bardakci:1969cs,Chan:1969ex,Goebel:1969qw,Mandelstam:1974fq}, which we keep \textit{a priori} unfixed along with the spacetime dimension $d$. For the problem at-hand, where we wish to consider half-ladder diagrams, we take $\mathcal{T}$ to be the ``ray-like'' triangulation (shown at $n = 8$ in Fig.~\ref{fig:ray-tri}.) originating at point $1$. The object Eq.~\eqref{eq:string-amp} is known to have singularities at $X_{i,j} = -M_N^2$, with spectrum given by $M_N^2 = N + \alpha_0$ for level $N = 0, 1, 2, \ldots$. The external states are scalars with masses at the $N = 0$ level, that is, at $m_{\textrm{ext}}^2 = \alpha_0$. 

There are two special values of the Regge intercept where a worldsheet interpretation is known: at $\alpha_0 = 0$ below $d \leq 10$ dimensions, we have the so-called ``$Z$-theory'' amplitudes (with massless external scalars)~\cite{Broedel:2013tta,Carrasco:2016ldy}, whereas the famous open bosonic string resides at $\alpha_0 = -1, d \leq 26$ (external tachyons)~\cite{Lovelace:1971fa,Brower:1972wj,Goddard:1972iy,Goddard:1973qh,Green:1987sp,Arkani-Hamed:2022gsa,Mansfield:2025gca,Chen:2026zmz}. In the field-theory limit $\alpha' \to 0$, $Z$-theory reduces to tree-level $\tr (\phi^3)$ amplitudes~\cite{Arkani-Hamed:2017mur}. Whether these two particular amplitudes are singled out by consistency alone has been studied extensively at four points~\cite{Caron-Huot:2016icg,Cheung:2022mkw,Haring:2023zwu,Albert:2024yap,Berman:2024wyt,Cheung:2024uhn,Cheung:2025tbr,Elvang:2026pmc} and more recently through higher-point constraints \cite{Arkani-Hamed:2023jwn,Cheung:2025nhw,Berman:2025owb,Basile:2026gnd}. 

The family of amplitudes~\eqref{eq:string-amp} constitutes simple deformations of standard string theory amplitudes, where we allow the Regge intercept and the spacetime dimension to vary. Deformations of the four-point string amplitude compatible with two-to-two unitarity are plentiful~\cite{Coon:1969yw,Gross:1969db,Cheung:2022mkw,Cheung:2023adk,Cheung:2023uwn,Geiser:2022exp,Maldacena:2022ckr,Figueroa:2022onw,Bhardwaj:2022lbz,Rigatos:2024beq,Rigatos:2023asb,Mansfield:2024wjc,Jepsen:2025baw,Cheung:2024uhn}. By making use of the requirement of consistent factorization at higher-points~\cite{Arkani-Hamed:2023jwn,Basile:2026gnd} and multi-positivity~\cite{Cheung:2025nhw,Bhardwaj:2024klc,Jeong:2026xzk}, many of them can be excluded. For the family~\eqref{eq:string-amp} in particular, Ref.~\cite{Arkani-Hamed:2023jwn} finds every $\alpha_0 \geq -1$ consistent with factorization through six-points, while Ref.~\cite{Basile:2026gnd} fixes $\alpha_0 = -1$ from seven-point factorization under the assumption of a non-degenerate spectrum.

The introduction of the $u$-variables allows us to straightforwardly extract residues of Eq.~\eqref{eq:string-amp}. This was done in Ref.~\cite{Arkani-Hamed:2024nzc} for the ray-like triangulations in the following manner. One first solves the $u$-equations~\eqref{eq:u-eq} in terms of the chord variables $y_{1,j}$, $j=3,\dots,n-1$. Then, the integrand of Eq.~\eqref{eq:string-amp} becomes a product of nested $F$-polynomials of Ref.~\cite{Arkani-Hamed:2024nzc}:
\begin{equation}\label{eq:F-polynomial-string}
    \mathcal A_n=\int_0^\infty\prod_{j=3}^{n-1}\frac{dy_{1,j}}{y_{1,j}}\,y_{1,j}^{\,X_{1,j} + \alpha_0}
    \prod_{\substack{1\le i<j\le n-1\\ j\ge i+2}}\mathcal F_{i,j}^{\,-c_{i,j}},\qquad
    \mathcal F_{i,j}=1+\sum_{l=i+2}^{j}\prod_{k=l}^{j}y_{1,k},
\end{equation}
with $c_{i,j}=X_{i,j}+X_{i+1,j+1}-X_{i,j+1}-X_{i+1,j}=-2\,p_i\cdot p_j$.\footnote{Note that we have also set the Regge slope $\alpha' = 1$, which we will continue to do for the remainder of the paper.} The residue on all $n-3$ poles $X_{1,j}=-M^2_{N_j}$ simultaneously is the coefficient of $\prod_j y_{1,j}^{N_j}$ in the expansion of the product of $\mathcal F$'s: 
\begin{equation}\label{eq:F-expansion}
    \prod_{\substack{1\le i<j\le n-1\\ j\ge i+2}}\mathcal F_{i,j}^{\,-c_{i,j}}
    =\sum_{\{N_j\}}\ \prod_{j=3}^{n-1}y_{1,j}^{\,N_j}\;R_{N_3,\dots,N_{n-1}},
    \qquad
    \operatorname*{Res}_{X_{1,3}=-M_{N_3}^2}\cdots\operatorname*{Res}_{X_{1,n-1}=-M_{N_{n-1}}^2}\mathcal A_n=R_{N_3,\dots,N_{n-1}}.
\end{equation}
This is because the singular part of the integral gives
\begin{equation}
    \int_0^1 \prod_{j=3}^{n-1}\frac{dy_{1,j}}{y_{1,j}}\,y_{1,j}^{\,X_{1,j}+\alpha_0} y_{1,j}^{N_j} \to \prod_{j=3}^{n-1} \frac{1}{X_{1,j} + N_j + \alpha_0},
\end{equation}
so whatever multiplies this is the residue of the half-ladder diagram.

This is the object we will now expand in terms of our $n$-point half-ladder partial wave basis.

\subsection{Moments}\label{sec:moms}

First, let us consider the following situation. We take an $n$-point half-ladder at generic $d$, $\alpha_0$, with $n - 3$ internal particles at identical mass $M_i = M_N$ and identical spin $s_i = s$. We additionally take each \textit{internal} three-point amplitude (which is not uniquely known) to be labeled by the same vertex quantum number $\mu_i = \mu$. In this particular set-up, the diagram is of course proportional to precisely one basis element in our partial wave basis, whose coefficient we refer to as $m_{n - 4}^{(N, s, \mu)}$. 

In string theory, this configuration corresponds to selecting a level $N = 0, 1, 2, \ldots$ with $M_N^2 = N + \alpha_0$, and choosing one of the available spins at that level: $0 \leq s \leq N$. However, in these string amplitudes, there is generally a \textit{degeneracy} of states at each level and spin $(N, s)$, which is obscured when taking residues of these string amplitude, since the internal states are already summed-over. From unitarity, we know that each moment $m_{n - 4}^{(N, s, \mu)}$ can be computed from sewing together three-point amplitudes and summing over all particles flowing through the internal lines: hence we have
\begin{equation}\label{eq:mom-expansion}
    m_0 = g_i g_i, \quad m_1 = g_i G_{ij} g_j, \quad m_2 = g_i G_{ij} G_{jk} g_k, \quad \ldots,
\end{equation}
where $g_i$ is the coupling of external-external-internal $i$ vertices on either end of the ladder, and $G_{ij}$ is the coupling of the internal $i$ - external - internal $j$ vertex in the middle of the diagram.

From the identifications in Eq.~\eqref{eq:mom-expansion}, we can derive an infinite set of \textit{positivity conditions}~\cite{Chandrasekaran:2018qmx}. At $n = 4$, we recover the famous four-point condition $m_0 \ge 0$. Our next condition occurs at $n = 6$, where, defining vectors $v_i = g_i$ and $w_i = G_{ij} g_j$, we have
\begin{equation}
    (v_i w_i)^2 \leq (v_i v_i) (w_i w_i)
\end{equation}
by Cauchy-Schwarz. In terms of moments, this is $m_0 m_2 - m_1^2 \geq 0$, which is equivalent to the condition that
\begin{equation}
    H_1 =
    \begin{pmatrix}
    m_0 & m_1 \\
    m_1 & m_2
    \end{pmatrix}
    \succeq 0.
\end{equation}
As it turns out, if we form the matrix
\begin{equation}
    H_p = [m_{i + j}]_{i,j = 0}^p
\end{equation}
which requires knowledge of $m_k$ up to $k = 2p$, then unitarity guarantees it is a positive semi-definite matrix:
\begin{equation}\label{eq:positivity-conds}
    H_p \succeq 0.
\end{equation}
These $H_p$ are known as the \textit{Hankel matrices}, and we give a review of them in App.~\ref{sec:hankel}.\footnote{In soft kinematics, and summed over the spins of the exchanged states, the same matrices are the residue matrices of Ref.~\cite{Cheung:2025nhw}, whose positivity is manifest there because in and out states are complex conjugates of one another. With massless external scalars, their construction and the one laid out here agree after summing over the internal spins and vertex quantum numbers. The spin-resolved form in $d = 4$, for a single factorization channel, is the subject of Ref.~\cite{Jeong:2026xzk}.}

As it turns out, by applying the conditions in Eq.~\eqref{eq:positivity-conds} to our string amplitude for \textit{only} scalar internal particles, we are able to derive that the only intercepts consistent with this equal-level test are precisely those belonging to $Z$-theory and to the open bosonic string. To do this, we only needed to compute up to $m_4$ (that is, using up to and including the requirement of $H_2$ semi-definite positivity) for levels $N = 1,2,3$. We elaborate on this --- and give explicit results --- in the next subsection Sec.~\ref{sec:unitarity}.

However, it has long been suspected that, at finite $\alpha'$, $Z$-theory does not describe a string theory consistent with unitarity. In Sec.~\ref{sec:violation}, we discuss the reason $Z$-theory always satisfies the equal-$(N,s)$ unitarity test: all moments $m_k$ for any $(N,s)$ and internal vertex choice are kinematically forced to be partial waves of the four-point problem, which are known to be consistent with unitarity in dimension $d \leq 10$. So, to find the unitarity violation, we must be slightly more creative. We examine the ``twisted'' half-ladder (where adjacent rungs are on opposite sides of the diagram) and place alternating states $(N_1,s_1)$, $(N_2,s_2)$ on the internal lines. Doing so for states $(2,1)$ and $(1,0)$ \textit{fails} the Cauchy-Schwarz inequality test for all $d$, thus proving the folklore that $Z$-theory is inconsistent with unitarity.

In addition to constraining the deformed string amplitude~\eqref{eq:string-amp}, we can also use knowledge of the moments to reconstruct how each individual degenerate particle at $(N, s)$ contributes to the half-ladder diagrams. For the bosonic string (the only remaining theory consistent with unitarity), we give some explicit results in Sec.~\ref{sec:degen-1}, where we pull from the data shown in App.~\ref{sec:data}. The discussion in Sec.~\ref{sec:degen-1} will also rely on the Hankel matrix technology discussed in App.~\ref{sec:hankel}; the last paragraph therein translates the general mathematical formalism into ready-to-use string-theory language. 

To derive these results, we use the functions contained within the ancillary file \texttt{combwaves.py}, which is an implementation of the physics described in Sec.~\ref{sec:halfladder}. We discuss some of the important functions in App.~\ref{sec:code}. 

\subsection{Multi-particle positivity from unitarity}\label{sec:unitarity}

Let us begin by considering the simplest possible configuration of internal modes in the half-ladder: scalars at the same level $N$, with $M_N^2 = N + \alpha_0$. At the first level $N = 1$, we find that 
\begin{equation}
    \det H_0^{(1)} = \frac{\alpha_0 + 1}{2}, \quad \det H_1^{(1)} = 0,
\end{equation}
with $H_p^{(N)}$ the Hankel matrix $H_p$ coming from the scalar(s) at level $N$. These results imply that $r = 1$ (i.e., that there is only one distinguishable scalar flowing through the half-ladder), and thus we find that the moments are all captured by the geometric formula
\begin{equation}\label{eq:N-1-m}
    m_k^{(1)} = \frac{\alpha_0 + 1}{2} \left( \frac{\alpha_0 + 2}{2}\right)^k,
\end{equation}
with shorthand $m_k^{(N)} = m_k^{(N,0,0)}$. Enforcing $\det H_0^{(1)} \geq 0$ implies that we need $\alpha_0 \geq -1$. Since this constraint only requires information from the four-point partial wave expansion, it is certainly not a new finding; see, e.g., Refs.~\cite{Arkani-Hamed:2023jwn,Cheung:2022mkw,Arkani-Hamed:2022gsa}.

At level $N = 2$, things get more interesting. We find that for generic values of $(\alpha_0, d)$, there are $r = 3$ scalar states contributing to the diagram. This is because we derive the following Hankel matrix determinants:\footnote{These expressions hold verbatim for all $d \geq 5$. For $d = 3,4$, they must first be multiplied by Eq.~\eqref{eq:limit-factor} and then the limit $d_\star \to 3,4$, respectively, must be taken.}
\begin{align}\label{eq:N-2-dets}
    \det H^{(2)}_0 &= \frac{(d+8)\,\alpha_0^2+(2d-14)\,\alpha_0+4}{8\,(d-1)}\,,\nonumber\\[4pt]
    \det H^{(2)}_1 &= -\,\frac{\alpha_0^2\,(\alpha_0+1)^2\;C(\alpha_0,d)}{512\,(d-1)^3\,(\alpha_0+2)^4}\,,\\[4pt]
    \det H^{(2)}_2 &= -\,\frac{\alpha_0^6\,(\alpha_0+1)^4\;Q(\alpha_0,d)^2}{16384\,(d-1)^5\,(\alpha_0+2)^7}\,\nonumber,
\end{align}
which requires knowledge of $m_k^{(2)}$ up to $k = 4$ (i.e., $n = 8$). Note that the expressions above contain the functions $C$ and $Q$, polynomials in $\alpha_0$, $d$ whose particular values are quite long, and so we do not write them here. Requiring first that $\det H_1^{(2)} \geq 0$, we need either $\alpha_0 = -1,0$, or $C(\alpha_0,d) \leq 0$. Further, requiring $\det H_2^{(2)} \geq 0$, it is clear (if already restricted to $\alpha_0 \geq -1$ from four-point positivity) that we can again have $\alpha_0 = 0,-1$, or now $Q(\alpha_0, d) = 0$. So, the two special values $\alpha_0 = -1, 0$ are allowed, but there may be more values if we can simultaneously satisfy $Q(\alpha_0, d) = 0$ and $C(\alpha_0,d) \leq 0$. However, one can show that, whenever $Q(\alpha_0,d) = 0$, $C(\alpha_0, d)$ must be \textit{positive} for all $d > 1, \alpha_0 \geq -1$. Hence, we are restricted from this test to $\alpha_0 = -1,0$. In either of these allowed cases, it is interesting to note that $\det H_1^{(2)} = 0$, and thus the count of distinguishable scalar modes falls from $3$ to $1$. Additionally, when $\alpha_{0} = -1$, we have that
\begin{equation}
    \det H_0^{(2)}(\alpha_0 = -1) = \frac{26 - d}{8(d-1)} \geq 0 \implies d \leq 26,
\end{equation}
which is again a result obtained from four-point partial waves~\cite{Arkani-Hamed:2022gsa,Lovelace:1971fa,Brower:1972wj,Goddard:1972iy,Chen:2026zmz}. However, when $\alpha_0 = 0$, $\det H_0^{(2)} \geq 0$ for all $d > 1$. So as of the $N=2$ constraints $\alpha_0 = 0$ is consistent in all possible spacetime dimensions.

The situation changes when we go to level $N = 3$, where we have for general $\alpha_0, d$ that
\begin{equation}
    m_0^{(3)} = \frac{(\alpha_0 - 1)(\alpha_0 + 1)(\alpha_0 d + 26 \alpha_0 + 3d - 30)}{48(d-1)}.
\end{equation}
For $\alpha_0 = 0$, this becomes
\begin{equation}
    m_0^{(3)}(\alpha_0 = 0) = \frac{10 - d}{16(d-1)} \geq 0 \implies d \leq 10,
\end{equation}
which completes this equal-$(N,s)$ bootstrap procedure~\cite{Arkani-Hamed:2022gsa,Mansfield:2025gca}. So, we find that, from applying multi-particle positivity up to $n = 8$ to the internal scalar modes at $N = 1,2,3$, we are very quickly able to show that the only two amplitudes consistent with the equal-$(N,s)$ half-ladder unitarity test in Eq.~\eqref{eq:string-amp} are those belonging to $Z$-theory ($\alpha_0 = 0, d \leq 10$) and open bosonic string theory ($\alpha_0 = -1, d \leq 26$). This is complementary to the factorization results of Refs.~\cite{Arkani-Hamed:2023jwn,Basile:2026gnd} and to the multi-positivity bounds of Ref.~\cite{Cheung:2025nhw}, which are only formulated at $\alpha_0 = 0$. Note, additionally, that in the present section we have made no assumptions on the string spectrum, and we have been able to resolve contributions to the half-ladder diagram by their STT irreps (which is complementary to the results given in Ref.~\cite{Jeong:2026xzk}).

\subsection{$Z$-theory unitarity violation}\label{sec:violation}

Now, as referenced previously, $Z$-theory is strongly suspected to be inconsistent with unitarity at finite $\alpha'$. Then why does it pass the multi-particle unitarity tests we pursued in the previous subsection?

Let us take a closer look at our findings at $\alpha_0 = 0$. As it turns out, the structure is very general and very simple: for all $(N, s, \vec{\mu})$ with generically different $\mu_i$, the moments admit the following closed-form expression, where the vertices factorize:
\begin{equation}\label{eq:alpha0-rank-1}
    m_k^{(N,s,\vec\mu)} = w_{N,s} \prod_{i=1}^{k} a_{s,\mu_i},
\end{equation}
so that for any fixed vertex label $\mu_i = \mu$ the sequence in $k$ is geometric $m_k^{(N,s,\mu)} = w_{N,s}\,(a_{s,\mu})^k$, and $r = 1$ always\footnote{This rank-one structure corresponds precisely to the saturation of the multi-positivity bounds by the open string observed in Ref.~\cite{Cheung:2025nhw}.}. Here, $w_{N,s}$ is the four-point partial wave coefficient of the level-$N$ residue given implicitly by
\begin{equation}
    P_N(\cos\theta_{2,3}) \equiv \binom{\tfrac{N}{2}(1+\cos\theta_{2,3})}{N}
    = \sum_{s=0}^{N}\, w_{N,s}(d)\, C^{(\nu_0)}_s(\cos\theta_{2,3}),
\end{equation}
and $a_{s,\mu}$ is exactly~\eqref{eq:a-k-mu-form} for $b = s$ in the limit as external mass $m_{i + 2} \to 0$ with $M_i = M_{i+1} = M$. We have checked this formula at all levels $N = 0,1, \ldots, 6$ and at all spins and vertex quantum numbers allowed per level, up to $k = 2$.

There are several interesting questions raised by this expression. The first is the rank-$1$ observation: does this result indicate that, at $\alpha_0 = 0$, there is \textit{one} state flowing through the half-ladder with these couplings, or is it the result of multiple contributing states which are truly degenerate, that is, which have the \textit{same} three-point couplings? The two situations can be differentiated by deriving the moments when $\alpha_0$ is slightly away from $0$ and then taking the limit as $\alpha_0 \to 0$. There are two possibilities in this limit. Suppose we start with rank $r > 1$ at $\alpha_0 \neq 0$. If all $w$'s except one go to zero as we take $\alpha_0 \to 0$, that means that all but one state is genuinely decoupling at the point $\alpha_0 = 0$. If, however, all the eigenvalues converge to one value, then that means that all the states are still contributing but they are individually invisible at the point $\alpha_0 = 0 $ since they all develop the same coupling. 

We can test what is happening explicitly in some examples. As discussed in Sec.~\ref{sec:unitarity}, we explicitly computed the scalar moments at level $N=2$, whose determinants are shown in Eq.~\eqref{eq:N-2-dets}. At $\alpha_0 \neq 0$, we see that the spectrum generically is rank $3$ (with one negative weight $w$, indicating a ghost, and the rest of the weights and eigenvalues positive), whereas at $\alpha_0 = 0$ it degenerates into rank $1$. We have explicitly checked that indeed, in this case, the eigenvalues merge together as one approaches $\alpha_0 \to 0$. So, here we find that there are truly $3$ states present, but they all degenerate to have the same eigenvalue.\footnote{This degeneracy is what exempts $\alpha_0 = 0$ from the minimal-degeneracy factorization argument of Ref.~\cite{Basile:2026gnd}.}

There is a simple kinematic way to understand why the result at $\alpha_0 = 0$ is always rank $1$. Let us start by examining the first internal vertex in an $n$-point half-ladder diagram, with particles $I_1, p_3, I_2$. We see, for equal-level states on both internal lines, that
\begin{equation}
    p_3 \cdot q_1 = \frac{1}{2}(q_2^2 - q_1^2 - p_3^2) = \frac{1}{2}(-M_2^2 + M_1^2 - p_3^2) = 0,
\end{equation}
since $M_1 = M_2$ and $p_3^2 = -\alpha_0 = 0$. In the rest frame of $I_1$ with $q^\mu_1 = (M_1,0, \ldots)$, this forces that particle $3$ has $E_3 = 0$, which means that $p_3^\mu$ must be the \textit{zero} vector. Of course, this applies to all rungs of the ladder, and so we have that $p_i^\mu = 0$ for all $i = 3,\ldots, n-2$. In terms of our Lorentz-invariant kinematic data, this fact implies that all planar invariants $X_{i,j} = 0$ except for $X_{2,n} = 2p_2 \cdot p_{n-1}$ (and, of course, those fixed by the residue $X_{1,i} = -M_{i-2}^2$). So, the rungs of the ladder are ``decoupling'' in the equal-$(N,s)$ half-ladder, leaving us with just a four-point half-ladder between incoming pair $(1,2)$ and outgoing pair $(n-1,n)$. Just as at four-points, this single $X$ variable can be parameterized by a polar angle $\Theta$ between these pairs:
\begin{equation}
    \cos(\Theta) = 1 - \frac{2X_{2,n}}{M^2}.
\end{equation}
So, in terms of this angular variable, the residue should just be given by the known four-point partial wave basis element:
\begin{equation}\label{eq:4pt-npt}
    R^{(n,d)} = w_{N,s} C_s^{(\nu_0)}(\cos\Theta).
\end{equation}
This angle $\Theta$ does not correspond to any $\theta_{i,j}$ of the general $n$-point expansion. Rather it is some combination of them: for example, at $n=5$ it is given by
\begin{equation}
    \cos \Theta = -\cos\theta_{2,3} \cos \theta_{3,4} + \sin\theta_{2,3} \sin\theta_{3,4} \cos\theta_{2,4},
\end{equation}
which is just how we recover the general partial-wave expansion out of Eq.~\eqref{eq:4pt-npt}.

While we now see how the four-point partial wave coefficients appear in Eq.~\eqref{eq:alpha0-rank-1}, Eq.~\eqref{eq:4pt-npt} doesn't quite seem to agree with Eq.~\eqref{eq:alpha0-rank-1}. However, they do agree, which one can show as follows. The fact that $p_i^\mu = 0$ for all internal $i$ implies that all internal three-point amplitudes~\eqref{eq:3pt-expansion} collapse to a unique value:
\begin{equation}
    \mathcal{A}_3(\epsilon_{I_i}, 0, \epsilon_{I_{i+1}}) = (\epsilon_{I_i} \cdot \epsilon_{I_{i+1}})^s,
\end{equation}
where $s = s_i = s_{i+1}$. If the spins are unequal $s_i \neq s_{i+1}$, the three-point vertex must vanish if $p^\mu = 0$, as one can see directly from Eq.~\eqref{eq:3pt-expansion}.\footnote{This implies that the internal vertices are diagonal in spin $\delta_{s_i, s_{i+1}}$ in the equal-mass half-ladder at $\alpha_0 = 0$.} In the $b$-basis, the moments~\eqref{eq:alpha0-rank-1} should therefore only be nonzero at $b = s$. The relationship between these two bases, which directly derives from Eq.~\eqref{eq:k-to-mu-npt}, is given by
\begin{equation}
    \widetilde{m}_k^{(N,s,b)} = \sum_{\mu_1,\dots,\mu_k = 0}^{s}
    \Big[\prod_{i=1}^{k} \big(a^{-1}\big)_{\mu_i,b}\Big]\,
    m_k^{(N,s;\vec{\mu})},
\end{equation}
so that we find that
\begin{equation}\label{eq:p-m-k}
    \widetilde{m}_k^{(N,s,b)} = w_{N,s} \delta_{b,s},
\end{equation}
which is exactly what is shown in Eq.~\eqref{eq:4pt-npt}! So, at $\alpha_0 = 0$ with equal-$(N,s)$ internal lines, the rungs of the ladder decouple, the internal three-particle amplitudes are exactly known, and the partial wave extraction returns exactly the four-point partial wave coefficients.

As a result, due to the kinematics of having \textit{massless} external particles, we can resolve no more information than what we found at four-points, meaning the spectrum is always rank-$1$. Additionally, we never find a unitarity violation in this test, since the $\alpha_0 = 0$ point is completely consistent with unitarity at $n$-points so long as it is consistent with unitarity at four-points, which only requires $d \leq 10$.

\begin{figure}[t]
    \centering
    \includegraphics[width=1.0\linewidth]{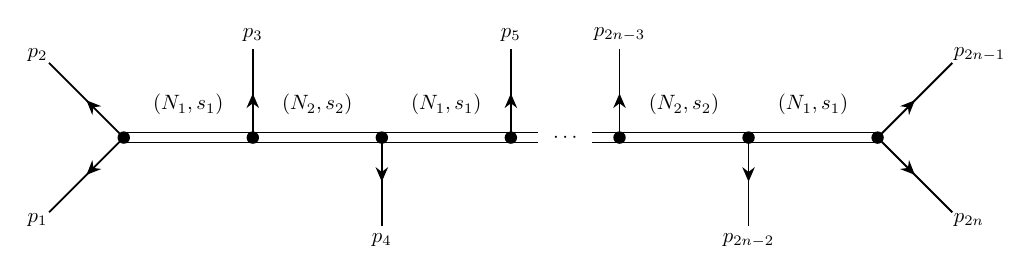}
    \caption{The so-called ``twisted'' half-ladder diagram at $2n$-points. The internal lines alternate between the string states $(N_1,s_1)$ and $(N_2,s_2)$.}
    \label{fig:twist-lad}
\end{figure}

So, we do not learn very much about $Z$-theory from the equal $(N,s)$ half-ladder diagrams. However, there are many other interesting situations we can investigate. For example, we can consider the ``twisted'' half-ladder configuration, as shown in Fig~\ref{fig:twist-lad}, where we place alternating string states $(N_1,s_1)$ and $(N_2,s_2)$ on the internal lines. For this Feynman diagram, the form of the partial wave expansion is completely unchanged, but the $F$-polynomials given in Eqs.~\eqref{eq:F-polynomial-string} and~\eqref{eq:F-expansion} for the ray-like triangulation must be modified to fit the twisted half-ladder triangulation (which corresponds to a ``zigzag'' on the momentum polygon). 

Let us start by considering the $n = 4$ version, which is the same as for the regular half-ladder and gives
\begin{equation}
\fourpt \;=\; \overline m_0 \;=\; g_i\, g_i,
\end{equation}
for some internal string state $(N_1,s_1)$. At $n = 6$, we have
\begin{equation}
\sixpt \;=\; \overline{m}_2 \;=\; g_i\, A_{ik}\, B_{kj}\, g_j,
\end{equation}
where the three-point couplings $A_{ik}, B_{kj}$ are given by
\begin{equation}
\vertexA \;=\; A_{ij}\,,\qquad\qquad \vertexB \;=\; B_{ij}.
\end{equation}
If we rotate the three-particle amplitude corresponding to $B$ by $\pi$ we see clearly that $A_{ij} = B_{ji}$. So, the full combination $G_{ij} = A_{ik}B_{kj}$ is symmetric:
\begin{equation}
    G_{ji} = A_{jk}B_{ki} = B_{kj}A_{ik} = A_{ik}B_{kj} = G_{ij}.
\end{equation}
This means that we can use these $2n$-point twisted half-ladders to build a unitarity test which, notably, does not suffer from the extreme degeneracy problem we saw with the equal-level half-ladders.\footnote{This same argument does not work with the ordinary half-ladder with alternating internal states, since in this case the internal three-point vertices live in different color-orderings.} So, with the eight-point given by
\begin{equation}
\eightpt \;=\; \overline{m}_4 \;=\; g_i\, G_{ik}\,G_{kj}\, g_j,
\end{equation}
we know using the Cauchy-Schwarz inequality that unitarity demands:
\begin{equation}
    \overline{m}_0 \overline{m}_4 - (\overline{m}_2)^2 \geq 0.
\end{equation}
We now have a new test of $Z$-theory. To put it to use, let us begin by selecting low-level states for $N_1 \neq N_2$. We cannot select the leading Regge states $N_i = s_i$, since we know these are unique and thus the inequality turns into an equality. We instead select $(N_1,s_1) = (2,1)$ and $(N_2,s_2) = (1,0)$. Notably, the fact that $s_2 = 0$ means there is a unique internal three-point structure, so we do not need to worry about $b$'s or $\mu$'s. From an explicit computation using \texttt{combwaves.py} (which works equally well for the twisted half-ladder), we find that
\begin{equation}
    \overline{m}_0 \overline{m}_4 - (\overline{m}_2)^2 = -\frac{1}{8(d-3)^2},
\end{equation}
which is \textit{negative} for all $d$! So we find that $Z$-theory is indeed inconsistent with unitarity in all spacetime dimensions $d$.

\subsection{Degeneracies of the open bosonic string}\label{sec:degen-1}

Having thoroughly investigated $Z$-theory, we are finally ready to turn to the only remaining unitary theory in these string amplitude deformations: the open bosonic string at $\alpha_0 = -1$. Here, we will work with the moments in the $b$-basis $\widetilde{m}_k^{(N,s,b)}$. In App.~\ref{sec:data} we give expressions for these moments through level $N = 6$, and this is the data we will use throughout this subsection. 

We will begin the discussion here with the first non-tachyonic massive level, the scalar at $N = 2$, where we have\footnote{When the internal states are scalars, the internal three-point coupling is just a constant, and so $\widetilde{m} = m$.}
\begin{equation}
    \widetilde{m}_k^{(2,0,0)} = -\frac{d - 26}{8(d-1)}\left( \frac{9d - 34}{8(d-1)} \right)^k.
\end{equation}
Hence, there is one distinguishable state with
\begin{equation}
    \widetilde{w}_1^{(2,0,0)} = -\frac{d - 26}{8(d-1)}, \quad \widetilde{\lambda}_1^{(2,0,0)} = \frac{9d - 34}{8(d-1)}.
\end{equation}
Just as at level $1$, the vector decouples also at level $2$ from the half-ladder. However, for the leading state at $s = N = 2$ in the Regge trajectory, we find
\begin{equation}
    \widetilde m_k^{(2,2,0)} = \frac{25}{4(d-3)(d-1)}\,2^{k},\qquad
    \widetilde m_k^{(2,2,1)} = \frac{25}{4(d-3)(d-1)}\,(-4)^{k},\qquad
    \widetilde m_k^{(2,2,2)} = \frac{25}{4(d-3)(d-1)},
\end{equation}
which are all describing one distinguishable state,\footnote{This, of course, is the well-known fact that the leading state $N = s$ in the spectrum is unique.} whose three-point coupling to the tachyon is, up to normalization, the $d$-independent vertex:
\begin{equation}
2(\epsilon_{I_i}\!\cdot p_{i+2})^2(\epsilon_{I_{i+1}}\!\cdot p_{i+2})^2 - 4(\epsilon_{I_i}\!\cdot\epsilon_{I_{i+1}})(\epsilon_{I_i}\!\cdot p_{i+2})(\epsilon_{I_{i+1}}\!\cdot p_{i+2}) + (\epsilon_{I_i}\!\cdot\epsilon_{I_{i+1}})^2,
\end{equation}
i.e.\ $(\widetilde{\lambda}_0,\widetilde{\lambda}_1,\widetilde{\lambda}_2) = (2,-4,1)$. Note that at the critical dimension $d = 26$, this is the first massive state that contributes to the half-ladder.

At $N = 3$, we find more of the same: the scalars and the tensors decouple, whereas the vector and spin-$3$ states are all rank $1$: for the latter, we find
\begin{equation}
    \widetilde m_k^{(3,3,b)} = \frac{27}{(d-3)(d-1)(d+1)}\,\big(\widetilde{\lambda}_b\big)^k,\qquad (\widetilde{\lambda}_0,\widetilde{\lambda}_1,\widetilde{\lambda}_2,\widetilde{\lambda}_3) = \big(-\tfrac43,\,6,\,-6,\,1\big),
\end{equation}
which again give $d$-independent couplings:
\begin{equation}
\begin{split}
    -\tfrac{4}{3}\,(\epsilon_{I_i}\!\cdot p_{i+2})^3(\epsilon_{I_{i+1}}\!\cdot p_{i+2})^3
    &+ 6\,(\epsilon_{I_i}\!\cdot\epsilon_{I_{i+1}})(\epsilon_{I_i}\!\cdot p_{i+2})^2(\epsilon_{I_{i+1}}\!\cdot p_{i+2})^2\\
    &- 6\,(\epsilon_{I_i}\!\cdot\epsilon_{I_{i+1}})^2(\epsilon_{I_i}\!\cdot p_{i+2})(\epsilon_{I_{i+1}}\!\cdot p_{i+2})
    + (\epsilon_{I_i}\!\cdot\epsilon_{I_{i+1}})^3 .
\end{split}
\end{equation}
The vector has
\begin{equation}
    \widetilde m_k^{(3,1,b)} = -\frac{d-26}{2(d-3)(d+1)}\,\big(\widetilde{\lambda}_b\big)^k,\qquad
    \widetilde{\lambda}_0 = -\frac{101d-34}{48(d+1)},\qquad \widetilde{\lambda}_1 = \frac{35d-46}{32(d+1)},
\end{equation}
which decouples at the critical dimension just as at level $N = 2$. Note that now the couplings are $d$-dependent, yielding three-point vertex:
\begin{equation}
    -\frac{101d-34}{48(d+1)}\,(\epsilon_{I_i}\!\cdot p_{i+2})(\epsilon_{I_{i+1}}\!\cdot p_{i+2})
+\frac{35d-46}{32(d+1)}\,(\epsilon_{I_i}\!\cdot\epsilon_{I_{i+1}}).
\end{equation}
Things start to get interesting at $N = 4$. To illustrate what happens, let us begin by considering the scalar states. We compute moments $m_i$ for this case up to $i = 4$, and we find
\begin{align}\label{eq:Hankel-N-4}
    \det \widetilde{H}^{(4)}_0 &= \frac{9d^2-490d+6704}{384\,(d-1)(d+1)}\,,\nonumber\\[4pt]
    \det \widetilde{H}^{(4)}_1 &= \frac{(d-26)(5d-152)\,\big(27d^2-910d+7344\big)^2}{84934656\,(d-1)^4\,(d+1)^3}\,,\\[4pt]
    \det \widetilde{H}^{(4)}_2 &= 0\,.\nonumber
\end{align}
Hence, for $d < 26$, the space has rank $r = 2$, and there are genuinely two distinguishable scalars flowing in the half-ladder. Following the procedure outlined in App.~\ref{sec:hankel}, we determine that
\begin{gather}\label{eq:N-4-couplings}
\widetilde{m}^{(4)}_k=\widetilde{w}_0\,\widetilde{\lambda}_0^{\,k}+\widetilde{w_1}\,\widetilde{\lambda}_1^{\,k},
\end{gather}
where $\widetilde{w}_{0,1}, \widetilde{\lambda}_{0,1}$ can be found in App.~\ref{sec:data}. So, at $N = 4$, we get our first genuine degeneracy between particles which have different couplings $\widetilde{\lambda}_i$ and $\widetilde{w}_i$. Interestingly, as one approaches $d \to 26$ from below, we see in App.~\ref{sec:data} that $\widetilde{w}_1 \to 0$, which indicates that the second state decouples right at the critical dimension. (That the rank has to fall to $1$ can be gleaned from Eq.~\eqref{eq:Hankel-N-4}.) Hence, we are left with
\begin{equation}
    \widetilde{m}_k^{(4)}(d \to 26) = \widetilde{w}_0 \widetilde{\lambda}_0^k
\end{equation}
with
\begin{equation}
    \widetilde{\lambda}_0 = \frac{1631}{1800}, \qquad \widetilde{w}_0 = \frac{1}{5400}.
\end{equation}
For particles with spin, we find that $s = 1,3$ states decouple completely at $N = 4$. We are therefore left with spin-$2$ and spin-$4$. Beginning at spin-$2$, we find that, just like for the scalar modes, it is rank-$2$ for $d < 26$, where the last nonzero determinants are
\begin{align}
    \det \widetilde H^{(4,2,0)}_1 &= -\frac{2401\,(d-26)(17d-250)^2}{497664\,(d-3)^2(d-1)^2(d+3)^3}\,,\nonumber\\[4pt]
    \det \widetilde H^{(4,2,1)}_1 &= -\frac{2401\,(d-26)(5d-342)^2}{497664\,(d-3)^2(d-1)^2(d+3)^3}\,,\\[4pt]
    \det \widetilde H^{(4,2,2)}_1 &= -\frac{2401\,(d-26)(3d-82)^2}{7962624\,(d-3)^2(d-1)^2(d+3)^3}\, ,\nonumber
\end{align}
and at $d = 26$ it drops to rank $1$. One can find expressions for each $b = 0, 1, 2$ in App.~\ref{sec:data}. We only note here that, at $d = 26$, the $\widetilde{w}_1$'s all vanish and we recover rank-$1$ behavior:
\begin{align}
    \underline{b=0}:&\qquad \widetilde{\lambda}_0 = \frac{89}{174}, \qquad \widetilde{w}_0 = \frac{49}{800400}, \nonumber\\
    \underline{b=1}:&\qquad \widetilde{\lambda}_0 = -\frac{66}{29}, \qquad \widetilde{w}_0 = \frac{49}{800400}, \\
    \underline{b=2}:&\qquad \widetilde{\lambda}_0 = \frac{86}{87}, \qquad \widetilde{w}_0 = \frac{49}{800400}. \nonumber
\end{align}
Note that all $\widetilde{w}_0$ are equal in this case at rank-$1$, just as they should be given their interpretation as the coupling squared of the external - external - internal vertex on the far-left and far-right of the half-ladder.

With this data, we can now explicitly write out the three-point amplitudes involving these two tensor modes and an external tachyon. First, since we have these two states at $(N,s) = (4,2)$, we need to specify a basis. Let us select the basis where the $b = 2$ coupling is diagonal, that is, where we have for both states $1$:
\begin{equation}
    \frac{77d + 62}{72(d+3)} (\epsilon_{I_i} \cdot \epsilon_{I_{i+1}})^2 
\end{equation}
and for both states $2$ :
\begin{equation}
    (\epsilon_{I_i} \cdot \epsilon_{I_{i+1}})^2.
\end{equation}
In particular, we do not allow for state $1$ to couple to state $2$ for the $b = 2$ amplitude structure. As required, one can see that these coefficients are just $\widetilde{\lambda}_i$'s for $b = 2$. Then, we must rotate the $b = 0$ and $b=1$ results shown in App.~\ref{sec:data} so that they agree with this basis choice, and we find for the full answer that
\begin{align}\label{eq:V-4-2}
V_{00}&=\frac{409d^2-8596d-65804}{36\,(d+3)(5d-154)}\,M_0-\frac{127d^2-3476d-4980}{6\,(d+3)(5d-154)}\,M_1+\frac{77d+62}{72\,(d+3)}\,M_2,\\[6pt]
V_{11}&=\frac{d-314}{6\,(5d-154)}\,M_0-\frac{2\,(17d-586)}{3\,(5d-154)}\,M_1+M_2,\\[6pt]
V_{01}&=-\sqrt{\frac{26-d}{6(d+3)}}\;\frac{71d-346}{6\,(5d-154)}\,M_0+\sqrt{\frac{26-d}{6(d+3)}}\;\frac{2\,(7d+190)}{3\,(5d-154)}\,M_1 .
\end{align}
where we have
\begin{equation}
M_b\equiv\big(\epsilon_{I_i}\!\cdot\epsilon_{I_{i+1}}\big)^{b}\big(\epsilon_{I_i}\!\cdot p_{i+2}\big)^{2-b}\big(\epsilon_{I_{i+1}}\!\cdot p_{i+2}\big)^{2-b},\qquad b=0,1,2.
\end{equation}
$V_{i,j}$ gives the three-point amplitude between internal tensor $i$, internal tensor $j$, and the external tachyon at $N = 4$.

Finally, let us discuss the leading state at $N = s = 4$. There, we find (as given in App.~\ref{sec:data}) that:
\begin{equation}
    \widetilde m^{(4,4,b)}_k=\frac{2401}{16\,(d-3)(d-1)(d+1)(d+3)}\;\big(\widetilde\lambda_b\big)^k,\qquad
    \big(\widetilde\lambda_0,\widetilde\lambda_1,\widetilde\lambda_2,\widetilde\lambda_3,\widetilde\lambda_4\big)=\Big(\tfrac23,\,-\tfrac{16}{3},\,12,\,-8,\,1\Big).
\end{equation}
Again, we find that the couplings are $d$-independent pure numbers. The corresponding internal three-point amplitude is thus:
\begin{align}
&\tfrac{2}{3}\,(\epsilon_{I_i}\!\cdot p_{i+2})^4(\epsilon_{I_{i+1}}\!\cdot p_{i+2})^4
-\tfrac{16}{3}\,(\epsilon_{I_i}\!\cdot\epsilon_{I_{i+1}})(\epsilon_{I_i}\!\cdot p_{i+2})^3(\epsilon_{I_{i+1}}\!\cdot p_{i+2})^3\nonumber\\
&\quad+12\,(\epsilon_{I_i}\!\cdot\epsilon_{I_{i+1}})^2(\epsilon_{I_i}\!\cdot p_{i+2})^2(\epsilon_{I_{i+1}}\!\cdot p_{i+2})^2
-8\,(\epsilon_{I_i}\!\cdot\epsilon_{I_{i+1}})^3(\epsilon_{I_i}\!\cdot p_{i+2})(\epsilon_{I_{i+1}}\!\cdot p_{i+2})\nonumber\\
&\quad+(\epsilon_{I_i}\!\cdot\epsilon_{I_{i+1}})^4 .
\end{align}

Using standard covariant quantization for the open bosonic string~\cite{Polchinski:1998rq}, one can compute the number of physical states at each $(N,s)$ for any $d \leq 26$, as well as use the tachyon vertex operator to calculate how each of these states couples to the external tachyon. By doing this and fixing the normalization by matching to the four-point residue, we are able to check the results given in App.~\ref{sec:data}, and we find agreement. Further, we performed the same checks for the twisted half-ladder at $\alpha_0 = -1$. Parity of the underlying worldsheet theory requires that the twisted half-ladder equals the untwisted one up to a sign $(-1)^{N_l + N_r}$ for each twisted vertex, where $N_l, N_r$ are the levels of the adjacent internal modes. By plugging the twisted half-ladder open string residues into \texttt{combwaves.py}, we verified in a large number of examples that the answers indeed obey this expected sign flip, both for equal and alternating $(N,s)$ internal lines configurations.

We leave a more detailed, comprehensive investigation of these and related findings to future work. For the couplings of individual massive string states extracted by other means, see Refs.~\cite{Arkani-Hamed:2023jwn,Feng:2010yx,Lust:2009pz,Bianchi:2010es,Bianchi:2015yta,Bucciotti:2025dnh}.

\section{Discussion}\label{sec:discussion}

The partial wave expansion has a long and storied history, dating back to the early days of quantum mechanics~\cite{Faxen:1927,Jacob:1959at,Wick:1962zz}. In its classical form for $2\to 2$ scattering, it allows one to resolve how any given spin-$S$ exchanged particle contributes to the $s$-channel diagram. From imposing unitarity, we find that each coefficient in the partial wave expansion must be non-negative and bounded~\cite{Arkani-Hamed:2020blm,Chiang:2021ziz,Bellazzini:2020cot,Berman:2023jys}, which forbids the amplitude from growing too fast at high-energies~\cite{Froissart:1961ux,Martin:1962rt}. These observations underlie much of the modern S-matrix bootstrap program.

In this paper, we introduced a general partial wave expansion for the half-ladder Feynman tree diagram at any multiplicity and in any spacetime dimension, and proved it by enforcing that it is an eigenvector of each internal particle's angular momentum. By selecting an appropriate set of angular variables (and relating them in a Lorentz-invariant way to dot products of external momenta), we showed that the basis takes the form of a sum over products of special functions derived from the Gegenbauer polynomials. For both the full kinematics when $d \geq n - 1$ and the Gram-restricted kinematics otherwise, we gave a simple graphical rule based on upside-down triangles that allowed us to algorithmically write down the partial wave expansion at any $n, d$. We then applied this formalism to ``string theory'' amplitudes. We first introduced a tunable Regge intercept $\alpha_0$ and showed that, by enforcing multi-particle unitarity at equal-$(N,s)$ for just the very few first levels of internal scalars, positivity restricted $\alpha_0 = 0, d\leq 10$ ($Z$-theory) or $\alpha_0 = -1, d \leq 26$ (open bosonic string theory). Then, we constructed a positivity test for the ``twisted'' half-ladder with alternating internal particles $(2,1)$ and $(1,0)$, and showed that $Z$-theory failed it for all spacetime dimension $d$. This confirmed the expectation that $Z$-theory does not define a unitary string theory in any dimension at finite $\alpha'$. Finally, for the open bosonic string (the only unitary theory in this deformation), we took the first steps into using the partial wave expansion (combined with Hankel matrix technology) to compute how individual degenerate states at a given $(N,s)$ couple to one another in each three-particle vertex of the half-ladder.

There are several directions for future work suggested by the results of this paper. 

The first is to write down the partial wave basis for all $SO(d-1)$ irreps which generically appear inside the half-ladder diagrams. In the present paper, we considered only the symmetric-traceless tensor irreps, since for these the special function $T$ admitted a straightforward closed-form expression in terms of Gegenbauers~\eqref{eq:T-def}. However, starting at $n = 6$, the middle internal particle $I_2$ can be of mixed-symmetry type. That is, instead of being labeled by one quantum number $s$, it can be labeled by two $[\sigma,\tau]$. For example, when $\sigma = \tau = 1$, this would correspond to an internal anti-symmetric rank $2$ tensor. We suspect that the \textit{form} of the partial wave expansion would stay the same, but that the exact expressions for the $T$'s and potentially the coefficient matrices $Y$ would change. Note that an avatar of this question was asked in Refs.~\cite{Buric:2023ykg,Chowdhury:2019kaq,Hebbar:2020ukp,Caron-Huot:2022jli}, where the authors considered spinning external states in general $d$ for the four-point amplitude. (Similarly, we can consider external spinning states at general $n$ multiplicity.)

Secondly, starting at $n = 6$, the full set of contributing planar diagrams to a scattering procedure contains more topologies than just the half-ladder. At $n = 6$ itself, there is the ``Mercedes Benz,'' where all the three internal particles meet in the center and each couples to two external particles. We discussed this diagram in Sec.~\ref{sec:six-pt}, and there we saw some encouraging signs that there is nice structure. One could try to understand this particular example better. Or, more adventurously, one could explore the most generic topologies and try to write down a general partial wave expansion which works on \textit{any} tree-level Feynman diagram. We believe it is very likely that such an object exists, and the proof strategy from the text (enforcing that the residue is an eigenfunction of angular momentum in each internal particle's center-of-mass frame) should go through. One would need to work out the specifics, such as a general set of angular variables parameterizing the space as well as exactly what kind of special functions appear. 

When we discussed applying this technology to string amplitudes, we only just began to understand how individual states at a given $(N,s)$ contribute to the half-ladder for the open bosonic string; for example, in Sec.~\ref{sec:degen-1}, we found that, at level $N = 4$, there are two distinguishable scalar modes contributing in $d < 26$, and we gave their couplings in Eq.~\eqref{eq:N-4-couplings}. One could push this direction further, enumerating a more comprehensive table of these results for the string at $\alpha_{0} = -1$. For example, with more computational power, one could derive the famous Hagedorn growth of string theory in-action from this purely on-shell analysis.

However, the main result of this paper --- an explicit realization of the $n$-point partial wave expansion in generic dimension $d$ for the half-ladder diagram --- of course applies not just to string amplitudes but to \textit{any} amplitudes one can cook up. So, in its most adventurous incarnation, we can use this technology to constrain and understand new candidates for an $n$-point amplitude or amplitude-like object~\cite{Cheung:2023uwn,Bhardwaj:2024klc,Arkani-Hamed:2019plo,Arkani-Hamed:2023jwn,Geiser:2023qqq,Basile:2026gnd,Jepsen:2025baw,Bjerrum-Bohr:2024wyw}. From such an ansatz, we can test whether the ``amplitude'' can be built in terms of three-point vertices as shown in the half-ladder diagram, exactly which (distinguishable) particles are flowing through it, what the three-point on-shell vertices are--- and we can check, using our positivity conditions, whether it in fact describes a consistent, unitary amplitude.

\hspace{1cm}

\textbf{Acknowledgments}

I am greatly appreciative of enlightening conversations with Nima Arkani-Hamed, Justin Berman, Mathieu Giroux, Aidan Herderschee, and Sebastian Mizera. I also thank Nima Arkani-Hamed and Justin Berman for comments on the draft. I acknowledge the use of Claude and ChatGPT AI models for aspects of analysis, code-writing, checking formulae, and editing. This work was supported by the NSF Graduate Research Fellowship under Grant No. DGE-2444107. 

\appendix

\section{Orthogonality of the Partial Wave Basis}\label{sec:ortho}

The Gegenbauer polynomials are orthogonal in their lower index $n$ under measure $\sin^{2\nu_0}\theta$:
\begin{equation}\label{eq:gegen-ortho}
    \int_{0}^{\pi}d\theta\,\sin^{2\nu_0}\theta\;C^{(\nu_0)}_n(\cos\theta)\,C^{(\nu_0)}_{n'}(\cos\theta)
    =\delta_{nn'}\,h^{(\nu_0)}_n,
    \qquad
    h^{(\lambda)}_n=\frac{2^{1-2\lambda}\,\pi\,\Gamma(n+2\lambda)}{n!\,(n+\lambda)\,\Gamma(\lambda)^2}.
\end{equation}
Given this, one can show that the associated Gegenbauer polynomials obey
\begin{equation}\label{eq:A-ortho}
\begin{split}
    \int_{0}^{\pi}\!d\theta\,\sin^{2\nu_0}\theta\int_{0}^{\pi}\!d\varphi\,\sin^{2\nu_1}\varphi\;
    \Big[A^{d}_{s,\mu}(\theta)\,C^{(\nu_1)}_{\mu}(\cos\varphi)\Big]
    \Big[A^{d}_{s',\mu'}(\theta)\,C^{(\nu_1)}_{\mu'}(\cos\varphi)\Big]&\\[2pt]
    =\delta_{ss'}\,\delta_{\mu\mu'}\;h^{(\mu+\nu_0)}_{s-\mu}\,h^{(\nu_1)}_{\mu}&
\end{split}
\end{equation}
in both $s$ and $\mu$.

From these relations, one can derive that the (rescaled) $T$ polynomial given in Eq.~\eqref{eq:rescaled-T} obeys the following orthogonality condition in spin index $s$:
\begin{equation}
    \sum_{l=0}^{\min(\mu_1,\mu_2)} b_{\mu_1,l}(\nu_1)\, b_{\mu_2,l}(\nu_1)\,\|t_l\|^{2}
    \int_{-1}^{1}dx\,(1-x^2)^{\nu_1}\;
    T^{d}_{s,\mu_1,\mu_2,l}(x)\,T^{d}_{s',\mu_1,\mu_2,l}(x)
    \;=\;\delta_{ss'}\;I_{2\nu_0}\;\mathcal Q_{s;\mu_1,\mu_2},
\end{equation}
with
\begin{equation}
\begin{gathered}
    I_a\equiv\int_0^\pi d\theta\,\sin^a\theta,\qquad
    \|t_l\|^2=\mathcal A^{d-1}_{\mu_1,l}\,\mathcal A^{d-1}_{\mu_2,l}\,\mathcal C_l,\\[4pt]
    \mathcal A^{d-1}_{\mu,l}= \frac{h_{\mu-l}^{(l+\nu_1)}}{I_{2\nu_1}},\qquad
    \mathcal C_l=\frac{(2\nu_2)_l\;\nu_2}{l!\,(l+\nu_2)},
\end{gathered}
\end{equation}
and
\begin{align}
    \mathcal Q_{s;\mu_1,\mu_2}
    &=\Big[\frac{(2\nu_1)_p\,(2\nu_1)_q}{p!\,q!}\Big]^2\frac{1}{\mathcal N_s},
    \\[6pt]
    \mathcal N_{s}&=\frac{2\,(s+\nu_0)\,(2\nu_{-1})_{s-1}}{s!},
    \qquad p=\min(\mu_1,\mu_2),\qquad q=\max(\mu_1,\mu_2).
\end{align}
With these ingredients, one can then prove via a recursive argument that the partial-wave basis $R^{(n,d)}$ given explicitly in Sec.~\ref{sec:proof} is orthogonal. Defining the appropriate measure and inner product as
\begin{equation}
    \big\langle f,g\big\rangle_{n,d}
    \;=\;\int f\,g\;du^{(d)}_n,
    \qquad
    du^{(d)}_n=\prod_{2\le i<j\le n-1}\sin^{2\nu_{j-i-1}}\theta_{i,j}\;d\theta_{i,j},
\end{equation}
(an angle at depth $r=j-i-1$ carries the weight $\sin^{2\nu_r}$), the orthogonality condition reads
\begin{equation}\label{eq:ortho-con}
    \Big\langle R^{(n,d)}_{\vec s;\vec \mu},\,R^{(n,d)}_{\vec s\,';\vec \mu'}\Big\rangle_{n,d}
    =\;\delta_{\vec s\,\vec s\,'}\;\delta_{\vec \mu\,\vec \mu'}\;
    \big\|R^{(n,d)}_{\vec s;\vec \mu}\big\|^{2}
    \qquad
    \delta_{\vec s\vec s'}=\prod_{i=1}^{n-3}\delta_{s_is_i'},\quad
    \delta_{\vec \mu\vec \mu'}=\prod_{i=1}^{n-4}\delta_{\mu_i\mu_i'} .
\end{equation}
with norm given (recursively) by
\begin{equation}
    \big\|R^{(n,d)}_{\vec s;\vec \mu}\big\|^{2}
    =\big(N^{(n)}\big)^{2}\;
    h^{(\mu_1+\nu_0)}_{s_1-\mu_1}\;
    h^{(\mu_{n-4}+\nu_0)}_{s_{n-3}-\mu_{n-4}}
    \sum_{\vec l}\,Y_{\vec{l}}(\nu_1)^{\,2}\;
    \Big[\prod_{r=2}^{n-4}\tau^{\,d}_{s_r,\mu_{r-1},\mu_r,l_{r-1}}\Big]\;
    \big\|R^{(n-1,d-1)}_{\vec \mu;\vec l}\big\|^{2},
\end{equation}
with
\begin{equation}
    \tau^{d}_{s,\mu_1,\mu_2,l}\equiv\int_0^\pi d\theta\,\sin^{2\nu_0}\theta\,\Big[T^{d}_{s,\mu_1,\mu_2,l}(\cos\theta)\Big]^2,
\end{equation}
seeded by the basic five-point result
\begin{equation}
    \big\|R^{(5,d)}_{(s_1,s_2);\mu}\big\|^{2}=h^{(\mu+\nu_0)}_{s_1-\mu}\,h^{(\mu+\nu_0)}_{s_2-\mu}\,h^{(\nu_1)}_{\mu}.
\end{equation}
In fact, at six-points as well the norm has closed form
\begin{equation}\label{eq:six-pt-ortho}
    \big\|R^{(6,d)}_{(s_1,s_2,s_3);(\mu_1,\mu_2)}\big\|^{2}
    =\big(N^{(6)}\big)^{2}\;
    h^{(\mu_1+\nu_0)}_{s_1-\mu_1}\;
    h^{(\mu_2+\nu_0)}_{s_3-\mu_2}\;
    I_{2\nu_0}\,I_{2\nu_1}^{\,2}\,I_{2\nu_2}\;\mathcal Q_{s_2;\mu_1,\mu_2},
\end{equation}
because there is only one instance of $T$. Thus, we can extract the partial-wave coefficients $y_{\vec{s},\vec{\mu}}$ of a given half-ladder residue $\mathcal{R}$ by
\begin{equation}\label{eq:coeff-extract}
    y_{\vec s,\vec \mu}=\frac{\big\langle R^{(n,d)}_{\vec s;\vec \mu},\mathcal R\big\rangle_{n,d}}{\big\|R^{(n,d)}_{\vec s;\vec \mu}\big\|^2}.
\end{equation}
Note that we implicitly assumed that all momenta are independent except for the constraint from momentum conservation, that is, that $d \geq n - 1$. However, as we detailed in Sec.~\ref{sec:gram}, we can define a partial wave basis in $d < n - 1$ as a modification of the fully-general expansion in higher $d$. The inner product in these restricted kinematics truncates at the depth $r_\star = 2\nu_0$ where the kinematics end:
\begin{equation}\label{eq:restrict-inner}
\big\langle f,g\big\rangle_{n,d}
=\frac{1}{2^{\,n-4-r_\star}}\sum_{\epsilon_1,\dots,\epsilon_{n-4-r_\star}=\pm1}\;
\int\;\prod_{r=0}^{r_\star}\;\prod_{p=1}^{n-3-r}\sin^{\,2\nu_r}\!\theta_{p+1,p+2+r}\;d\theta_{p+1,p+2+r}\;\;
f\,g\;
\end{equation}
with $f,g$ taken to have $\cos\theta_{p+1,p+3+r_\star}=\epsilon_p,\;\sin\theta_{p+1,p+3+r_\star}=0$. Then, with this in-hand we can extract the coefficients $y_{\vec{s}, \vec{\mu}}$ just as we did in Eq.~\eqref{eq:coeff-extract}. Orthogonality of the expansion on restricted kinematics holds over from our previous discussion at $d \geq n - 1$.

\section{Lightning Review of Hankel Matrices}\label{sec:hankel}

We will now give a brief review of the Hankel matrix technology~\cite{Schmudgen:2017}.

We have already introduced the moments $m_k$ in Eq.~\eqref{eq:mom-expansion} that we will work with in this appendix. In matrix notation, they may be written as
\begin{equation}\label{eq:m-mat-def}
    m_k = g^T G^k g.
\end{equation}
Let us take $G$ to be a square $n \times n$ matrix, and perform a spectral decomposition:
\begin{equation}
    G = \sum_{a = 1}^r \lambda_a P_a, \quad G v_i = \lambda_i v_i, \quad P_i = \sum_j v_j v_j^T,
\end{equation}
with $v_i$ orthonormal $v_i \cdot v_j = \delta_{i,j}$ and the sum in $P_i$ over all eigenvectors $v_j$ with the same eigenvalue $\lambda_i$; hence, $\lambda_i \neq \lambda_j$ unless $i = j$. The rank $r \le n$ is therefore the number of \textit{distinguishable} eigenvalues $\lambda_i$. We then derive
\begin{equation}\label{eq:Gk-g}
    G^k g = \sum_{i = 1}^r \lambda_i^k P_i g, 
\end{equation}
which means that 
\begin{equation}
    G^k g \in \mathrm{span} (P_1 g, P_2 g, \ldots, P_r g).
\end{equation}
Inverting Eq.~\eqref{eq:Gk-g} under the Vandermonde matrix $V_{ij} = \lambda_i^j$ (which is invertible since all eigenvalues are unequal), we have the result that any $P_i g$ can be expressed in terms of the $G^i g$ up to $i = r-1$, and thus
\begin{equation}
    G^k g \in \mathrm{span}( g, G g, \ldots, G^{r-1} g) = \mathcal{K},
\end{equation}
which is known as the \textit{Krylov space} with $\mathrm{rank} (\mathcal{K}) = r$. Thus, we find\footnote{The extraction of the $\lambda_a$, $w_a$ given below is Prony's method~\cite{Prony:1795}.}
\begin{equation}\label{eq:Gr-g-cs}
    G^r g = \sum_{a = 0}^{r-1} c_a G^a g \implies J(G) g = 0.
\end{equation}
for unknown constants $c_a$ and
\begin{equation}
    J(x) = x^r - \sum_{j = 0}^{r-1} c_j x^j.
\end{equation}
This implies that
\begin{equation}\label{eq:hankel-mat-eq}
    m_{r + k} = g^T G^{r +k} g = \sum_{a = 0}^{r-1} c_a m_{a+k}
\end{equation}
for all $k \geq 0$. Additionally, we have that $m_k$ can be written as
\begin{equation}\label{eq:mk-dep}
    m_k = \sum_{a = 1}^r w_a \lambda_a^k, \quad w_a = g^T P_a g.
\end{equation}
Now, let us define the $\textit{Hankel matrix}$
\begin{equation}
    H_p = [m_{i+j}]_{i,j = 0}^p,
\end{equation}
which requires knowledge of $m_k$ up to $k = 2p$. From Eq.~\eqref{eq:hankel-mat-eq}, it is clear that, when $p = r$, the columns and rows of $H_p$ are no longer linearly independent. That is,
\begin{equation}
    \det H_r = 0.
\end{equation}
So the procedure goes as follows: obtain all $m_k$ up to $k = 2p$. At each $p$, construct $H_p$ and compute its determinant. If it is nonzero, keep going; if it is zero, then $r = p$. We can then explicitly compute the coefficients $c_a$ by setting up the matrix equation using Eq.~\eqref{eq:hankel-mat-eq}:
\begin{equation}\label{eq:c-solve}
    \begin{pmatrix}
        m_0     & m_1   & \cdots & m_{r-1}\\
        m_1     & m_2   & \cdots & m_{r}\\
        \vdots  & \vdots & \ddots & \vdots\\
        m_{r-1} & m_{r} & \cdots & m_{2r-2}
    \end{pmatrix}
    \begin{pmatrix} c_0\\ c_1\\ \vdots\\ c_{r-1}\end{pmatrix}
    =
    \begin{pmatrix} m_r\\ m_{r+1}\\ \vdots\\ m_{2r-1}\end{pmatrix}.
\end{equation}
One can indeed invert this equation, since the left-hand matrix is just $H_{r-1}$, which has nonzero determinant by assumption. With the $c_a$ in-hand, we can extract the distinguishable eigenvalues $\lambda_a$ of the matrix $G$ as follows: we first note that, because we have Eq.~\eqref{eq:Gk-g} and Eq.~\eqref{eq:Gr-g-cs}, the following relation holds:
\begin{equation}
    J(G) g = \sum_{a = 1}^r J(\lambda_a) P_a g = 0.
\end{equation}
However, because the $P_a g$ are independent vectors, $J(\lambda_a) = 0$ for every $a$, and hence the roots of $J(x)$ are given by the eigenvalues:
\begin{equation}\label{eq:eigen-solv}
    J(x) = \prod_{a = 1}^r (x - \lambda_a) = x^r - \sum_{j = 0}^{r-1} c_j x^j.
\end{equation}
To obtain the weights $w_a$, we can use Eq.~\eqref{eq:mk-dep} for all $k = 0, 1, \ldots, r-1$ to form another matrix equation:
\begin{equation}\label{eq:w-solve}
    \begin{pmatrix}
        1               & 1               & \cdots & 1\\
        \lambda_1       & \lambda_2       & \cdots & \lambda_r\\
        \vdots          & \vdots          & \ddots & \vdots\\
        \lambda_1^{r-1} & \lambda_2^{r-1} & \cdots & \lambda_r^{r-1}
    \end{pmatrix}
    \begin{pmatrix} w_1\\ w_2\\ \vdots\\ w_r\end{pmatrix}
    =
    \begin{pmatrix} m_0\\ m_1\\ \vdots\\ m_{r-1}\end{pmatrix},
\end{equation}
where the left-hand matrix $V$ is just the Vandermonde $V_{ij} = \lambda_i^j$. As mentioned before, this matrix is invertible as long as all $\lambda$'s are distinguishable, because its determinant takes the clean form
\begin{equation}
    \det V = \prod_{a < b} (\lambda_b - \lambda_a).
\end{equation}
Hence, we can invert to obtain $\vec{w} = V^{-1} \vec{m}$ once we have knowledge of the eigenvalues.

Finally, we can show very succinctly that $H_p$ is always a positive semi-definite matrix:
\begin{equation}
    x^T H_p x = \sum_{i,j} x_i x_j m_{i+j} = \sum_{a = 1}^r w_a \left( \sum_i x_i \lambda_a^i\right)^2 \geq 0,
\end{equation}
since $w_a \geq 0$ by its definition in Eq.~\eqref{eq:mk-dep}.

For our string-theory purposes in Sec.~\ref{sec:string-theory}, at a particular $(N, s, \mu)$, $r$ is the number of distinguishable particles (that is, particles with different internal three-point couplings) contributing to the half-ladder, $\lambda_i$ is the coupling of the three-point vertex internal $i$ - external - internal $i$, and $w_i$ is the coupling squared of the internal $i$ - external - external vertex. (The $m_k$ moments have an $O(r)$ symmetry, so we are free to choose a basis which diagonalizes $G_{i,j} = \lambda_i \delta_{i,j}$.)

\section{Brief Introduction to the Ancillary Code}\label{sec:code}

In this appendix, we give a brief introduction to \texttt{combwaves.py}, the ancillary code which allows one to compute, given an arbitrary residue as a polynomial in dot products of the external $p_i^\mu$, the coefficient $y_{\vec{s},\vec{\mu}}$ for any given $\vec{s}$, $\vec{\mu}$, at arbitrary multiplicity and dimension. As in the entire paper, we take all momenta to be outgoing and pick the mostly-plus metric convention.

We will review here a few of its critical functions. The first is \texttt{planar\_X(n, M2, m2)}, which takes as input the multiplicity $n$, the list of internal masses squared $M2 = (M_1^2, \ldots, M_{n-3}^2)$, and the list of external masses squared $m2 = (m_1^2, \ldots, m_n^2)$, and returns every $X_{i,j}$ written in angular variables $c_{i,j} = \cos \theta_{i,j}$ and $s_{i,j} = \sin \theta_{i,j}$. This gives us the map between the residue as a polynomial in dot products of momenta and as a polynomial in $c_{i,j}, s_{i,j}$, the variables we use to write the partial wave expansion.

Another important function is \texttt{wave(n, spins, vers, dd=None)}. This function takes as input the multiplicity $n$, a list of internal ``spins'' $(s_1, s_2, \ldots, s_{n-3})$, and a list of vertex quantum numbers ``vers'' $(\mu_1, \ldots, \mu_{n-4})$, and returns the partial wave basis element as a polynomial in $c_{i,j}, s_{i,j}$. The optional argument is the dimension dd, which is kept symbolic if one doesn't list a particular numeric dimension.

Then, we have \texttt{project(n, spins, vers, R, V, dd=None)}, which returns the coefficient of one basis element labeled by its spins and vertex quantum numbers in dimension dd in the expansion of the residue $R$ written in terms of $c_{i,j}, s_{i,j}$. V is the output of \texttt{chart(n)}, which is the map from the labels $(i,j)$ to its pair of symbols $c_{i,j}, s_{i,j}$. One can also perform \texttt{project} on every label set with spins up to smax all at once by running \texttt{expandresidue(n,R,V,dd=None,smax=None)}. Alternatively, one can compute the expansion coefficients in the $b$-basis defined by Eq.~\eqref{eq:p-basis-3pt} by calling the function \texttt{expandresidue\_p(n, R, V, M2, m2, dd=None, smax=None)}.

All of these functions work regardless of whether $d \geq n - 1$ or not. Additionally, the script itself contains other, less important functions, which one can read about in the ancillary file. 

\section{Data: Half-Ladder Moments at $\alpha_0 = -1$ for Levels $N \leq 6$}\label{sec:data}

Below we give a list of half-ladder moments $\widetilde{m}_k$ at the massive levels $2 \leq N \leq 6$ at the open bosonic point $\alpha_0 = -1$. This follows the strategy given in Sec.~\ref{sec:string-theory} and App.~\ref{sec:hankel}. If an entry is missing, this means it does not couple to the half-ladder.

\allowdisplaybreaks
\subsection*{Level $N=2$}
\medskip
\noindent\underline{$(N,s,b)=(2,0,0)$, $r=1$}
\begin{align*}
\widetilde m_k^{(2,0,0)} &= \Big(- \frac{d - 26}{8 \left(d - 1\right)}\Big)\,\Big(\frac{9 d - 34}{8 \left(d - 1\right)}\Big)^{k}
\end{align*}
\medskip
\noindent\underline{$(N,s,b)=(2,2,0)$, $r=1$}
\begin{align*}
\widetilde m_k^{(2,2,0)} &= \Big(\frac{25}{4 \left(d - 3\right) \left(d - 1\right)}\Big)\,2^{k}
\end{align*}
\medskip
\noindent\underline{$(N,s,b)=(2,2,1)$, $r=1$}
\begin{align*}
\widetilde m_k^{(2,2,1)} &= \Big(\frac{25}{4 \left(d - 3\right) \left(d - 1\right)}\Big)\,\big(-4\big)^{k}
\end{align*}
\medskip
\noindent\underline{$(N,s,b)=(2,2,2)$, $r=1$}
\begin{align*}
\widetilde m_k^{(2,2,2)} &= \Big(\frac{25}{4 \left(d - 3\right) \left(d - 1\right)}\Big)\qquad(\widetilde\lambda=1)
\end{align*}
\subsection*{Level $N=3$}
\medskip
\noindent\underline{$(N,s,b)=(3,1,0)$, $r=1$}
\begin{align*}
\widetilde m_k^{(3,1,0)} &= \Big(- \frac{d - 26}{2 \left(d - 3\right) \left(d + 1\right)}\Big)\,\Big(- \frac{101 d - 34}{48 \left(d + 1\right)}\Big)^{k}
\end{align*}
\medskip
\noindent\underline{$(N,s,b)=(3,1,1)$, $r=1$}
\begin{align*}
\widetilde m_k^{(3,1,1)} &= \Big(- \frac{d - 26}{2 \left(d - 3\right) \left(d + 1\right)}\Big)\,\Big(\frac{35 d - 46}{32 \left(d + 1\right)}\Big)^{k}
\end{align*}
\medskip
\noindent\underline{$(N,s,b)=(3,3,0)$, $r=1$}
\begin{align*}
\widetilde m_k^{(3,3,0)} &= \Big(\frac{27}{\left(d - 3\right) \left(d - 1\right) \left(d + 1\right)}\Big)\,\big(- \frac{4}{3}\big)^{k}
\end{align*}
\medskip
\noindent\underline{$(N,s,b)=(3,3,1)$, $r=1$}
\begin{align*}
\widetilde m_k^{(3,3,1)} &= \Big(\frac{27}{\left(d - 3\right) \left(d - 1\right) \left(d + 1\right)}\Big)\,6^{k}
\end{align*}
\medskip
\noindent\underline{$(N,s,b)=(3,3,2)$, $r=1$}
\begin{align*}
\widetilde m_k^{(3,3,2)} &= \Big(\frac{27}{\left(d - 3\right) \left(d - 1\right) \left(d + 1\right)}\Big)\,\big(-6\big)^{k}
\end{align*}
\medskip
\noindent\underline{$(N,s,b)=(3,3,3)$, $r=1$}
\begin{align*}
\widetilde m_k^{(3,3,3)} &= \Big(\frac{27}{\left(d - 3\right) \left(d - 1\right) \left(d + 1\right)}\Big)\qquad(\widetilde\lambda=1)
\end{align*}
\subsection*{Level $N=4$}
\medskip
\noindent\underline{$(N,s,b)=(4,0,0)$, $r=2$}
\begin{align*}
\widetilde m_k^{(4,0,0)} &= \widetilde w_+\,\widetilde\lambda_+^{\,k} + \widetilde w_-\,\widetilde\lambda_-^{\,k}\\
\widetilde\lambda_\pm &= \frac{825 d^{2} - 2730 d + 1008 \pm \sqrt{Q}}{768 \left(d - 1\right) \left(d + 1\right)},\qquad \widetilde w_\pm = \frac{9 d^{2} - 490 d + 6704 \pm \mathcal C/\sqrt{Q}}{768 \left(d - 1\right) \left(d + 1\right)}\\
Q &= 3249 d^{4} - 94900 d^{3} + 862916 d^{2} - 3553472 d + 16306432\\
\mathcal C &= 513 d^{4} - 39060 d^{3} + 986116 d^{2} - 9156672 d + 22684928
\end{align*}
\medskip
\noindent\underline{$(N,s,b)=(4,2,0)$, $r=2$}
\begin{align*}
\widetilde m_k^{(4,2,0)} &= \widetilde w_+\,\widetilde\lambda_+^{\,k} + \widetilde w_-\,\widetilde\lambda_-^{\,k}\\
\widetilde\lambda_\pm &= \frac{83 d + 464 \pm \sqrt{Q}}{72 \left(d + 3\right)},\qquad \widetilde w_\pm = \frac{6468 - 245 d \mp \big(12887 d^{2} - 237748 d - 2377872\big)/\sqrt{Q}}{192 \left(d - 3\right) \left(d - 1\right) \left(d + 3\right)}\\
Q &= 1657 d^{2} + 43568 d + 162016
\end{align*}
\medskip
\noindent\underline{$(N,s,b)=(4,2,1)$, $r=2$}
\begin{align*}
\widetilde m_k^{(4,2,1)} &= \widetilde w_+\,\widetilde\lambda_+^{\,k} + \widetilde w_-\,\widetilde\lambda_-^{\,k}\\
\widetilde\lambda_\pm &= \frac{- 39 d - 78 \pm \sqrt{Q}}{12 \left(d + 3\right)},\qquad \widetilde w_\pm = \frac{6468 - 245 d \pm \big(2695 d^{2} - 70658 d + 44688\big)/\sqrt{Q}}{192 \left(d - 3\right) \left(d - 1\right) \left(d + 3\right)}\\
Q &= \frac{355 d^{2} + 988 d + 4332}{3}
\end{align*}
\medskip
\noindent\underline{$(N,s,b)=(4,2,2)$, $r=2$}
\begin{align*}
\widetilde m_k^{(4,2,2)} &= \widetilde w_1\,\widetilde\lambda_1^{\,k} + \widetilde w_2\,\widetilde\lambda_2^{\,k}\\
\widetilde\lambda_1 &= \frac{77 d + 62}{72 \left(d + 3\right)},\qquad \widetilde w_1 = - \frac{49 \left(3 d - 82\right)^{2}}{32 \left(d - 3\right) \left(d - 1\right) \left(d + 3\right) \left(5 d - 154\right)}\\
\widetilde\lambda_2 &= 1,\qquad \widetilde w_2 = \frac{49 \left(d - 26\right)}{48 \left(d - 3\right) \left(d - 1\right) \left(5 d - 154\right)}
\end{align*}
\medskip
\noindent\underline{$(N,s,b)=(4,4,0)$, $r=1$}
\begin{align*}
\widetilde m_k^{(4,4,0)} &= \Big(\frac{2401}{16 \left(d - 3\right) \left(d - 1\right) \left(d + 1\right) \left(d + 3\right)}\Big)\,\big(\frac{2}{3}\big)^{k}
\end{align*}
\medskip
\noindent\underline{$(N,s,b)=(4,4,1)$, $r=1$}
\begin{align*}
\widetilde m_k^{(4,4,1)} &= \Big(\frac{2401}{16 \left(d - 3\right) \left(d - 1\right) \left(d + 1\right) \left(d + 3\right)}\Big)\,\big(- \frac{16}{3}\big)^{k}
\end{align*}
\medskip
\noindent\underline{$(N,s,b)=(4,4,2)$, $r=1$}
\begin{align*}
\widetilde m_k^{(4,4,2)} &= \Big(\frac{2401}{16 \left(d - 3\right) \left(d - 1\right) \left(d + 1\right) \left(d + 3\right)}\Big)\,12^{k}
\end{align*}
\medskip
\noindent\underline{$(N,s,b)=(4,4,3)$, $r=1$}
\begin{align*}
\widetilde m_k^{(4,4,3)} &= \Big(\frac{2401}{16 \left(d - 3\right) \left(d - 1\right) \left(d + 1\right) \left(d + 3\right)}\Big)\,\big(-8\big)^{k}
\end{align*}
\medskip
\noindent\underline{$(N,s,b)=(4,4,4)$, $r=1$}
\begin{align*}
\widetilde m_k^{(4,4,4)} &= \Big(\frac{2401}{16 \left(d - 3\right) \left(d - 1\right) \left(d + 1\right) \left(d + 3\right)}\Big)\qquad(\widetilde\lambda=1)
\end{align*}
\subsection*{Level $N=5$}
\medskip
\noindent\underline{$(N,s,b)=(5,1,0)$, $r\ge2$}
\begin{align*}
\widetilde m_0^{(5,1,0)} &= \frac{2 d^{2} -  112 d + 1566}{15 \left(d - 3\right) \left(d + 1\right) \left(d + 3\right)}\\
\widetilde m_1^{(5,1,0)} &= -\frac{1}{460800 \left(d - 3\right) \left(d + 1\right)^{2} \left(d + 3\right)^{2}}\Big(\\
&\qquad 131121 d^{4} -  7166752 d^{3} + 93220742 d^{2}\\
&\qquad + 163764384 d - 965467431\Big)\\
\widetilde m_2^{(5,1,0)} &= \frac{1}{113246208000 \left(d - 3\right) \left(d - 2\right) \left(d + 1\right)^{3} \left(d + 3\right)^{3}}\Big(\\
&\qquad 68631786189 d^{7} -  3778094597480 d^{6}\\
&\qquad + 50046694090402 d^{5} + 90350317031648 d^{4}\\
&\qquad -  1055716252155379 d^{3} + 907701121082904 d^{2}\\
&\qquad + 6946221065749524 d - 11652414878760336\Big)
\end{align*}
\medskip
\noindent\underline{$(N,s,b)=(5,1,1)$, $r\ge2$}
\begin{align*}
\widetilde m_0^{(5,1,1)} &= \frac{2 d^{2} -  112 d + 1566}{15 \left(d - 3\right) \left(d + 1\right) \left(d + 3\right)}\\
\widetilde m_1^{(5,1,1)} &= \frac{1}{184320 \left(d - 3\right) \left(d + 1\right)^{2} \left(d + 3\right)^{2}}\Big(\\
&\qquad 27621 d^{4} -  1555872 d^{3} + 22053182 d^{2}\\
&\qquad -  5474016 d + 10733229\Big)\\
\widetilde m_2^{(5,1,1)} &= \frac{1}{18119393280 \left(d - 3\right) \left(d - 2\right) \left(d + 1\right)^{3} \left(d + 3\right)^{3}}\Big(\\
&\qquad 3050355429 d^{7} -  178773071400 d^{6}\\
&\qquad + 2818062117522 d^{5} -  5950368407072 d^{4}\\
&\qquad + 2903879440421 d^{3} -  1464599674536 d^{2}\\
&\qquad + 30977613745524 d - 29112986097936\Big)
\end{align*}
\medskip
\noindent\underline{$(N,s,b)=(5,3,0)$, $r\ge2$}
\begin{align*}
\widetilde m_0^{(5,3,0)} &= -\frac{16 d - 432}{\left(d - 3\right) \left(d - 1\right) \left(d + 1\right) \left(d + 5\right)}\\
\widetilde m_1^{(5,3,0)} &= \frac{52 d^{2} -  488 d - 23500}{3 \left(d - 3\right) \left(d - 1\right) \left(d + 1\right) \left(d + 5\right)^{2}}\\
\widetilde m_2^{(5,3,0)} &= -\frac{1063 d^{3} + 19764 d^{2} -  842525 d - 10674750}{72 \left(d - 3\right) \left(d - 1\right) \left(d + 1\right) \left(d + 5\right)^{3}}
\end{align*}
\medskip
\noindent\underline{$(N,s,b)=(5,3,1)$, $r\ge2$}
\begin{align*}
\widetilde m_0^{(5,3,1)} &= -\frac{16 d - 432}{\left(d - 3\right) \left(d - 1\right) \left(d + 1\right) \left(d + 5\right)}\\
\widetilde m_1^{(5,3,1)} &= -\frac{274 d^{2} -  5156 d - 57902}{3 \left(d - 3\right) \left(d - 1\right) \left(d + 1\right) \left(d + 5\right)^{2}}\\
\widetilde m_2^{(5,3,1)} &= -\frac{147343 d^{3} -  1531788 d^{2} -  53603253 d - 251220814}{288 \left(d - 3\right) \left(d - 1\right) \left(d + 1\right) \left(d + 5\right)^{3}}
\end{align*}
\medskip
\noindent\underline{$(N,s,b)=(5,3,2)$, $r\ge2$}
\begin{align*}
\widetilde m_0^{(5,3,2)} &= -\frac{16 d - 432}{\left(d - 3\right) \left(d - 1\right) \left(d + 1\right) \left(d + 5\right)}\\
\widetilde m_1^{(5,3,2)} &= \frac{298 d^{2} -  6980 d - 26438}{3 \left(d - 3\right) \left(d - 1\right) \left(d + 1\right) \left(d + 5\right)^{2}}\\
\widetilde m_2^{(5,3,2)} &= -\frac{177439 d^{3} -  3501396 d^{2} -  29978589 d - 56002654}{288 \left(d - 3\right) \left(d - 1\right) \left(d + 1\right) \left(d + 5\right)^{3}}
\end{align*}
\medskip
\noindent\underline{$(N,s,b)=(5,3,3)$, $r\ge2$}
\begin{align*}
\widetilde m_0^{(5,3,3)} &= -\frac{16 d - 432}{\left(d - 3\right) \left(d - 1\right) \left(d + 1\right) \left(d + 5\right)}\\
\widetilde m_1^{(5,3,3)} &= -\frac{17 d^{2} -  410 d - 1319}{\left(d - 3\right) \left(d - 1\right) \left(d + 1\right) \left(d + 5\right)^{2}}\\
\widetilde m_2^{(5,3,3)} &= -\frac{2311 d^{3} -  49044 d^{2} -  340085 d - 519534}{128 \left(d - 3\right) \left(d - 1\right) \left(d + 1\right) \left(d + 5\right)^{3}}
\end{align*}
\medskip
\noindent\underline{$(N,s,b)=(5,5,0)$, $r=1$}
\begin{align*}
\widetilde m_k^{(5,5,0)} &= \Big(\frac{1024}{\left(d - 3\right) \left(d - 1\right) \left(d + 1\right) \left(d + 3\right) \left(d + 5\right)}\Big)\,\big(- \frac{4}{15}\big)^{k}
\end{align*}
\medskip
\noindent\underline{$(N,s,b)=(5,5,1)$, $r=1$}
\begin{align*}
\widetilde m_k^{(5,5,1)} &= \Big(\frac{1024}{\left(d - 3\right) \left(d - 1\right) \left(d + 1\right) \left(d + 3\right) \left(d + 5\right)}\Big)\,\big(\frac{10}{3}\big)^{k}
\end{align*}
\medskip
\noindent\underline{$(N,s,b)=(5,5,2)$, $r=1$}
\begin{align*}
\widetilde m_k^{(5,5,2)} &= \Big(\frac{1024}{\left(d - 3\right) \left(d - 1\right) \left(d + 1\right) \left(d + 3\right) \left(d + 5\right)}\Big)\,\big(- \frac{40}{3}\big)^{k}
\end{align*}
\medskip
\noindent\underline{$(N,s,b)=(5,5,3)$, $r=1$}
\begin{align*}
\widetilde m_k^{(5,5,3)} &= \Big(\frac{1024}{\left(d - 3\right) \left(d - 1\right) \left(d + 1\right) \left(d + 3\right) \left(d + 5\right)}\Big)\,20^{k}
\end{align*}
\medskip
\noindent\underline{$(N,s,b)=(5,5,4)$, $r=1$}
\begin{align*}
\widetilde m_k^{(5,5,4)} &= \Big(\frac{1024}{\left(d - 3\right) \left(d - 1\right) \left(d + 1\right) \left(d + 3\right) \left(d + 5\right)}\Big)\,\big(-10\big)^{k}
\end{align*}
\medskip
\noindent\underline{$(N,s,b)=(5,5,5)$, $r=1$}
\begin{align*}
\widetilde m_k^{(5,5,5)} &= \Big(\frac{1024}{\left(d - 3\right) \left(d - 1\right) \left(d + 1\right) \left(d + 3\right) \left(d + 5\right)}\Big)\qquad(\widetilde\lambda=1)
\end{align*}
\subsection*{Level $N=6$}
\medskip
\noindent\underline{$(N,s,b)=(6,0,0)$, $r\ge2$}
\begin{align*}
\widetilde m_0^{(6,0,0)} &= -\frac{25 d^{3} -  2256 d^{2} + 67196 d - 663168}{5120 \left(d - 1\right) \left(d + 1\right) \left(d + 3\right)}\\
\widetilde m_1^{(6,0,0)} &= -\frac{1}{655360000 \left(d - 1\right)^{2} \left(d + 1\right)^{2} \left(d + 3\right)^{2}}\Big(\\
&\qquad 3705625 d^{6} -  347019000 d^{5} + 11087206600 d^{4}\\
&\qquad -  131944972320 d^{3} + 346565785936 d^{2}\\
&\qquad -  233334008832 d - 1307376451584\Big)\\
\widetilde m_2^{(6,0,0)} &= -\frac{1}{419430400000000 \left(d - 1\right)^{3} \left(d + 1\right)^{3} \left(d + 3\right)^{3}}\Big(\\
&\qquad 2745160048125 d^{9} -  266191766252000 d^{8}\\
&\qquad + 9054129493910500 d^{7} -  124374527316488960 d^{6}\\
&\qquad + 575573573288130928 d^{5} -  1053710572098930176 d^{4}\\
&\qquad -  1113363766480053056 d^{3} + 2628185216072269824 d^{2}\\
&\qquad -  5052170129784569856 d - 12699729253267144704\Big)
\end{align*}
\medskip
\noindent\underline{$(N,s,b)=(6,2,0)$, $r\ge2$}
\begin{align*}
\widetilde m_0^{(6,2,0)} &= \frac{2331 d^{2} -  134442 d + 1926720}{2560 \left(d - 3\right) \left(d - 1\right) \left(d + 3\right) \left(d + 5\right)}\\
\widetilde m_1^{(6,2,0)} &= \frac{1}{20480000 \left(d - 3\right) \left(d - 1\right) \left(d + 3\right)^{2} \left(d + 5\right)^{2}}\Big(\\
&\qquad 34068025 d^{4} -  1452946700 d^{3} -  243825980 d^{2}\\
&\qquad + 375297103680 d + 408425710464\Big)\\
\widetilde m_2^{(6,2,0)} &= \frac{1}{7372800000000 \left(d - 3\right)^{2} \left(d - 1\right)^{2} \left(d + 3\right)^{3} \left(d + 5\right)^{3}}\Big(\\
&\qquad 19999712653175 d^{8} -  565108391494850 d^{7}\\
&\qquad -  13872450685753935 d^{6} + 293805209374944310 d^{5}\\
&\qquad + 2851487352518737148 d^{4} -  7199008999451415960 d^{3}\\
&\qquad -  12292536401819648640 d^{2} -  1000753237458938880 d\\
&\qquad + 16862740035132782592\Big)
\end{align*}
\medskip
\noindent\underline{$(N,s,b)=(6,2,1)$, $r\ge2$}
\begin{align*}
\widetilde m_0^{(6,2,1)} &= \frac{2331 d^{2} -  134442 d + 1926720}{2560 \left(d - 3\right) \left(d - 1\right) \left(d + 3\right) \left(d + 5\right)}\\
\widetilde m_1^{(6,2,1)} &= -\frac{1}{10240000 \left(d - 3\right) \left(d - 1\right) \left(d + 3\right)^{2} \left(d + 5\right)^{2}}\Big(\\
&\qquad 39450225 d^{4} -  2066169300 d^{3} + 21175212180 d^{2}\\
&\qquad + 159709880160 d - 112627209024\Big)\\
\widetilde m_2^{(6,2,1)} &= \frac{1}{204800000000 \left(d - 3\right)^{2} \left(d - 1\right)^{2} \left(d + 3\right)^{3} \left(d + 5\right)^{3}}\Big(\\
&\qquad 3332824337075 d^{8} -  169433558220950 d^{7}\\
&\qquad + 1499389336743885 d^{6} + 18596383499556970 d^{5}\\
&\qquad -  30249732332868748 d^{4} -  201193743616355400 d^{3}\\
&\qquad + 455228137384679520 d^{2} -  609692106325242240 d\\
&\qquad + 357444898502819328\Big)
\end{align*}
\medskip
\noindent\underline{$(N,s,b)=(6,2,2)$, $r\ge2$}
\begin{align*}
\widetilde m_0^{(6,2,2)} &= \frac{2331 d^{2} -  134442 d + 1926720}{2560 \left(d - 3\right) \left(d - 1\right) \left(d + 3\right) \left(d + 5\right)}\\
\widetilde m_1^{(6,2,2)} &= \frac{1}{8192000 \left(d - 3\right) \left(d - 1\right) \left(d + 3\right)^{2} \left(d + 5\right)^{2}}\Big(\\
&\qquad 8252325 d^{4} -  444820500 d^{3} + 5037532740 d^{2}\\
&\qquad + 24261585696 d + 29527364160\Big)\\
\widetilde m_2^{(6,2,2)} &= \frac{1}{131072000000 \left(d - 3\right)^{2} \left(d - 1\right)^{2} \left(d + 3\right)^{3} \left(d + 5\right)^{3}}\Big(\\
&\qquad 145996638075 d^{8} -  7896038060070 d^{7}\\
&\qquad + 88932609214533 d^{6} + 488371204692954 d^{5}\\
&\qquad -  272616667727724 d^{4} -  4196011189497480 d^{3}\\
&\qquad -  3231944137687968 d^{2} -  615066988740480 d\\
&\qquad + 23353147584000\Big)
\end{align*}
\medskip
\noindent\underline{$(N,s,b)=(6,4,0)$, $r\ge2$}
\begin{align*}
\widetilde m_0^{(6,4,0)} &= \frac{- 15309 d + 424278}{128 \left(d - 3\right) \left(d - 1\right) \left(d + 1\right) \left(d + 3\right) \left(d + 7\right)}\\
\widetilde m_1^{(6,4,0)} &= -\frac{686475 d^{2} + 5477220 d - 652689252}{12800 \left(d - 3\right) \left(d - 1\right) \left(d + 1\right) \left(d + 3\right) \left(d + 7\right)^{2}}\\
\widetilde m_2^{(6,4,0)} &= -\frac{1}{1280000 \left(d - 3\right) \left(d - 1\right) \left(d + 1\right) \left(d + 3\right) \left(d + 7\right)^{3}}\Big(\\
&\qquad 12786525 d^{3} + 1950527790 d^{2} -  21828022884 d\\
&\qquad - 1057180787352\Big)
\end{align*}
\medskip
\noindent\underline{$(N,s,b)=(6,4,1)$, $r\ge2$}
\begin{align*}
\widetilde m_0^{(6,4,1)} &= \frac{- 15309 d + 424278}{128 \left(d - 3\right) \left(d - 1\right) \left(d + 1\right) \left(d + 3\right) \left(d + 7\right)}\\
\widetilde m_1^{(6,4,1)} &= \frac{856575 d^{2} -  7494120 d - 439336224}{1600 \left(d - 3\right) \left(d - 1\right) \left(d + 1\right) \left(d + 3\right) \left(d + 7\right)^{2}}\\
\widetilde m_2^{(6,4,1)} &= -\frac{1}{20000 \left(d - 3\right) \left(d - 1\right) \left(d + 1\right) \left(d + 3\right) \left(d + 7\right)^{3}}\Big(\\
&\qquad 42233400 d^{3} + 701192295 d^{2} -  32982747696 d\\
&\qquad - 474595080768\Big)
\end{align*}
\medskip
\noindent\underline{$(N,s,b)=(6,4,2)$, $r\ge2$}
\begin{align*}
\widetilde m_0^{(6,4,2)} &= \frac{- 15309 d + 424278}{128 \left(d - 3\right) \left(d - 1\right) \left(d + 1\right) \left(d + 3\right) \left(d + 7\right)}\\
\widetilde m_1^{(6,4,2)} &= -\frac{8802675 d^{2} -  156064320 d - 2351234952}{6400 \left(d - 3\right) \left(d - 1\right) \left(d + 1\right) \left(d + 3\right) \left(d + 7\right)^{2}}\\
\widetilde m_2^{(6,4,2)} &= -\frac{1}{320000 \left(d - 3\right) \left(d - 1\right) \left(d + 1\right) \left(d + 3\right) \left(d + 7\right)^{3}}\Big(\\
&\qquad 5015228400 d^{3} -  38241871065 d^{2} -  2200156187184 d\\
&\qquad - 13095363599712\Big)
\end{align*}
\medskip
\noindent\underline{$(N,s,b)=(6,4,3)$, $r\ge2$}
\begin{align*}
\widetilde m_0^{(6,4,3)} &= \frac{- 15309 d + 424278}{128 \left(d - 3\right) \left(d - 1\right) \left(d + 1\right) \left(d + 3\right) \left(d + 7\right)}\\
\widetilde m_1^{(6,4,3)} &= \frac{3152925 d^{2} -  69823620 d - 453534228}{3200 \left(d - 3\right) \left(d - 1\right) \left(d + 1\right) \left(d + 3\right) \left(d + 7\right)^{2}}\\
\widetilde m_2^{(6,4,3)} &= -\frac{1}{80000 \left(d - 3\right) \left(d - 1\right) \left(d + 1\right) \left(d + 3\right) \left(d + 7\right)^{3}}\Big(\\
&\qquad 648852525 d^{3} -  10692017010 d^{2} -  168979076964 d\\
&\qquad - 504119913192\Big)
\end{align*}
\medskip
\noindent\underline{$(N,s,b)=(6,4,4)$, $r\ge2$}
\begin{align*}
\widetilde m_0^{(6,4,4)} &= \frac{- 15309 d + 424278}{128 \left(d - 3\right) \left(d - 1\right) \left(d + 1\right) \left(d + 3\right) \left(d + 7\right)}\\
\widetilde m_1^{(6,4,4)} &= -\frac{3225825 d^{2} -  73526940 d - 438388524}{25600 \left(d - 3\right) \left(d - 1\right) \left(d + 1\right) \left(d + 3\right) \left(d + 7\right)^{2}}\\
\widetilde m_2^{(6,4,4)} &= -\frac{1}{5120000 \left(d - 3\right) \left(d - 1\right) \left(d + 1\right) \left(d + 3\right) \left(d + 7\right)^{3}}\Big(\\
&\qquad 679446225 d^{3} -  12133322910 d^{2} -  168421106196 d\\
&\qquad - 454720042728\Big)
\end{align*}
\medskip
\noindent\underline{$(N,s,b)=(6,6,0)$, $r=1$}
\begin{align*}
\widetilde m_k^{(6,6,0)} &= \Big(\frac{531441}{64 \left(d - 3\right) \left(d - 1\right) \left(d + 1\right) \left(d + 3\right) \left(d + 5\right) \left(d + 7\right)}\Big)\,\big(\frac{4}{45}\big)^{k}
\end{align*}
\medskip
\noindent\underline{$(N,s,b)=(6,6,1)$, $r=1$}
\begin{align*}
\widetilde m_k^{(6,6,1)} &= \Big(\frac{531441}{64 \left(d - 3\right) \left(d - 1\right) \left(d + 1\right) \left(d + 3\right) \left(d + 5\right) \left(d + 7\right)}\Big)\,\big(- \frac{8}{5}\big)^{k}
\end{align*}
\medskip
\noindent\underline{$(N,s,b)=(6,6,2)$, $r=1$}
\begin{align*}
\widetilde m_k^{(6,6,2)} &= \Big(\frac{531441}{64 \left(d - 3\right) \left(d - 1\right) \left(d + 1\right) \left(d + 3\right) \left(d + 5\right) \left(d + 7\right)}\Big)\,10^{k}
\end{align*}
\medskip
\noindent\underline{$(N,s,b)=(6,6,3)$, $r=1$}
\begin{align*}
\widetilde m_k^{(6,6,3)} &= \Big(\frac{531441}{64 \left(d - 3\right) \left(d - 1\right) \left(d + 1\right) \left(d + 3\right) \left(d + 5\right) \left(d + 7\right)}\Big)\,\big(- \frac{80}{3}\big)^{k}
\end{align*}
\medskip
\noindent\underline{$(N,s,b)=(6,6,4)$, $r=1$}
\begin{align*}
\widetilde m_k^{(6,6,4)} &= \Big(\frac{531441}{64 \left(d - 3\right) \left(d - 1\right) \left(d + 1\right) \left(d + 3\right) \left(d + 5\right) \left(d + 7\right)}\Big)\,30^{k}
\end{align*}
\medskip
\noindent\underline{$(N,s,b)=(6,6,5)$, $r=1$}
\begin{align*}
\widetilde m_k^{(6,6,5)} &= \Big(\frac{531441}{64 \left(d - 3\right) \left(d - 1\right) \left(d + 1\right) \left(d + 3\right) \left(d + 5\right) \left(d + 7\right)}\Big)\,\big(-12\big)^{k}
\end{align*}
\medskip
\noindent\underline{$(N,s,b)=(6,6,6)$, $r=1$}
\begin{align*}
\widetilde m_k^{(6,6,6)} &= \Big(\frac{531441}{64 \left(d - 3\right) \left(d - 1\right) \left(d + 1\right) \left(d + 3\right) \left(d + 5\right) \left(d + 7\right)}\Big)\qquad(\widetilde\lambda=1)
\end{align*}

\bibliographystyle{apsrev4-2}
\bibliography{GeneralBibliography}
\end{document}